\documentclass[a4paper,fleqn,usenatbib]{mnras}

\usepackage{newtxtext,newtxmath}
\usepackage[T1]{fontenc}
\usepackage{ae,aecompl}
\usepackage{pdflscape}

\usepackage{graphicx}	% Including figure files
\usepackage{amsmath}	% Advanced maths commands
\usepackage{amssymb}	% Extra maths symbols

\hypersetup{draft}		% prevents crash/failure to compile when reference spans pages and a two-column figure is also present (see https://tex.stackexchange.com/questions/235700/citation-link-not-split-for-page-break-again/368731#368731)
\usepackage{multirow,bigdelim}
\usepackage{chemformula}
\usepackage{adjustbox}

\usepackage[colorinlistoftodos,color=blue!20]{todonotes}

\newcommand\kms{km\,s$^{-1}$}
\def\e#1{$\times 10^{#1}$}

\newcommand\msol{{M}$_\odot$}

\newcommand\spy{\;\msol~yr$^{-1}$}
\def\so2{SO$_2$}
\def\h2{H$_2$}

\title[ALMA view of SO and SO$_2$ around O-rich AGB stars]{An ALMA view of SO and SO$_2$ around oxygen-rich AGB stars}

\author[T. Danilovich et al.]{
T. Danilovich,$^{1}$\thanks{E-mail: taissa.danilovich@kuleuven.be}\thanks{Postdoctoral Fellow of the Fund for Scientific Research (FWO), Flanders, Belgium}
A. M. S. Richards,$^{2}$
L. Decin,$^{1,3}$
M. Van de Sande$^{1}$
and
C. A. Gottlieb$^{4}$
\\
$^{1}$Department of Physics and Astronomy, Institute of Astronomy, KU Leuven, Celestijnenlaan 200D, 3001 Leuven, Belgium  \\
$^{2}$JBCA, Department Physics and Astronomy, University of Manchester, Manchester M13 9PL, UK\\
$^{3}$School of Chemistry, University of Leeds, Leeds LS2 9JT, UK\\
$^{4}$Harvard-Smithsonian Center for Astrophysics, 60 Garden Street, Cambridge, MA 02138, USA
}

\date{Accepted 2019 March 9. Received 2020 March 9; in original form 2019 September 16}

\pubyear{2020}

\begin{document}
\label{firstpage}
\pagerange{\pageref{firstpage}--\pageref{lastpage}}
\maketitle

\begin{abstract} %250 words max
We present and analyse SO and \so2, recently observed with high angular resolution and sensitivity in a spectral line survey with ALMA, for two oxygen-rich AGB stars: the low mass-loss rate R Dor and high mass-loss rate IK Tau. We analyse 8 lines of SO detected towards both stars, 78 lines of \so2 detected towards R Dor and 52 lines of \so2 detected towards IK Tau. We detect several lines of $^{34}$SO, $^{33}$SO and $^{34}$\so2 towards both stars, and tentatively S$^{18}$O towards R Dor, and hence derive isotopic ratios for these species. The spatially resolved observations show us that the two sulphur oxides are co-located towards R Dor and trace out the same wind structures in the { circumstellar envelope} (CSE). Much of the emission is well reproduced with a Gaussian abundance distribution { spatially centred on the star}. Emission from the higher energy levels of SO and \so2 towards R Dor provide evidence in support of a rotating inner region of gas identified in earlier work. The new observations allow us to refine the abundance distribution of SO in IK Tau derived from prior observations with single antennas, and confirm the distribution is shell-like with the peak in the fractional abundance not centred on the star.
%in the CSE, rather than close to the star as in R Dor. 
{ The confirmation of different types of SO abundance distributions will help fine-tune chemical models and allows for an additional method to discriminate between low and high mass-loss rates for oxygen-rich AGB stars.}
\end{abstract}


%%%%%%%%%%%%%%%%%%%%%%%%%%%%%%%%%%%%%%%%%%%%%%%%%%

%%%%%%%%%%%%%%%%% BODY OF PAPER %%%%%%%%%%%%%%%%%%

\section{Introduction}

Upon leaving the main sequence and passing through the red giant branch, low to intermediate mass stars \citep[$\sim0.8$ -- 8~\msol,][]{Hofner2018} enter the asymptotic giant branch (AGB). During the AGB phase, stars lose mass rapidly, eject matter in a stellar wind, and form an expanding circumstellar envelope (CSE) where molecules and dust are produced. The chemical type of an AGB star is classified according to the  photospheric C/O ratio, which in turn strongly affects the molecular composition of the CSE. The C/O ratio is  < 1 in oxygen-rich stars, > 1 in carbon-rich stars, and $\sim 1$ in S-type stars. Carbon-rich CSEs contain a variety of carbon-bearing molecules and oxygen-rich CSEs contain a greater { variety} of oxygen-bearing molecules { than do the other two chemical types}.

Spectral line observations of the rich molecular environment of the CSEs of AGB stars 
%at high angular resolution and sensitivity with millimeter-wave interferometers 
reveal much about the physical and dynamical conditions in these regions \citep{Hofner2018}.
In the past few years sulphur bearing molecules in the CSE of oxygen-rich AGB stars have been the subject of several extensive papers \citep[][\citeyear{Danilovich2019}]{Danilovich2016,Danilovich2017a,Danilovich2018}.
Sulphur is not synthesised in AGB stars or in their main sequence progenitors, but instead is produced in massive stars through 
oxygen burning and in Type II supernovae. 
Because sulphur is not depleted onto circumstellar dust in significant quantities, { as evidenced by the near-solar abundances seen in post-AGB stars, \citep{Waelkens1991,Reyniers2007,Kamath2019}, we can estimate the total sulphur abundance of nearby stars by assuming the solar abundance.}
The sulphur-bearing molecules that are most commonly observed in the CSE with radio telescopes are CS, SiS, SO, SO$_2$, and \h2S.
{ CS is most commonly seen towards carbon stars, SiS and \h2S are most commonly seen towards the highest mass-loss rate stars of all chemical types \citep{Danilovich2017a,Danilovich2018}, and SO and \so2 have only been detected towards oxygen-rich stars \citep[see][and references therein]{Danilovich2016}}.

Rotational lines of SO and \so2 are especially prominent towards oxygen-rich AGB stars.
In their study of SO and \so2 observed with the {\sl Herschel} and APEX single antennas, \cite{Danilovich2016} found that the { spatial }abundance distribution of SO towards five oxygen-rich stars stars (IK~Tau, R~Dor, TX~Cam, W~Hya, and R~Cas) differs between stars with low ($1-2 \times 10^{-7}$~\msol~yr$^{-1}$) and higher ($8-50  \times 10^{-7}$~\msol~yr$^{-1}$) mass-loss rates.
The circumstellar emission of SO in low mass-loss rate stars was well reproduced with a Gaussian abundance distribution centred on the star, whereas the relative abundance of SO in stars with higher mass-loss rates was greatest in the CSE some distance from the central star --- i.e., the abundance distribution is shell-like. Furthermore, the peak relative SO abundance for the higher mass-loss rate stars was lower than for the lower mass-loss rate stars.
The { radial }abundance distribution of SO in both high and low mass-loss rate stars correlated with the peak in the OH abundance derived from the \h2O abundance distribution and photodissociation radius.
The difference in the shapes of the SO abundance distributions (centred on the star vs shell-like) was attributed to OH which drives the formation of SO by the reaction of OH with S. Owing to the limitations of the observations, \cite{Danilovich2016} were unable to directly determine whether the abundance distribution of \so2 is similar to that of SO.
From the observations of H$_2$S towards five AGB stars with high mass-loss rates of 
$(5 - 40) \times 10^{-6}$~\msol~yr$^{-1}$ with APEX, \cite{Danilovich2017a} concluded H$_2$S could account for a significant fraction of the sulphur budget in the highest mass-loss rate sources, which may { explain} the lower SO abundances in these sources.
 
{ These significant chemical differences seem to be density-dependent, since mass-loss rate correlates with gas density. Despite this, any chemical network should be able to reproduce the observed abundances of sulphur molecules for all AGB stars. Hence, a more detailed understanding of how sulphur molecules behave will allow us to fine-tune chemical networks, which can, in turn, be applied to other astrophysical environments. Furthermore, the dependence of the abundances of these molecules on mass-loss rate means that they can act as a secondary diagnostic of mass-loss rate, especially in situations where there is some significant uncertainty (for example if the distance is not known).
%-fine-tune chemical network (which should be trasferrable to other astrophysical environments)
%- secondary diagnostic of mass-loss rate
}

{ Until recently, there were no spatially resolved observations comparing SO and \so2 towards low and high mass-loss rate AGB stars. Such observations ought to provide additional information about potential three-dimensional structural differences between sulphur oxides in low and high mass-loss rate AGB CSEs, beyond the one-dimensional difference in radial abundance distribution that was already found by \cite{Danilovich2016}.}
{A recently published spectral scan covering frequencies 335--362~GHz for R~Dor and IK~Tau \citep{Decin2018}
detected approximately 200 rotational lines towards R~Dor, and 168 lines towards IK~Tau from at least 15 molecular species %and some unidentified carriers 
--- including many from sulphur-bearing molecules. }
R~Dor has a low mass-loss rate of 1.6\e{-7}\spy{} and IK~Tau has a mass-loss rate about 30~times higher at 5.0\e{-6}\spy{} \citep{Maercker2016}.
{ Transition lines of the main isotopic species of SO and \so2 make up} roughly 40\% of the lines towards both stars. NS was also observed towards IK~Tau, and CS and SiS were observed towards both stars and analysed by \cite{Danilovich2019}
who derived the corresponding abundance distributions for R~Dor and IK~Tau. 
{
% Given the wealth of data in the literature for these two stars, they are ideal for testing 
%-fine-tune chemical network (which should be trasferrable to other astrophysical environments)
%- secondary diagnostic of mass-loss rate}
}

The emphasis in this paper is on the spatial distribution of the sulphur oxides, SO and \so2, towards R~Dor and IK~Tau \citep[observed by][]{Decin2018} and the determination of the abundance distributions with a radiative transfer model.{ The model was }previously used to analyse the two sulphur oxides in five oxygen-rich AGB stars observed at low angular resolution with single antennas by \cite{Danilovich2016}.
The line identifications of the normal and rare isotopic species of SO  and \so2 are presented in Sect.~\ref{obs}; 
the radiative transfer model is briefly described in Sect.~\ref{sect:mod}; 
the analysis of the spectra and maps of R~Dor and IK~Tau in Sections~\ref{rdoranalysis} and \ref{iktauanalysis}; 
a detailed comparison of the abundance distributions of both molecules towards the two sources, and the results of recent chemical models are discussed in Sect.~\ref{coloc}; 
the derived isotopic ratios are presented in Sect.~\ref{isotopediscussion}; 
and supporting tables and spectra of SO and \so2 are given in Appendices~\ref{app:soplots} and \ref{app:so2plots}.

\section{Observations}\label{obs}

The spectral line survey in Band~7 of R~Dor and IK~Tau between $335-362$~GHz was done by \cite{Decin2018} 
in August and September of 2015 (proposal 2013.1.00166.S, PI: L. Decin). 
The interferometer baselines of $0.04-1.6$~km allowed for imaging of structures with an angular resolution of $\sim150$~mas and angular scales of up to 2\arcsec{}.  
{ The channel $\sigma$ rms noise varied between spectral windows, as explained in \cite{Decin2018}. For IK~Tau this sensitivity range was 3--9 mJy and for R Dor it was 2.7--5.7 mJy, with the higher noise occurring towards the top end of the frequency range.}
As a result of the data reduction, the absolute flux scale is uncertain by up to 10\% but the relative flux scale and astrometry 
for each star is very well registered for all species, such that the map noise is the main source of uncertainty in the flux.
Eight transitions in the ground vibrational state and three in the vibrationally excited level of SO were observed towards R~Dor and IK~Tau, 
75 transitions of \so2 were observed towards R~Dor, 54 lines of \so2 were observed towards IK~Tau;
a few lines of \so2 in the $v_2 =1$ excited vibrational level were observed towards both stars,
the rare isotopic species $^{33}$SO and $^{34}$SO were observed in both stars,
and $^{34}$SO$_2$ was observed in R~Dor.

{
When we discuss 1D spectra of these ALMA observations, we refer to an aperture size, which is the size of a circular region, centred on the continuum peak, for which the spectrum has been extracted. Smaller apertures allow us to more easily examine spectra from regions close to the star, while larger apertures ensure more of the extended emission is included in the spectrum. Spectra extracted for larger apertures also have the tendency to be noisier, since larger noise-dominated areas (with low or no emission) are likely to be included.
The map rms on small scales close to 'cleaned' sources is dominated by thermal noise.  However,  on scales approaching the largest angular scale imageable (2\arcsec), partially resolved-out flux, including from dust continuum causes large-scale error and contributes to errors in the cleaning process.
}

\subsection{SO}\label{so_overview}

\begin{table}
\caption{SO transitions detected with ALMA.}\label{solines}
\begin{adjustbox}{width=\columnwidth}
\begin{tabular}{cccrcc}
\hline\hline
  & Transition & Frequency && $E_\mathrm{up}$ & Star\\
 & $N^{u}_{J^u}\to N^{l}_{J^l}$ & [GHz] && [K] &\\
\hline
    SO & $11_{10} \to 10_{10}$, $v=0$ &  336.5533$^a$ && 143 & Both\\
       & $ 8_{ 7} \to  7_{ 6}$, $v=1$ & \phantom{*}337.8862*$^a$ && 1681 & Both \\
       & $ 3_{ 3} \to  2_{ 3}$, $v=0$ & 339.3415$^b$ && 26 & Both \\
       & $ 8_{ 7} \to  7_{ 6}$, $v=0$ & 340.7142$^b$ && 81 & Both \\
       &  $ 8_{ 8} \to  7_{ 7}$, $v=1$ & 341.5591$^a$ && 1688 & Both \\
       & $ 8_{ 9} \to  7_{ 8}$, $v=1$ & 343.8285$^a$ && 1679 & Both\\
       & $ 8_{ 8} \to  7_{ 7}$, $v=0$ & 344.3106$^b$ && 88 & Both \\
       & $ 8_{ 9} \to  7_{ 8}$, $v=0$ & \phantom{*}346.5285*$^b$ && 79 & Both\\
%      \multicolumn{4}{c}{$^{33}$SO: 11} \\
\hline
\smallskip
$^{33}$SO & $ 8_{ 7} \to  7_{ 6}$, $F=\frac{15}{2}\to\frac{13}{2}$  & \phantom{*}337.1978*$^a$ & \rdelim\}{11.5}{1pt} & \\\smallskip
 & \phantom{$ 8_{ 7} \to  7_{ 6}$,} $F=\frac{13}{2}\to\frac{11}{2}$  & 337.1980$^a$ \\
\smallskip
 & \phantom{$ 8_{ 7} \to  7_{ 6}$,} $F=\frac{17}{2}\to\frac{15}{2}$  &337.1986$^a$ \\\smallskip
 & \phantom{$ 8_{ 7} \to  7_{ 6}$,} $F=\frac{11}{2}\to\frac{9}{2}$  &337.1994$^a$  \\\smallskip
 & \phantom{$ 8_{ 7} \to  7_{ 6}$,} $F=\frac{11}{2}\to\frac{11}{2}$  &337.2462$^a$ &&81 & Both\\ \smallskip
 & \phantom{$ 8_{ 7} \to  7_{ 6}$,} $F=\frac{13}{2}\to\frac{13}{2}$  &337.2498$^a$  \\\smallskip
 & \phantom{$ 8_{ 7} \to  7_{ 6}$,} $F=\frac{15}{2}\to\frac{15}{2}$  &337.2528$^a$  \\\smallskip
 & \phantom{$ 8_{ 7} \to  7_{ 6}$,} $F=\frac{11}{2}\to\frac{13}{2}$  &337.2979$^a$ \\\smallskip
 & \phantom{$ 8_{ 7} \to  7_{ 6}$,} $F=\frac{13}{2}\to\frac{15}{2}$  &337.3048$^a$ \\\smallskip
 & $ 8_{ 8} \to  7_{ 7}$, $F=\frac{13}{2}\to\frac{11}{2}$ &     340.8373$^a$ &\rdelim\}{9}{1pt} &   \\\smallskip
 & \phantom{$ 8_{ 8} \to  7_{ 7}$,} $F=\frac{15}{2}\to\frac{13}{2}$ &     340.8379$^a$ &&   \\\smallskip
 & \phantom{$ 8_{ 8} \to  7_{ 7}$,} $F=\frac{17}{2}\to\frac{15}{2}$ &     340.8387$^a$ &&   \\\smallskip
 & \phantom{$ 8_{ 8} \to  7_{ 7}$,} $F=\frac{19}{2}\to\frac{17}{2}$ &     340.8396$^a$ & & 87 & Both \\\smallskip
 & \phantom{$ 8_{ 8} \to  7_{ 7}$,} $F=\frac{17}{2}\to\frac{17}{2}$ &     340.8417$^a$ &&   \\\smallskip
 & \phantom{$ 8_{ 8} \to  7_{ 7}$,} $F=\frac{15}{2}\to\frac{15}{2}$ &     340.8446$^a$ & &  \\\smallskip
 & \phantom{$ 8_{ 8} \to  7_{ 7}$,} $F=\frac{13}{2}\to\frac{13}{2}$ &     340.8463$^a$ &  & \\\smallskip
 &$ 8_{ 9} \to  7_{ 8}$, $F=\frac{19}{2}\to\frac{19}{2}$& 343.0325$^a$ & \rdelim\}{9}{1pt}&\\\smallskip
 &\phantom{$ 8_{ 9} \to  7_{ 8}$,} $F=\frac{17}{2}\to\frac{17}{2}$& 343.0418$^a$\\\smallskip
 &\phantom{$ 8_{ 9} \to  7_{ 8}$,} $F=\frac{15}{2}\to\frac{15}{2}$& 343.0492$^a$\\\smallskip
 & \phantom{$ 8_{ 9} \to  7_{ 8}$,} $F=\frac{15}{2}\to\frac{13}{2}$ &     343.0861$^a$ &&  78 & R Dor \\\smallskip
 & \phantom{$ 8_{ 9} \to  7_{ 8}$,} $F=\frac{17}{2}\to\frac{15}{2}$ &     343.0873$^a$ & & &IK Tau* \\\smallskip
 & \phantom{$ 8_{ 9} \to  7_{ 8}$,} $F=\frac{19}{2}\to\frac{17}{2}$ &     343.0881$^a$ &  & & \\\smallskip
 & \phantom{$ 8_{ 9} \to  7_{ 8}$,} $F=\frac{21}{2}\to\frac{19}{2}$ &     343.0883$^a$ & & &  \\
%\hline
%\multicolumn{4}{c}{$^{34}$SO: 2} \\
\hline
$^{34}$SO & $ 8_{ 8} \to  7_{ 7}$ &     337.5801$^a$ && 86 & Both\\
 & $ 8_{ 9} \to  7_{ 8}$ &     339.8576$^a$ && 77 & Both \\
\hline
S$^{18}$O & $9_8 \to 8_7$ & 355.5711$^a$ & &93 & R Dor$^\dagger$ \\
 & $9_9 \to 8_8$ & 358.6457$^a$ & &99 & Neither \\
 & $9_{10} \to 8_9$ & 360.6379$^a$ && 91 & R Dor$^\dagger$ \\
\hline
\end{tabular}
\end{adjustbox}
\textbf{Notes:} An * indicates lines participating in overlaps and a $^\dagger$ indicates a tentative detection. Curly braces indicate (unresolved) hyperfine components. All listed isotopologue lines are in the ground vibrational state $v=0$. Frequency references: ($^a$) Calculation method from \cite{Amano1974}, calculated frequencies from CDMS, \cite{Muller2001,Muller2005}; ($^b$) Measured frequencies from \cite{Clark1976}.
\end{table}	

Generally speaking, the brightest transition lines of SO are those which follow the rule $N-J = N'-J'$ for an allowed transition $N_J \to N'_{J'}$ { where $J$ is the total angular momentum excluding nuclear spin and $N$ is the total angular momentum including electronic spin \citep{Hartquist1998}}. In the survey range, the three brightest SO lines are ($8_7\to7_6$), ($8_8\to7_7$), and ($8_7\to7_8$) { in the ground vibrational state, $v=0$,} with frequencies listed in Table \ref{solines}. In addition to these three lines, we also detected their rotational counterparts in the first vibrationally excited state, $v=1$, and two fainter lines in the ground vibrational state: ($11_{10}\to10_{10}$) and ($3_3\to2_3$). The frequencies for all these lines are listed in Table \ref{solines}. 

\begin{figure*}
\centering
\includegraphics[width=\textwidth]{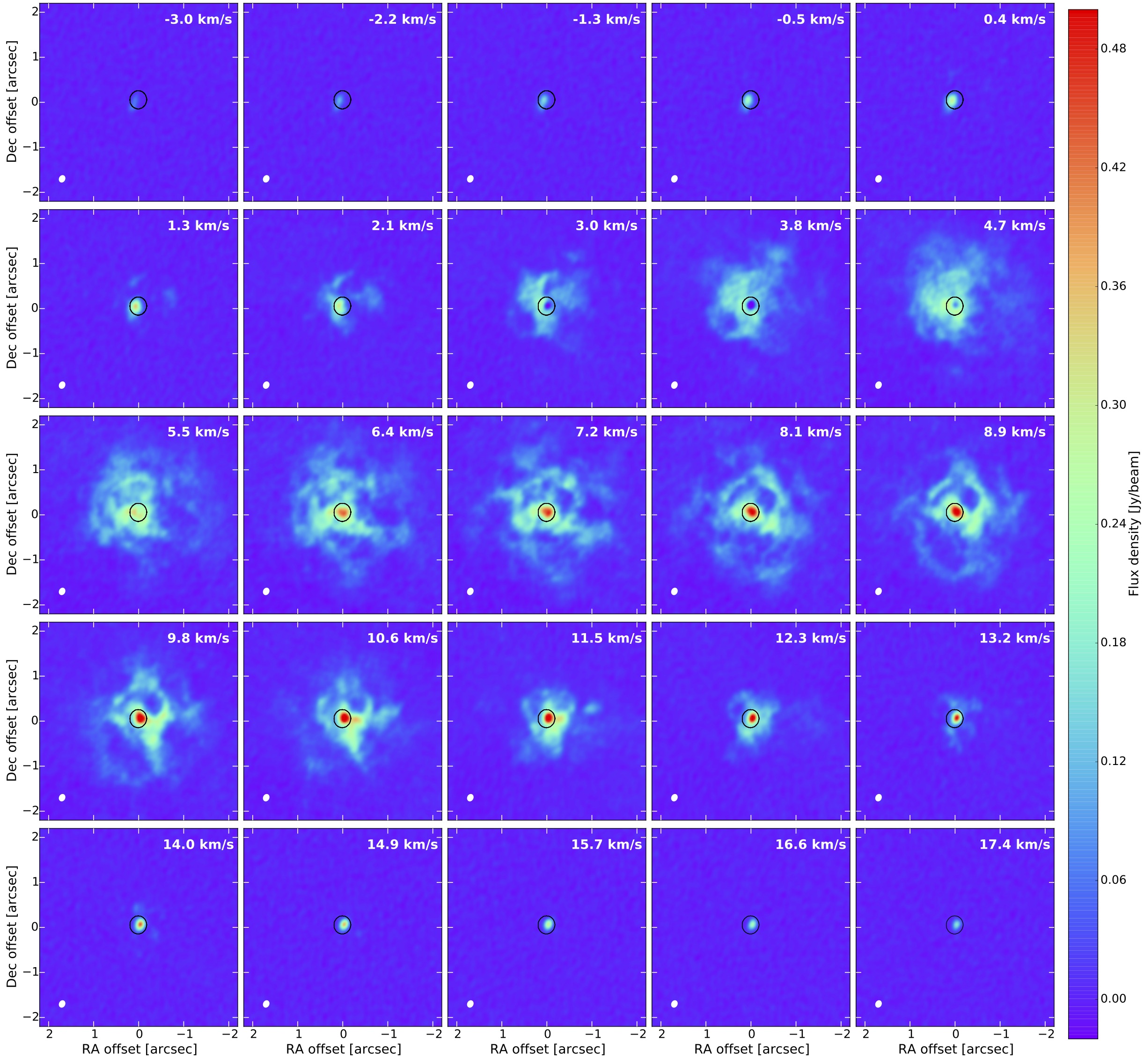} %final is pdf
\caption{Channel maps of SO ($8_8\to7_7$) towards R~Dor. The black contour is drawn for 1\% of the peak continuum flux and the beam is shown in white in the bottom left hand corners of each channel plot. Plots are best viewed on a screen.}
\label{rdorsochannels}
\end{figure*}

\begin{figure*}
\centering
\includegraphics[width=\textwidth]{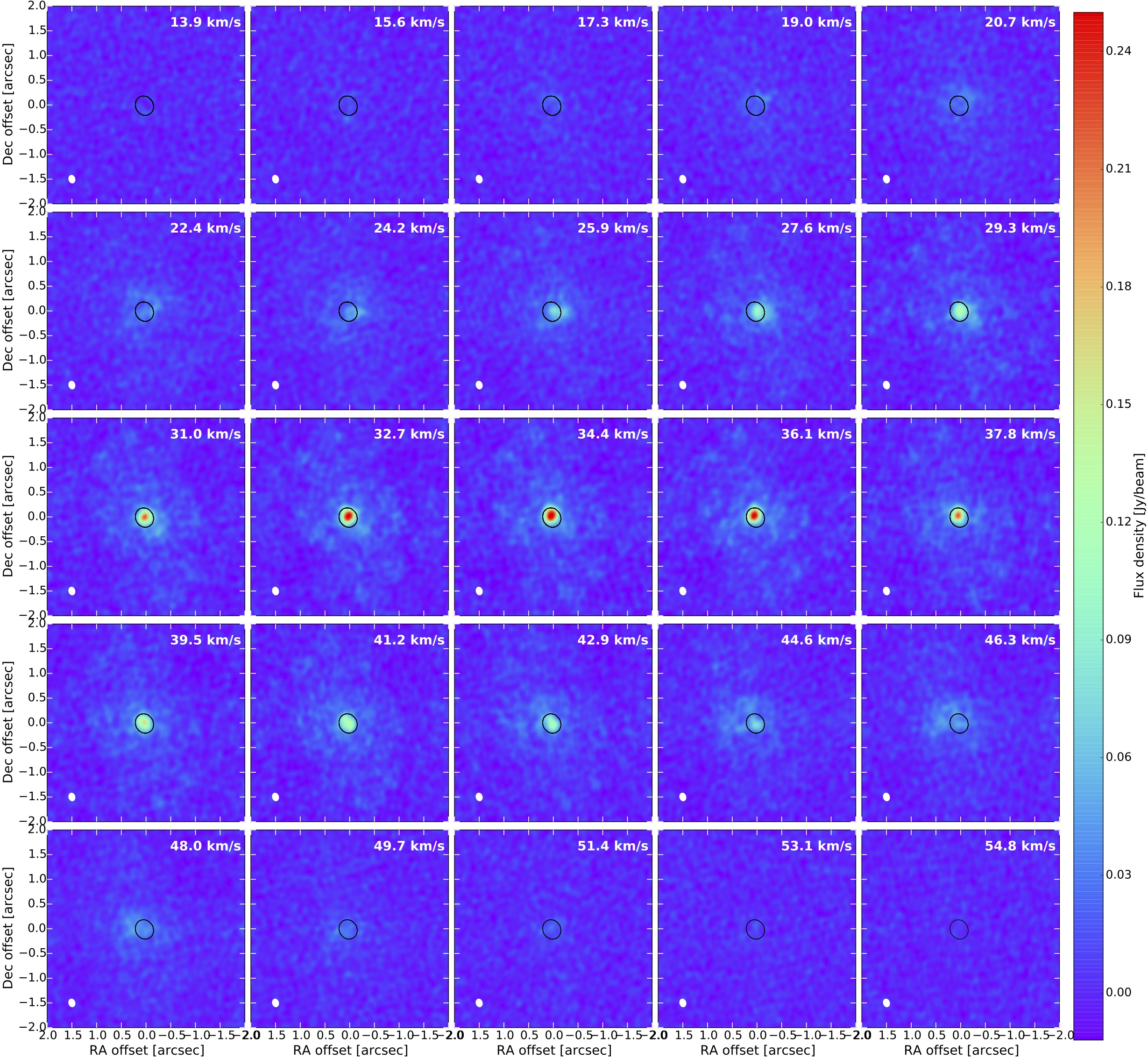} %final is pdf
\caption{Channel maps of SO ($8_8\to7_7$) towards IK~Tau. The black contour is drawn for 5\% of the peak continuum flux and the beam is shown in white in the bottom left hand corners of each channel plot. Plots are best viewed on a screen.}
\label{iktausochannels}
\end{figure*}

In Figures \ref{rdorsochannels} and \ref{iktausochannels} we show the channel maps for ($8_8\to7_7$), as a representative SO line, for R~Dor and IK~Tau, respectively. There is clear extended emission in both sets of channel maps, with the central emission peaking within the indicated continuum contours. 
{ R Dor shows an absorption feature towards the stellar position at a blue-shifted velocity of 3--4~\kms with respect to the LSR velocity.  This is evident due to the large angular size of R Dor,  60 mas \citep{Bedding1998a,Norris2012}, which is almost half the size of the 150~mas restoring beam. Absorption is not seen towards IK Tau, which has an angular diameter of 20 mas \citep{Decin2010a}, much smaller than the 170~mas restoring beam, so any absorption is blended with the surrounding emission.}
%There is a blue absorption feature in R~Dor at a velocity of 3--4 \kms with respect to the $V_{\rm{LSR}}$ (Fig.~\ref{rdorsochannels}), owing to the large ratio between the stellar angular diameter and the angular beam size \citep[see][]{Decin2018}, but the blue hole feature it is not visible in IK~Tau, because the star is more distant and has a smaller apparent stellar diameter. 
{ Structure can be seen in the SO emission towards R~Dor, with some higher flux density arcs/loops and lower flux density regions visible, especially in the central velocity channels from $\sim 5.5$~\kms to $\sim 10.6$~\kms, although there are some asymmetric features seen out to 1.3~\kms on the blue side and 14~\kms on the red side of the LSR velocity. Most of these arc-like features extend out to around 1.5\arcsec{} from the continuum peak, with lower-density regions located non-uniformly around the continuum peak at distances of around 0.5--1\arcsec{} from the centre.
In the case of IK~Tau, however, the SO emission shown in Fig. \ref{iktausochannels} is more uniform and diffuse{, extending up to 2\arcsec{} in the channels close to the LSR velocity of 34\kms, but with no clear structures visible above the noise}.}

\begin{figure}
%\minipage{12cm}
\centering
\includegraphics[width=0.35\textwidth]{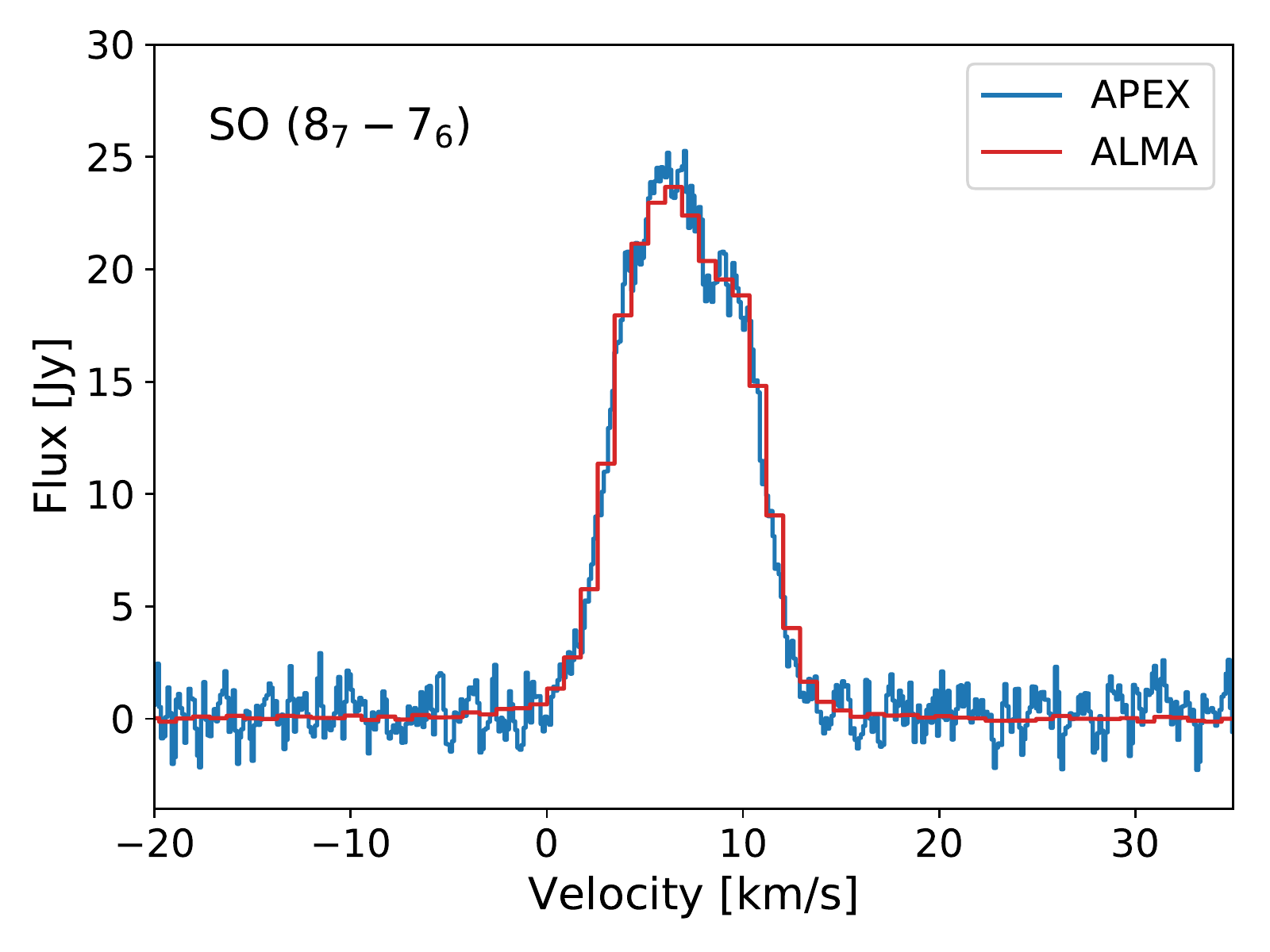}
\includegraphics[width=0.35\textwidth]{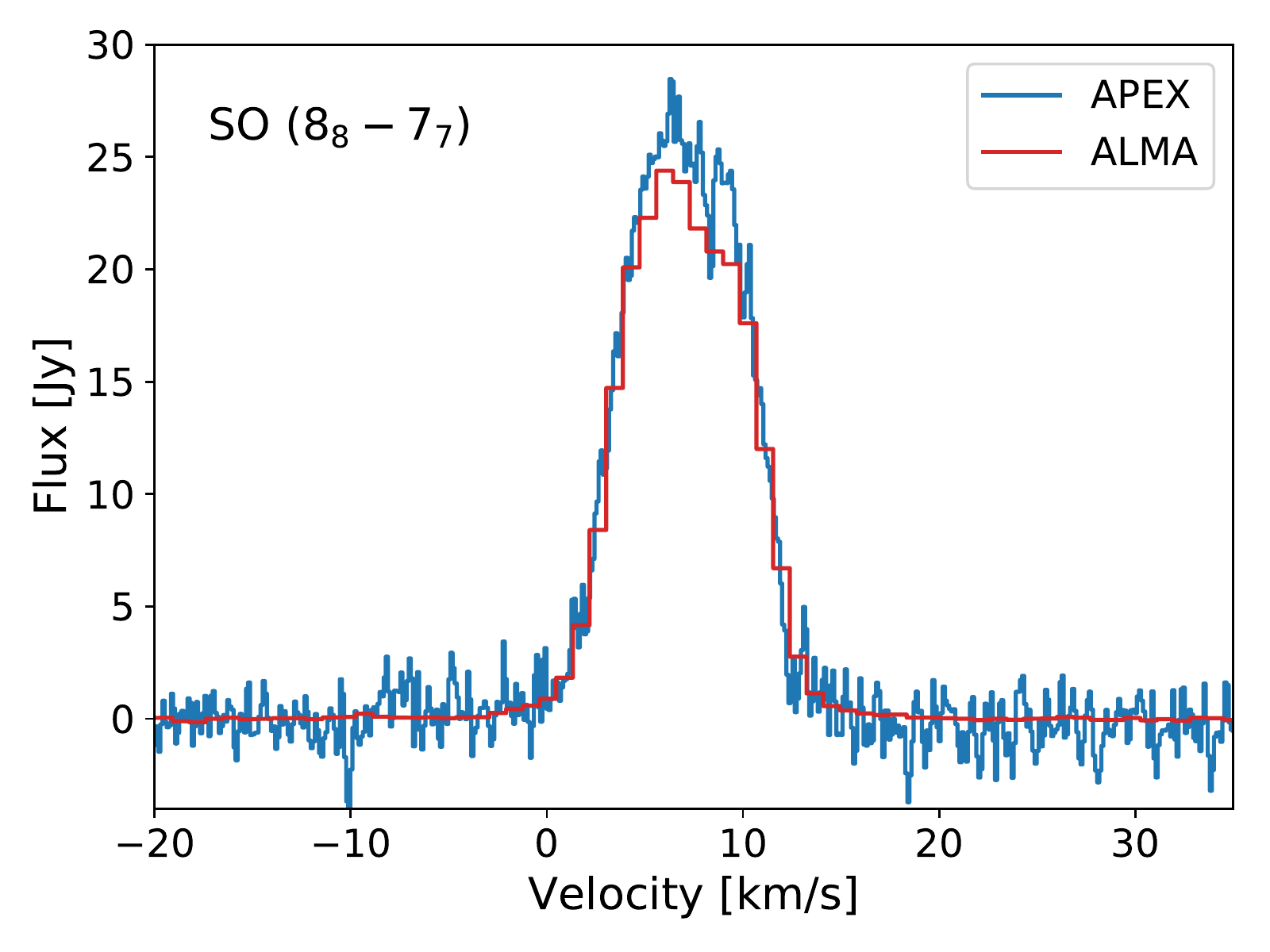}
\includegraphics[width=0.35\textwidth]{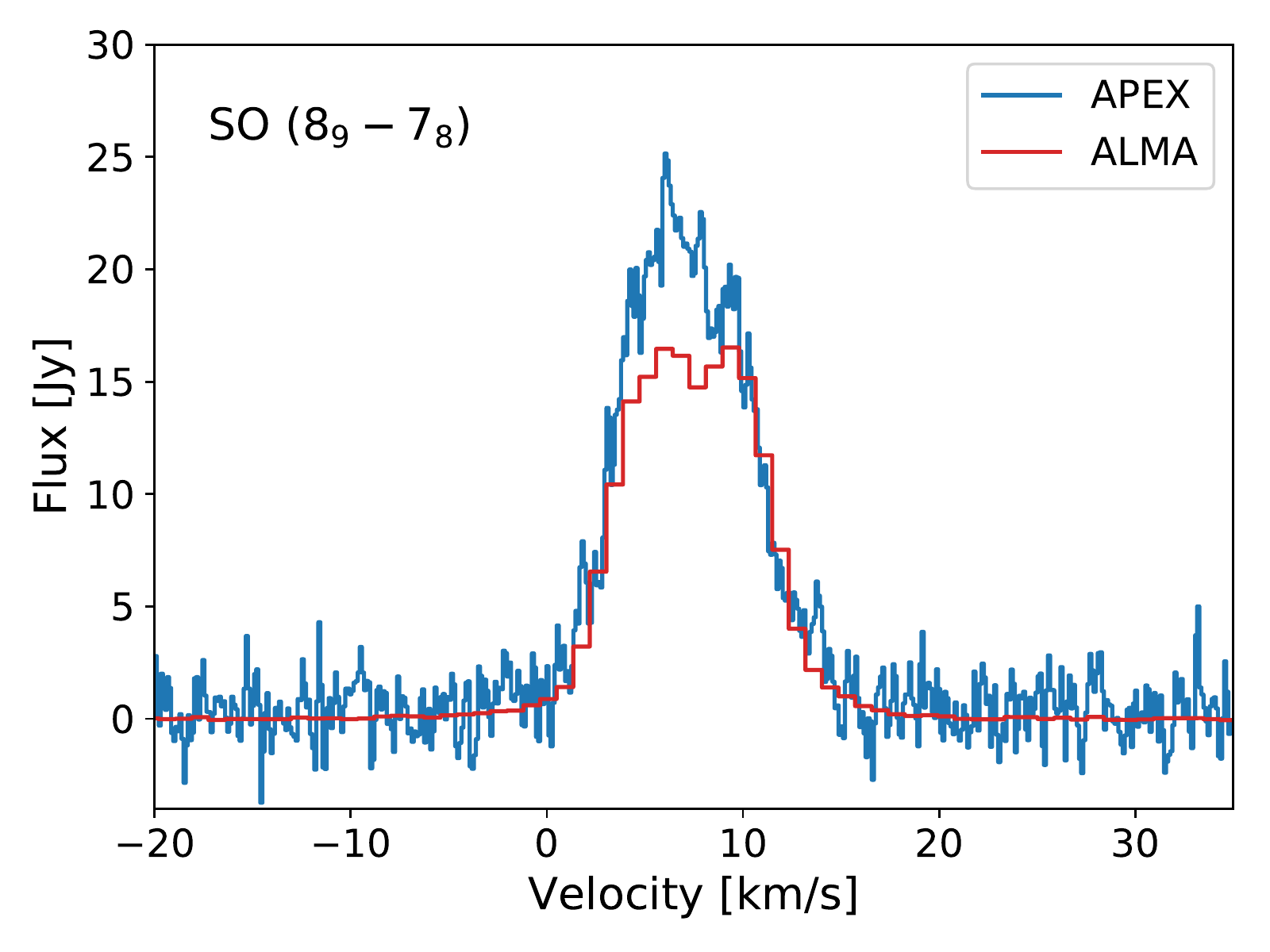}
\includegraphics[width=0.35\textwidth]{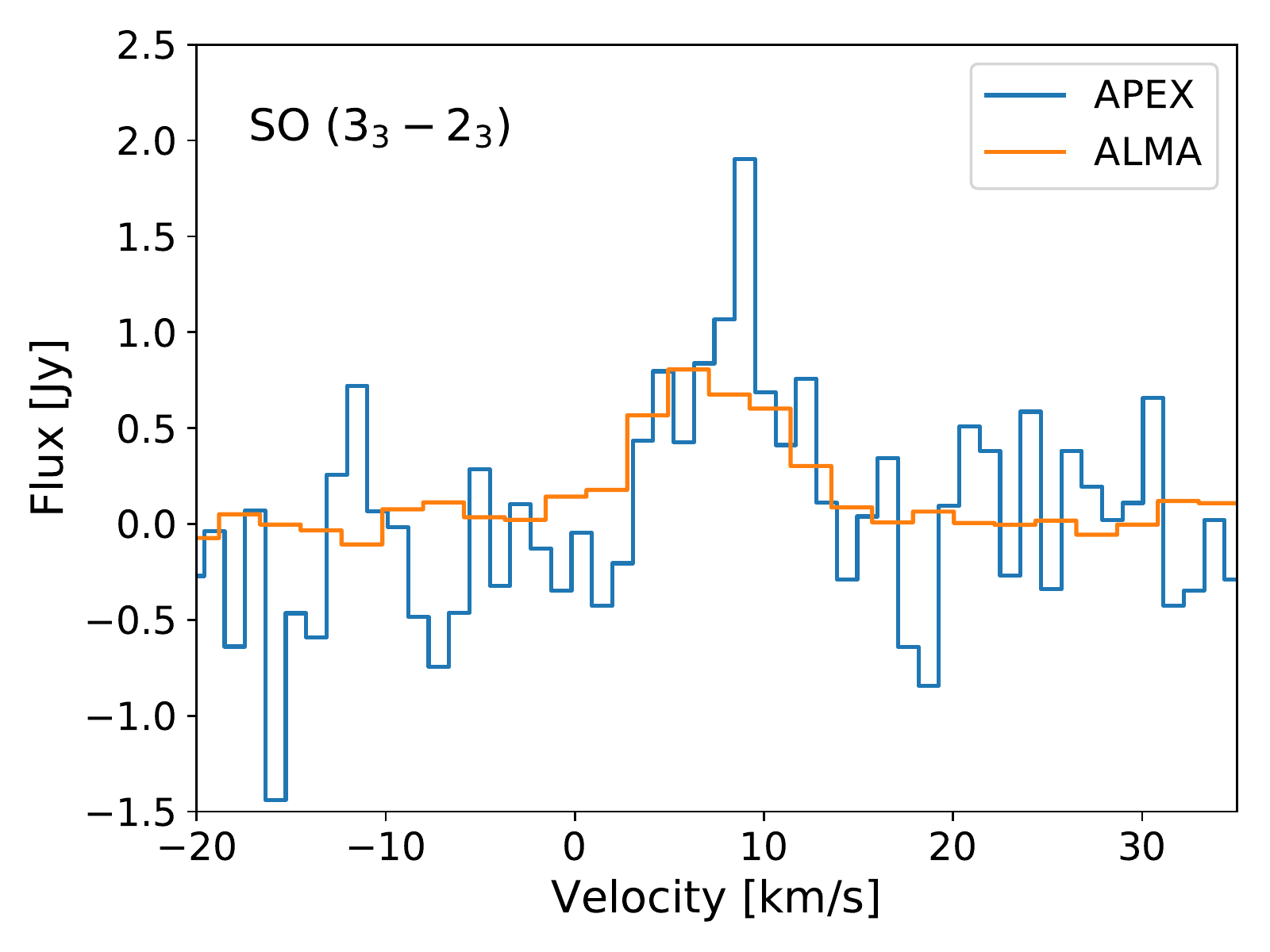}
%\endminipage
\caption{Comparison of the APEX (\textit{blue}) and ALMA (\textit{red} or \textit{orange}) observations of SO towards R~Dor. The ALMA spectra were extracted for a circular aperture with a radius of 5\arcsec{} (red lines) except for ($3_3 \to 2_3$) which was extracted for an aperture with a radius of 1\arcsec{} (orange line).}
\label{rdorsolostflux}
\end{figure}

\begin{figure}
\centering
\includegraphics[width=0.35\textwidth]{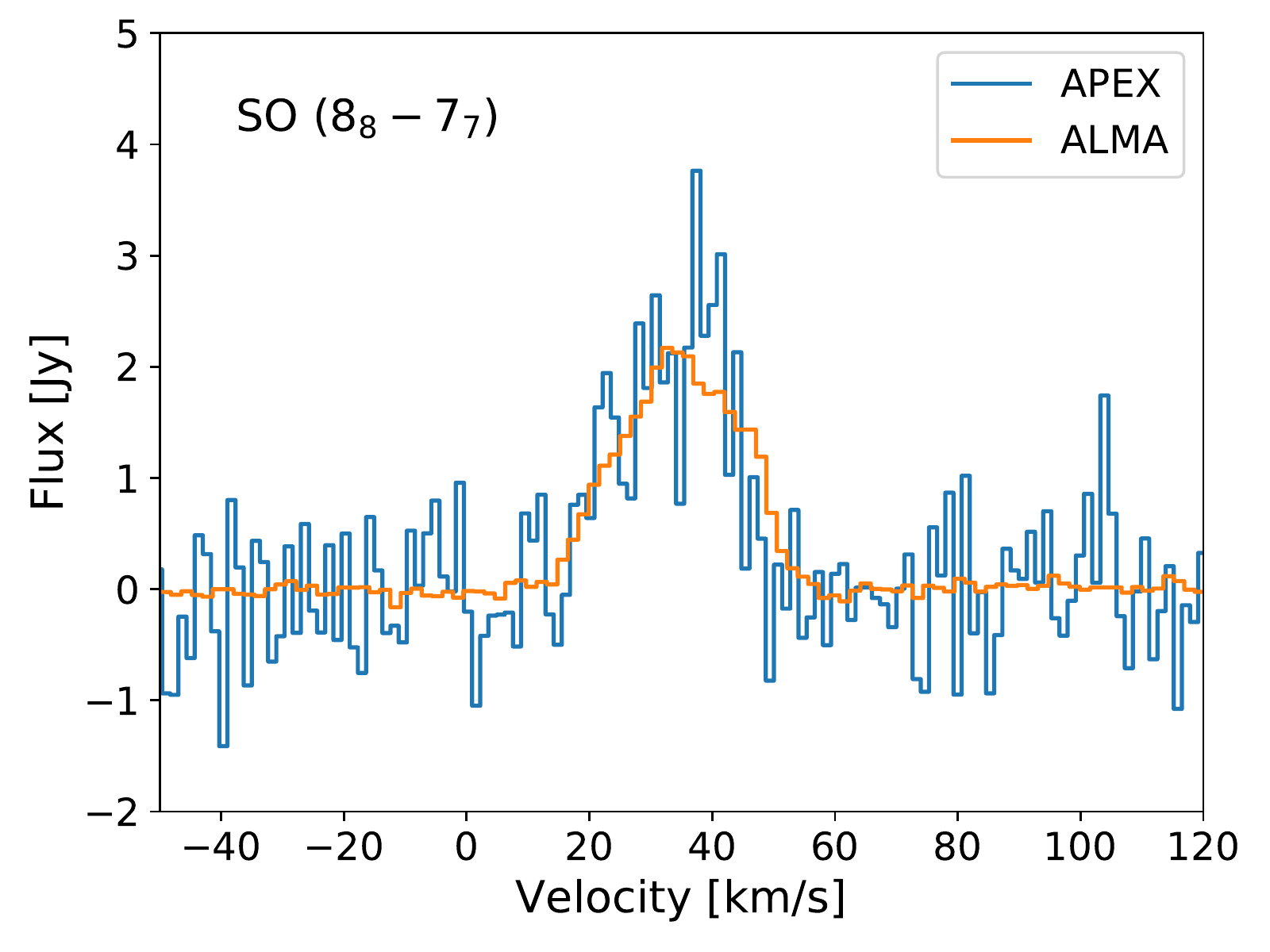}
\includegraphics[width=0.35\textwidth]{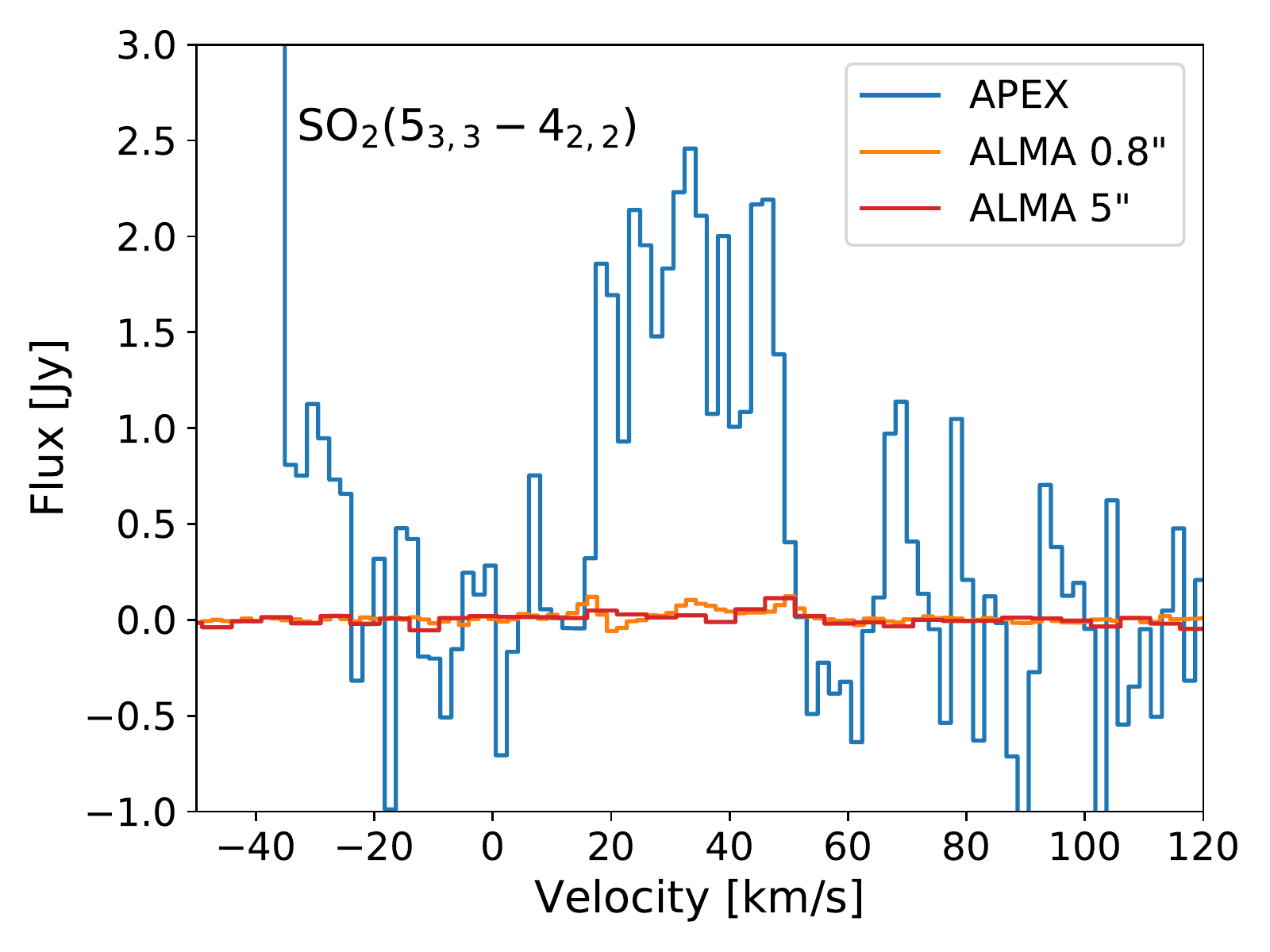}
\caption{Comparison of the APEX and ALMA observations of various sulphur-bearing molecules towards IK~Tau. From top to bottom: SO ($8_8\to7_7$), \so2($5_{3,3}\to4_{2,2}$). Orange lines indicate ALMA spectra extracted with an 800~mas radius aperture while red lines indicate ALMA spectra extracted using a 5\arcsec{} radius and with a coarser velocity resolution. Both spectra are shown for \so2 for which most of the flux has been resolved out.}% (The adjacent blue line is \up{29}SiO ($8\to7$).)}
\label{iktaulostflux}
\end{figure}

Of the SO lines detected with ALMA, all of the ground vibrational state lines except ($11_{10}\to10_{10}$) were previously detected with APEX and published by \cite{Danilovich2016} for R~Dor. For IK~Tau, the ground vibrational state ($8_8\to7_7$) line was previously published by \cite{Kim2010}. 
The single antenna observations allow us to compare the total flux recovered by ALMA with that observed by APEX\footnote{This publication is based on data acquired with the Atacama Pathfinder Experiment (APEX). APEX is a collaboration between the Max-Planck-Institut f\"ur Radioastronomie, the European Southern Observatory, and the Onsala Space Observatory.}. 
In Fig. \ref{rdorsolostflux} we plot the R~Dor ALMA spectra against the corresponding APEX spectra, both in Jy. As can be seen there, all of the flux has been recovered by ALMA for the ($8_7\to7_6$) line, a small amount may have been resolved out for the ($8_8\to7_7$) line (although the difference is small enough that this could be a calibration uncertainty { since the ALMA flux scale uncertainty in band 6 is $\le7$\% and the APEX uncertainty is $\sim20\%$}), some flux has been resolved out for the ($8_9\to7_8$) line and it is hard to make a conclusive statement about the ($3_3\to2_3$) line since the APEX spectrum is so much noisier than the ALMA spectrum. In the case of the ($8_9\to7_8$) line, it may be that its overlap with the \so2 ($16_{4,12}\to 16_{3,13}$) line contributes to the larger amount of flux being resolved out.
In Fig. \ref{iktaulostflux} we show the ALMA and APEX observations of the SO ($8_8\to7_7$) line towards IK~Tau, which indicates that no appreciable flux has been resolved out by ALMA.

\subsubsection{Isotopologues of SO}

\begin{figure*}
\centering
\includegraphics[height=3.9cm]{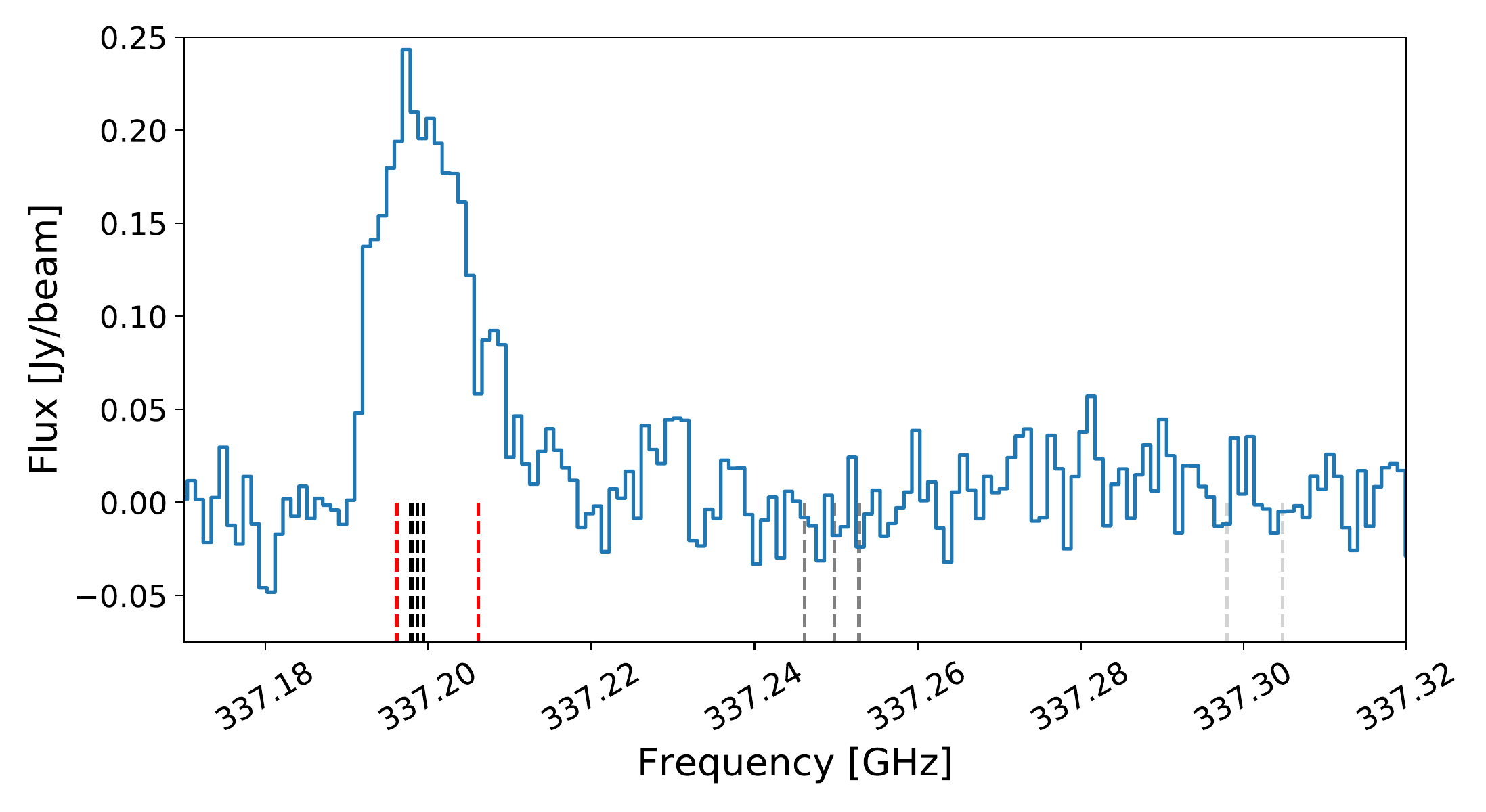}
\includegraphics[height=3.9cm]{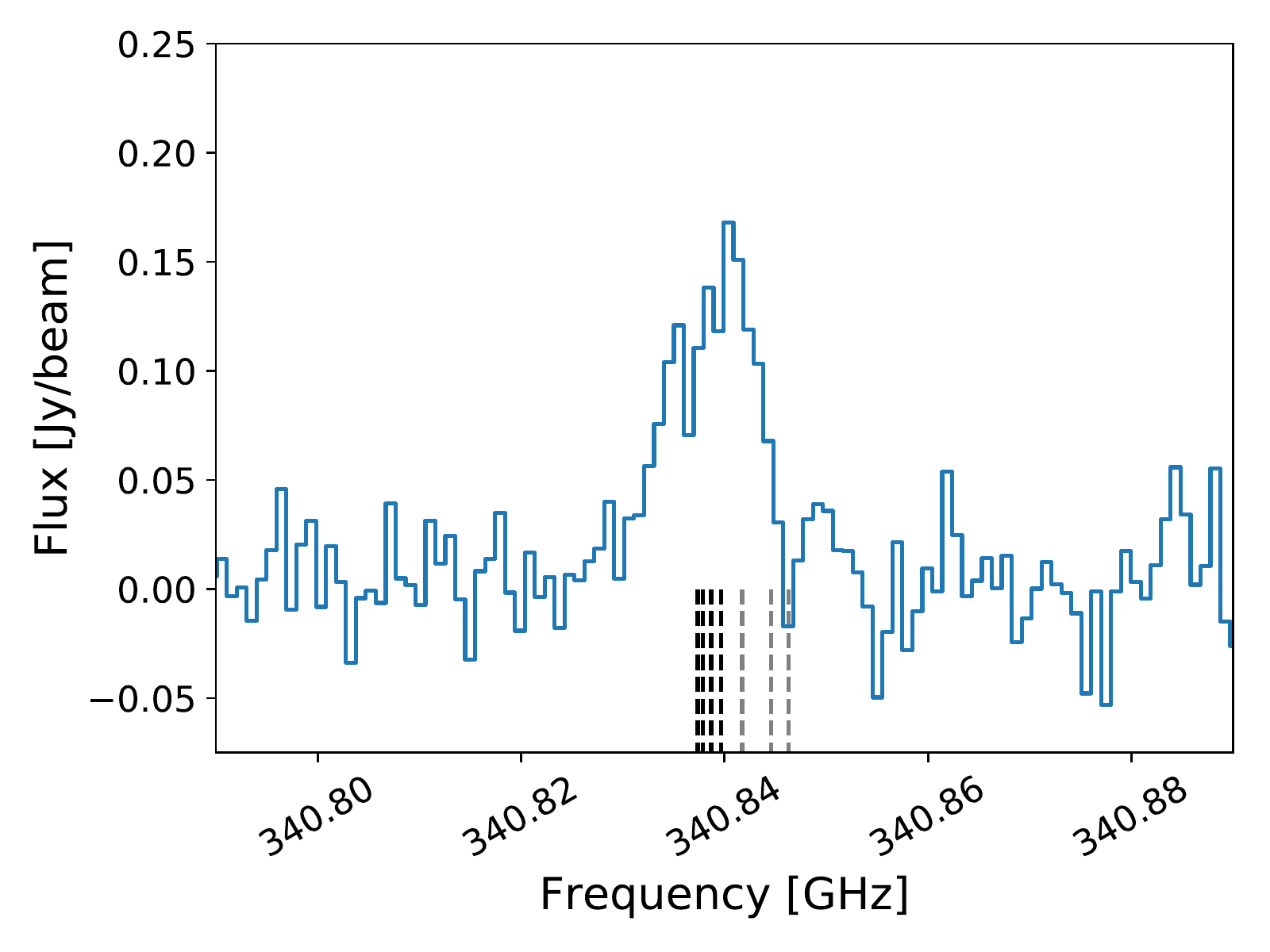}
\includegraphics[height=3.9cm]{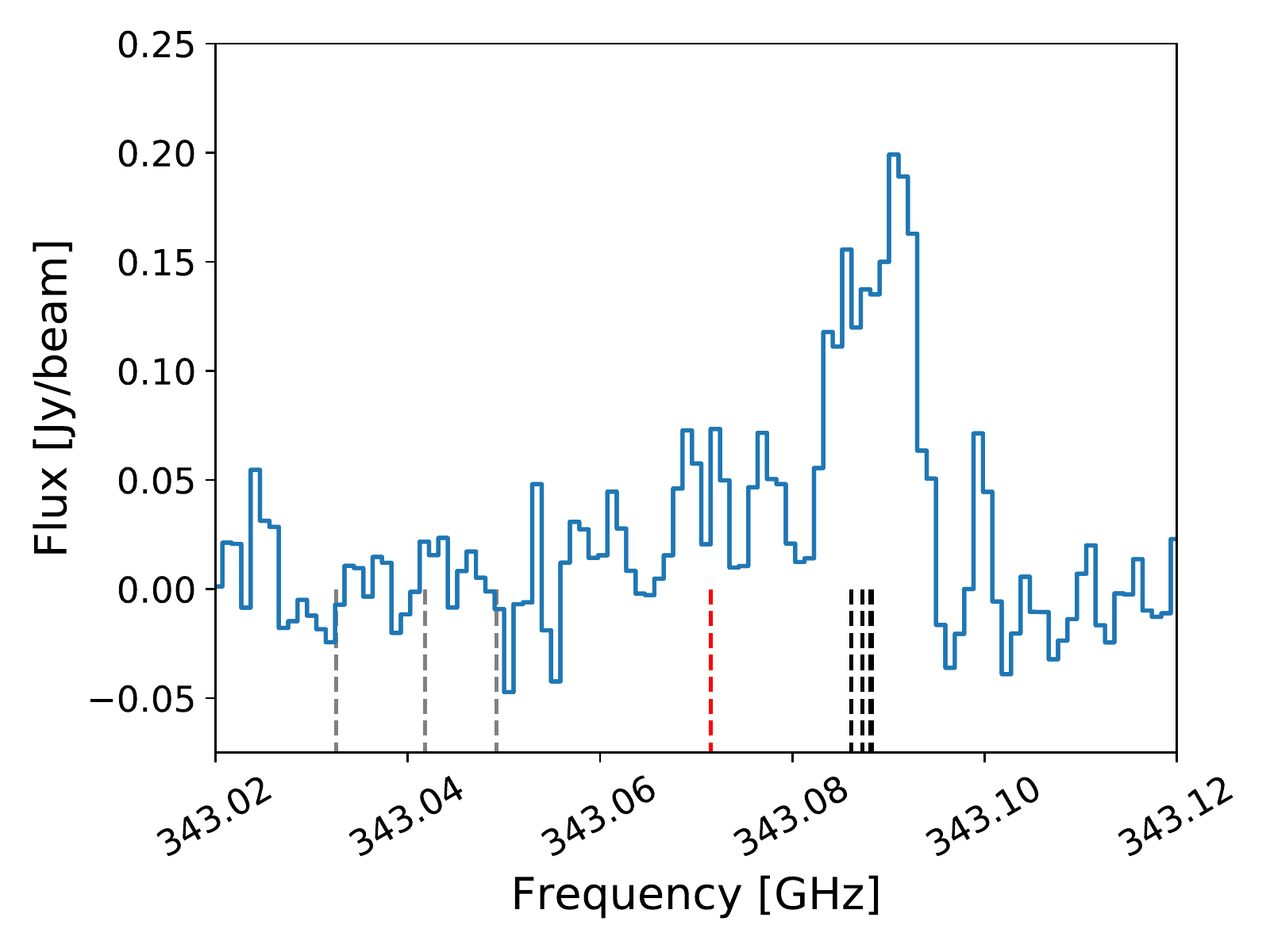}
\includegraphics[height=3.8cm]{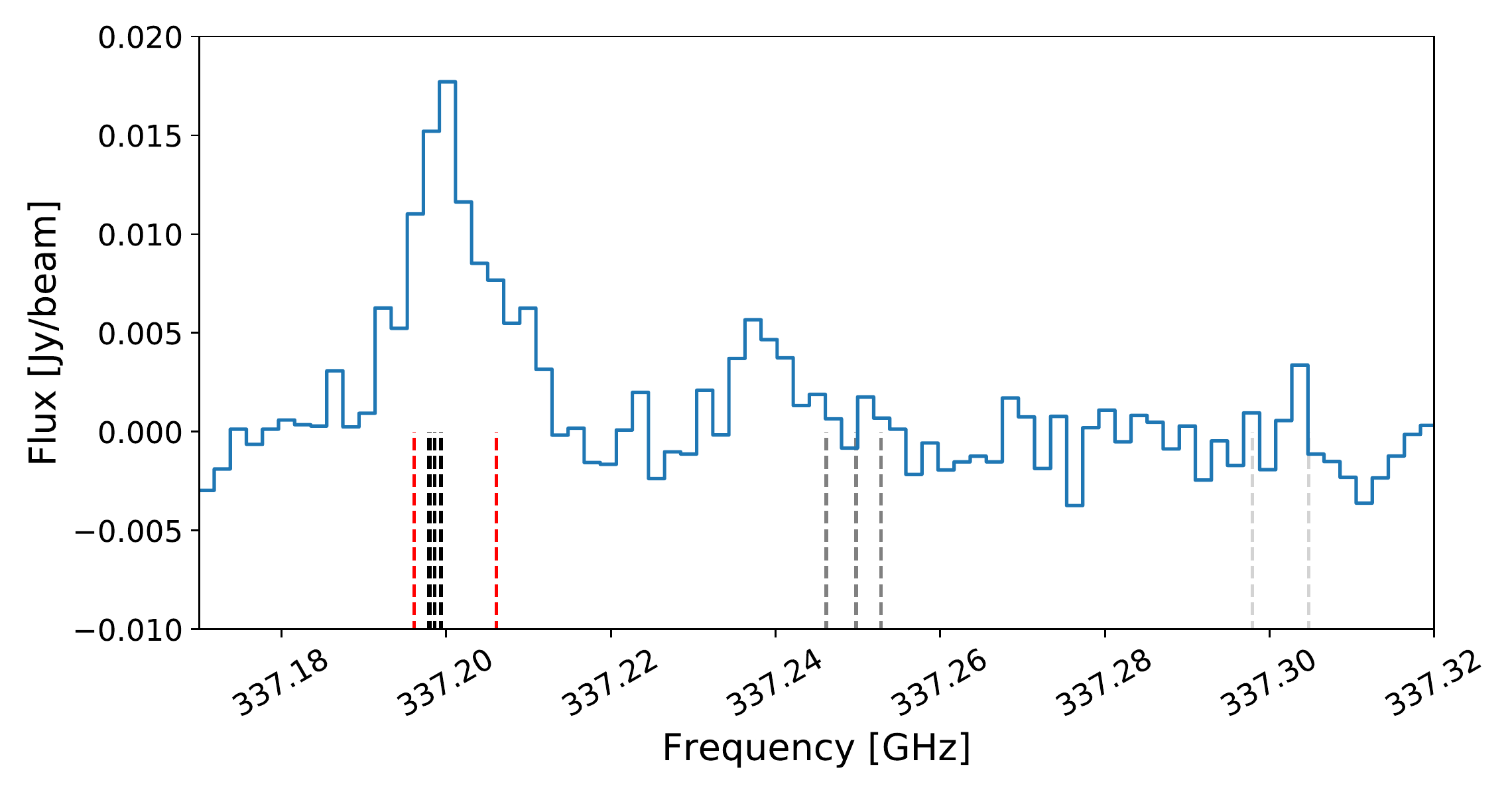}
\includegraphics[height=3.8cm]{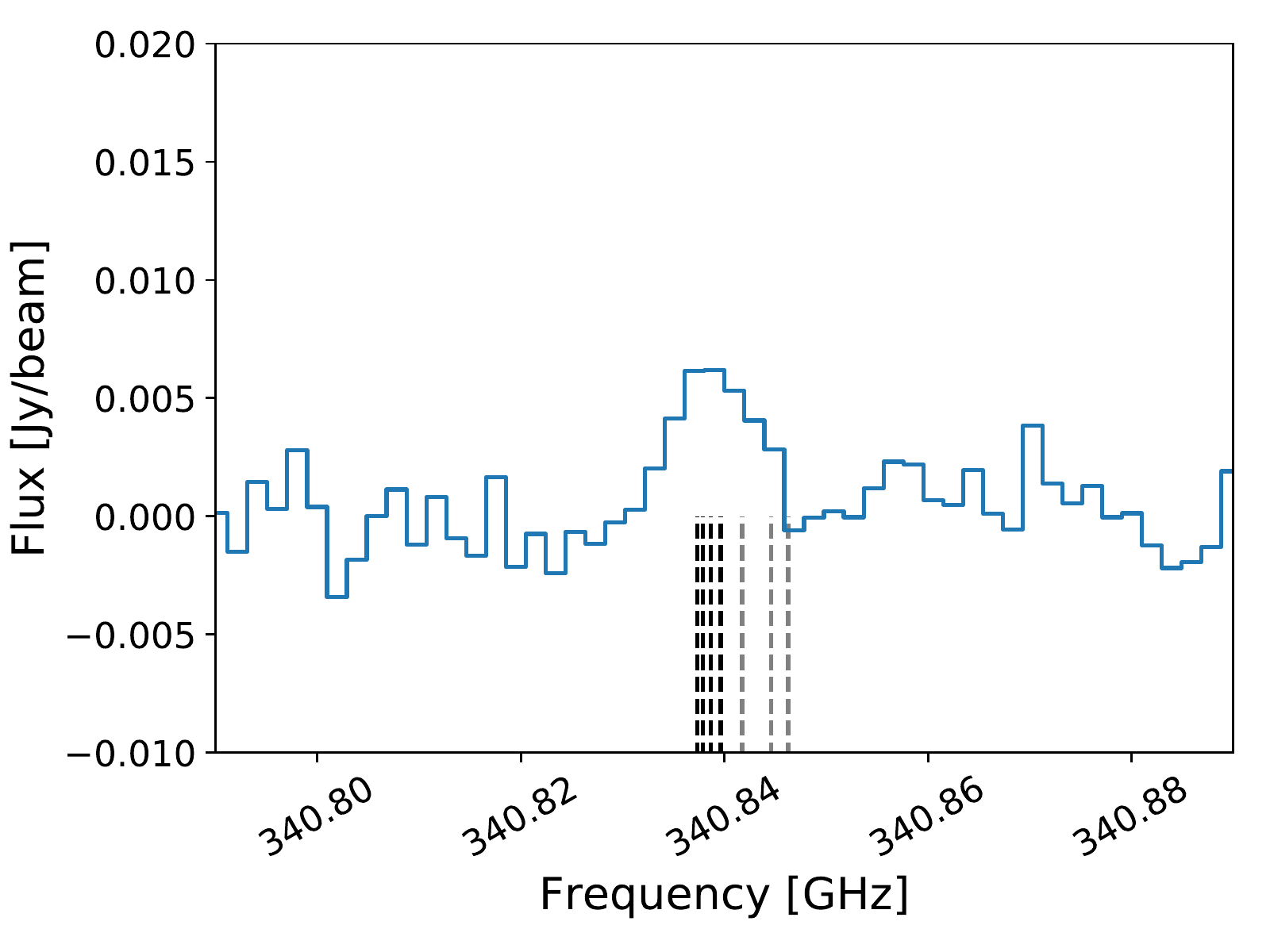}
\includegraphics[height=3.8cm]{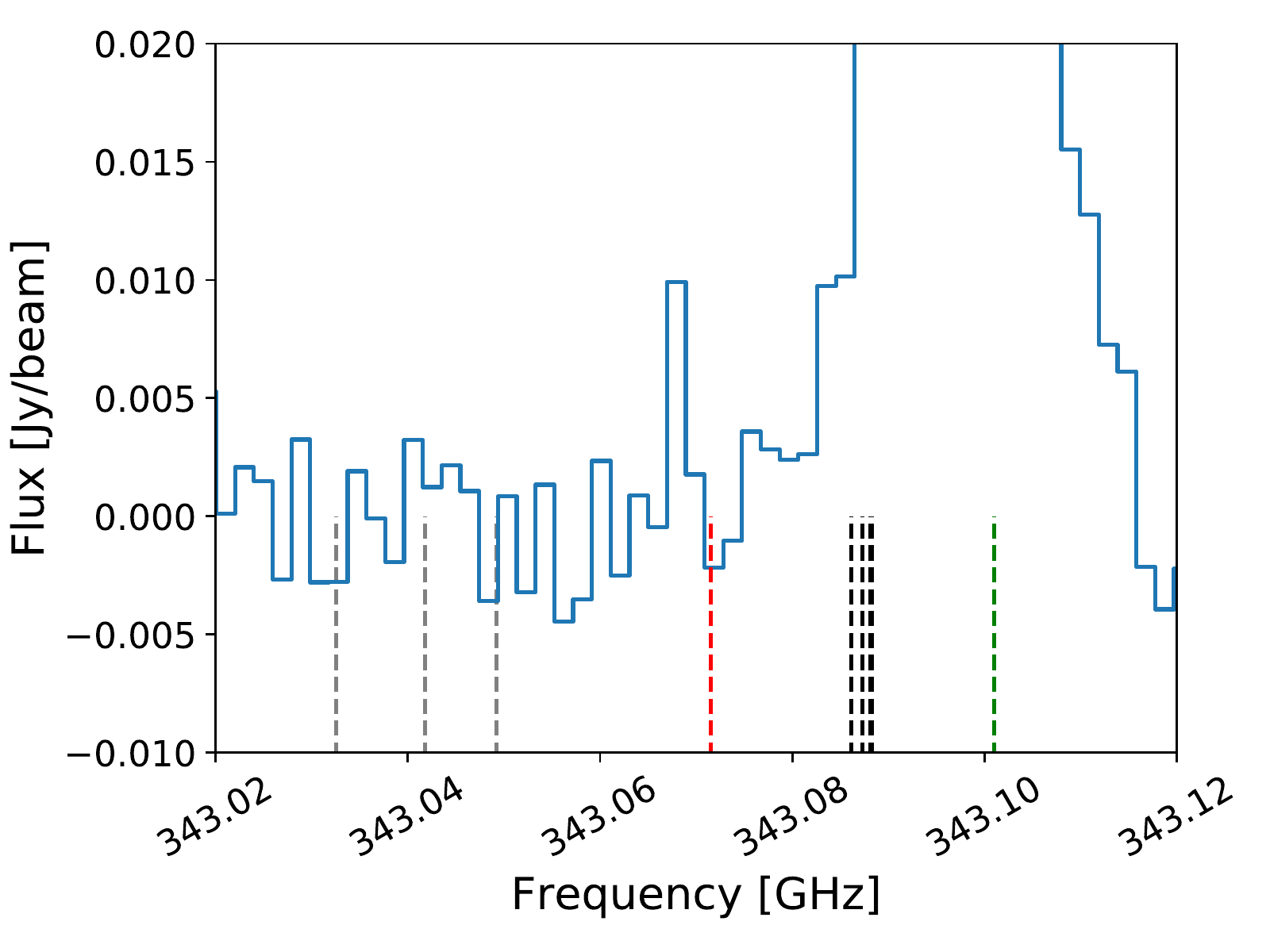}
\caption{$^{33}$SO lines towards R Dor (\textit{top}), extracted for a 1\arcsec{} radius aperture, and IK~Tau (\textit{bottom}), extracted for an 80~mas radius aperture. From left to right we show the ($ 8_{ 7} \to  7_{ 6}$), ($ 8_{ 8} \to  7_{ 7}$), and ($ 8_{ 9} \to  7_{ 8}$) line groups. Hyperfine components are indicated by the dashed lines: the black lines indicate the hyperfine components with the highest predicted intensities, the medium grey lines the hyperfine components with predicted intensities $\sim 1.5$~dex lower and the light grey lines hyperfine components with predicted intensities $\sim3.5$~dex lower. The red dashed lines indicate the frequencies of nearby TiO$_2$ lines and the green dashed line in the rightmost IK~Tau panel indicates the location of the SiS ($19\to18$, $v=1$), which overpowers the $^{33}$SO flux.}
\label{33SO}
\end{figure*}

For the isotopologues of SO, the two brightest lines of $^{34}$SO that fall in our range were detected towards both R~Dor and IK~Tau (see Table \ref{solines}). Lines of $^{33}$SO exhibit hyperfine structure due to the $^{33}$S nucleus \citep{Klaus1996} { and these are distinguished in Table \ref{solines} by the inclusion of $F$, the quantum number representing the total angular momentum including nuclear spin \citep{Townes2013}}. In Fig. \ref{33SO} we show the  $^{33}$SO lines observed towards R~Dor with all the hyperfine components marked on the spectrum. The ($ 8_{ 7} \to  7_{ 6}$) line group has a possible overlap with { tentative} lines of TiO$_2$ { ($25_{1,25}\to 24_{0,24}$) and ($24_{1,23}\to 23_{2,22}$) at 337.1961~GHz and 337.2061~GHz, respectively.} However, examining the channel maps, the emission { at that frequency has more morphological similarities with the }other isotopic species of SO than with other TiO$_2$ lines, so we assume the emission is dominated by $^{33}$SO. { Furthermore, the $^{33}$SO ($ 8_{ 9} \to  7_{ 8}$) line group is adjacent to TiO$_2$ ($23_{3,21} \to 22_{2,20}$), which is not detected above the noise in Fig. \ref{33SO}. All three of the mentioned TiO$_2$ lines have similar level energies (with the lower level energies around 192--198~K) and similar predicted intensities, so we can safely assume the two lines which overlap with $^{33}$SO ($ 8_{ 7} \to  7_{ 6}$) do not have a significant impact on the line intensity.  For completion, we indicate the location of the TiO$_2$ lines in Fig. \ref{33SO}.}
For IK Tau the $^{33}$SO lines are fainter than for R~Dor and are observed most clearly in the spectra extracted for the smallest radius. Also, the ($ 8_{ 9} \to  7_{ 8}$) line group coincides with SiS ($19\to18$, $v=1$), which is not detected towards R~Dor {\cite[see][for a comprehensive discussion of SiS towards R~Dor]{Danilovich2019}. We plot the IK~Tau $^{33}$SO lines in the lower portion of Fig. \ref{33SO}.} %For both of these reasons, we will focus on R~Dor rather than IK~Tau when discussing $^{33}$SO.

We also checked for any possible lines from S$^{17}$O (which has hyperfine components, like $^{33}$SO, due to the $^{17}$O nucleus), S$^{18}$O, and $^{36}$SO. From the line lists provided by the Cologne Database for Molecular Spectroscopy \citep[CDMS\footnote{\url{https://cdms.astro.uni-koeln.de}}][]{Muller2001,Muller2005,Endres2016}, we determined that the brightest lines of S$^{17}$O and $^{36}$SO fall outside of the frequency range surveyed by ALMA. Three bright lines of S$^{18}$O fall within our frequency range. Although they were not noted as detections by \citet{Decin2018}, they are listed in Table \ref{solines} and were checked carefully for weak signatures. We find no additional detections towards IK~Tau but found very tentative detections of the ($9_{8}\to8_{7}$) and possibly ($9_{10}\to8_{9}$) lines in the spectrum of R~Dor, although they cannot be discerned in the channel maps or zeroth moment maps. We do not find any evidence of the other possible S$^{18}$O line, which may be because it has a lower predicted integrated intensity at 300~K than the other two lines, according to CDMS, and falls on the edge of an observing window.

\subsection{SO$_2$}
\subsubsection{Overview}\label{so2overview}

\so2 is a near prolate asymmetric rotor ($\kappa = -0.94$) with $C_{2v}$ symmetry whose permanent electric dipole moment lies along the intermediate ($b$) inertial axis and whose rotational levels are labeled $J_{K_a,K_c}$, { where $J$ is the total angular momentum (excluding nuclear spin) and $K_a$ and $K_c$ are the projections of the angular momentum along the $a$ and $c$ molecular axes, both of which are orthogonal to the dipole moment of the molecule \citep{Hartquist1998}}. Because the two equivalent off-axis $^{16}$O atoms are bosons, half of the rotational levels --- i.e., those with $K_a + K_c = \mathrm{\it odd}$ --- are missing.
The sensitivity of ALMA allows us to detect a large number of lines, some of which are impossible to see with less sensitive telescopes such as APEX. Indicated in Table~\ref{so2linelist} is whether the lines were detected towards R~Dor or both stars, and any unusual features such as overlapping lines. 
A more detailed discussion of \so2 towards the individual stars is given in Sections \ref{rdorso2text} and \ref{iktauso2text}.

Owing to the large number of observed \so2 transitions in the main isotopologue in the ground vibrational state, we have divided them into two categories to aid the analysis and refer to them as ``most favoured'' (brighter) and ``less favoured'' (fainter) transitions. All Q-branch transitions (where $\Delta J =0$) fall into the most favoured category and all R-branch transitions (where $\Delta J = +1$) fall into the less favoured category. P-branch transitions (where $\Delta J = -1$) are split between the two groups: the P$_{-1,+1}$ subbranch containing transitions with $\Delta K_a,\Delta K_c = -1,+1$ is in the most favoured category and the P$_{+1,-3}$ subbranch with $\Delta K_a, \Delta K_c = +1, - 3$ is in the less favoured category. For a more detailed discussion of subbranches for asymmetric top rotational transitions, we direct the interested reader to \cite{Cross1944} and chapter 4 of \cite{Townes2013}.
The most favoured transitions, which are listed in the upper portion of Table \ref{so2linelist}, produce the brightest lines and the less favoured transitions, which are listed in the lower portion of Table~\ref{so2linelist}, are fainter and generally produce compact emission that is either unresolved or barely resolved for both stars. Most of the vibrationally excited lines, which have $\nu_2 =1$ and are listed in Table \ref{so2v1linelist}, fall into the more favoured category (with the two exceptions being noted as very weak), but also exhibit compact emission, which is either unresolved or barely resolved.

\begin{figure}
\centering
\includegraphics[width=\columnwidth]{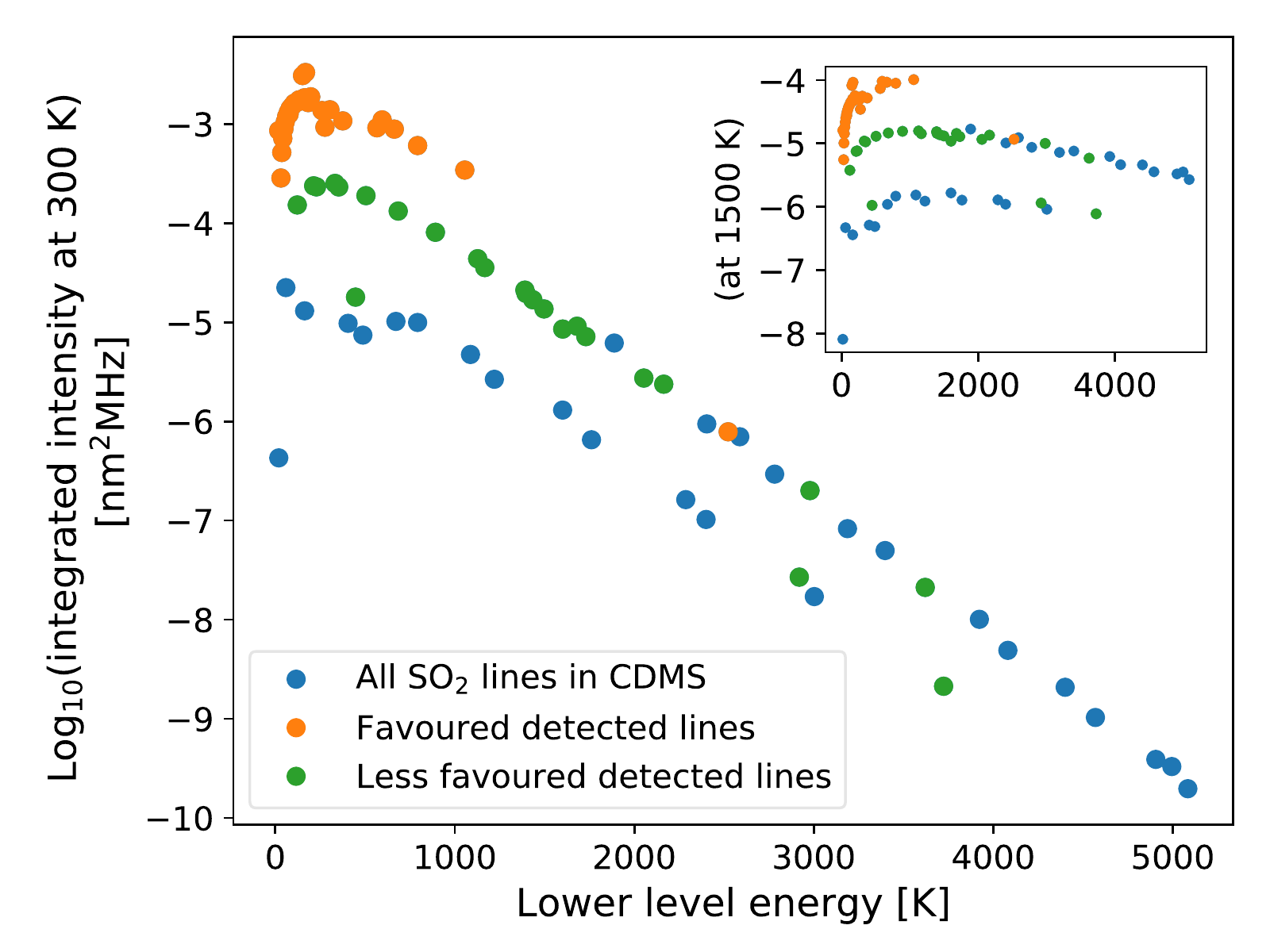}
\caption{The predicted integrated intensities at 300~K against the lower state energy levels for all the vibrational ground state \so2 lines listed in CDMS that fall within the frequency range of our line scan. The detections indicated are for R~Dor. The inset figure shows the same information but for 1500~K.}
\label{so2EvsI}
\end{figure}

{ To give an indication of which transitions are more likely to be detected, we use { theoretical} line intensities\footnote{Taken from CDMS. { The units of intensity, nm$^2$ MHz, are based on the integral of the absorption cross-section over the spectral line shape.} See \cite{Pickett1998} for further explanation.}. For the ground vibrational state, these intensities are plotted in Fig.~\ref{so2EvsI} for 300~K (corresponding to $\sim20 R_*$ for R~Dor) against the lower state energy levels for all} \so2 lines that lie within the frequency range of the spectral line survey. 
In the inset of Fig. \ref{so2EvsI} the same transitions are plotted for $T= 1500$~K, { representing gas temperatures in the inner regions of the CSE, within a few stellar radii of the star.}
The points are colour-coded to indicate whether the transitions were detected, and to denote the most favoured and less favoured transitions. 
We find that all of the most favoured transitions are detected, except for those from levels with very high excitation energies 
of $J\sim 90$ and $E > 4000$~K. This gives us a quick way to predict the likelihood of detecting an \so2 line based primarily on the transition's quantum numbers (in conjunction with the level energies).
Referring to the higher temperature plot in the inset in Fig.~\ref{so2EvsI}, it is apparent that the few outliers { at 300~K (detected lines with low predicted intensities) are not unrealistic.}
The most egregious outlier is the ($87_{9,79}\to88_{6,82}$) transition {(rightmost green point) which closely coincides in frequency} with the ($57_{6,52}\to56_{7,49}$) transition also of \so2.
In Table \ref{so2v1linelist} we list all the vibrationally excited \so2 lines, { all of which are excited to the first bending mode, $\nu_2=1$, which is the lowest-lying vibrational state of \so2}. An analogous plot to Fig. \ref{so2EvsI} is given in Fig. \ref{so2EvsIv1} for the vibrationally excited \so2 lines towards R~Dor.

\begin{table}
\caption{SO$_2$ lines detected with ALMA. Lines listed in the bottom section of the table are less favoured, as defined in Sect. \ref{so2overview}.}\label{so2linelist}
%\scalebox{0.97}{
\begin{adjustbox}{width=\columnwidth}
\begin{tabular}{ccrcl}
\hline\hline
 Frequency & Line  & $E_\mathrm{up}$ & Star & Notes\\
 $\mathrm{[GHz]}$ & & [K] & & I = IK~Tau, R = R~Dor\\
\hline
%SO$_2$& \\
     336.0892$^a$ &  $23_{ 3,21} \to 23_{ 2,22}$ & 276 & both & \so2 overlap\\
     338.3060$^a$ &  $18_{ 4,14} \to 18_{ 3,15}$ & 197 & both & \\
     338.6119$^a$ &  $20_{ 1,19} \to 19_{ 2,18}$ &199 & both & \\
     340.3164$^c$ &  $28_{ 2,26} \to 28_{ 1,27}$ & 392& both & I: SiS $v=2$ overlap\\
     341.1364$^c$ &  $70_{10,60} \to 69_{11,59}$ & 2539& both & \\
     341.4031$^a$ &  $40_{ 4,36} \to 40_{ 3,37}$ & 809& both & \\
     341.6740$^c$ &  $36_{ 5,31} \to 36_{ 4,32}$ & 679& both & \\
     342.7616$^c$ &  $34_{ 3,31} \to 34_{ 2,32}$ & 582& both & \\
     345.3385$^a$ &  $13_{ 2,12} \to 12_{ 1,11}$ & 93& both & H$^{13}$CN overlap\\
     346.5239$^c$ &  $16_{ 4,12} \to 16_{ 3,13}$ & 165& both & SO overlap \\
     346.6522$^a$ &  $19_{ 1,19} \to 18_{ 0,18}$ & 168& both & \\
     348.3878$^c$ &  $24_{ 2,22} \to 23_{ 3,21}$ & 293& both & R: TiO$_2$ overlap\\
&&& & I: $^{30}$SiS $v=1$ overlap\\
     349.7833$^c$ &  $46_{ 5,41} \to 46_{ 4,42}$ & 1072& both & \\
     351.2572$^a$ &  $ 5_{ 3, 3} \to  4_{ 2, 2}$ & 36 & both & R: \so2 $v=1$ overlap in\\
     &&& & wing, I: Si$^{34}$S overlap\\
     351.8739$^a$ &  $14_{ 4,10} \to 14_{ 3,11}$ & 136 & both & \\
     355.0455$^a$ &  $12_{ 4, 8} \to 12_{ 3, 9}$ &111 & both & \\
     356.7552$^a$ &  $10_{ 4, 6} \to 10_{ 3, 7}$ & 90& both & I: TiO$_2$ overlap \\
     357.1654$^a$ &  $13_{ 4,10} \to 13_{ 3,11}$ & 123& both & \\
     357.2412$^a$ &  $15_{ 4,12} \to 15_{ 3,13}$ &150 & both & \\
     357.3876$^a$ &  $11_{ 4, 8} \to 11_{ 3, 9}$ & 100& both & \\
     357.5814$^c$ &  $ 8_{ 4, 4} \to  8_{ 3, 5}$ & 72& both & \so2 $v=1$ overlap\\
     357.6718$^a$ &  $ 9_{ 4, 6} \to  9_{ 3, 7}$ & 81 & both & \\
     357.8924$^c$ &  $ 7_{ 4, 4} \to  7_{ 3, 5}$ & 65 & both & \\
     357.9258$^c$ &  $ 6_{ 4, 2} \to  6_{ 3, 3}$ & 59 & both & \\
     357.9629$^a$ &  $17_{ 4,14} \to 17_{ 3,15}$ & 180 & both & \\
     358.0131$^a$ &  $ 5_{ 4, 2} \to  5_{ 3, 3}$ & 53& R~Dor & \so2 overlap in wing\\
     358.0379$^a$ &  $ 4_{ 4, 0} \to  4_{ 3, 1}$ & 49& R~Dor & \so2 overlap in wing\\
     358.2156$^a$ &  $20_{ 0,20} \to 19_{ 1,19}$ & 185 & both & R: poss. Si$^{18}$O overlap\\
     359.1512$^a$ &  $25_{ 3,23} \to 25_{ 2,24}$ & 321 & both & \\
     359.7707$^a$ &  $19_{ 4,16} \to 19_{ 3,17}$ & 214 & both & \\
     360.2904$^c$ &  $34_{ 5,29} \to 34_{ 4,30}$ & 612& both & \\
\hline
     335.7732$^a$ &  $29_{ 5,25} \to 30_{ 2,28}$ & 463 & R~Dor & ID uncertain\\
     336.1135$^c$ &  $42_{12,30} \to 43_{11,33}$ & 1183 & both & \so2 overlap\\
     336.6696$^a$ &  $16_{ 7, 9} \to 17_{ 6,12}$ & 245 & R~Dor & \\
     338.8698$^c$ &  $47_{13,35} \to 48_{12,36}$ & 1451 & R~Dor & \\
     339.8909$^b$ &  $65_{ 9,57} \to 64_{10,54}$ & 2180 & R~Dor & ID uncertain\\
     341.2755$^a$ &  $21_{ 8,14} \to 22_{ 7,15}$ & 369 & both & \\
     341.3219$^c$ &  $52_{14,38} \to 53_{13,41}$ & 1746 & both & \so2 \& AlO v=1 overlap\\
     341.3233$^c$ &  $53_{ 6,48} \to 52_{ 7,45}$ & 1413 & both & \so2 \& AlO v=1 overlap\\
     343.4767$^b$ &  $57_{15,43} \to 58_{14,44}$ & 2070 & R~Dor & I: only see  Na$^{37}$Cl\\
     345.4490$^a$ &  $26_{ 9,17} \to 27_{ 8,20}$ & 521 & both & \\
     347.8276$^c$ &  $87_{ 9,79} \to 88_{ 6,82}$ & 3740 & R~Dor & \so2 overlap\\
     347.8292$^c$ &  $57_{ 6,52} \to 56_{ 7,49}$ & 1618 & R~Dor & \so2 overlap\\
     349.1914$^c$ &  $77_{19,59} \to 78_{18,60}$ & 3637 & R~Dor & ID uncertain\\
     349.2271$^c$ &  $31_{10,22} \to 32_{ 9,23}$ & 701 & both & \\
     350.1103$^b$ &  $55_{ 6,50} \to 54_{ 7,47}$ & 1514 & R~Dor & ID uncertain\\
     350.8628$^a$ &  $10_{ 6, 4} \to 11_{ 5, 7}$ & 139 & both & ID uncertain\\
     352.6390$^a$ &  $36_{11,25} \to 37_{10,28}$ & 909 & both & I: bad channels\\
353.1119$^c$ & $76_{11,65}\to 75_{12,64}$ & 2978 & R~Dor & ID uncertain\\
     355.7055$^b$ &  $41_{12,30} \to 42_{11,31}$ & 1144 & R~Dor & ID uncertain\\
     356.0406$^a$ &  $15_{ 7, 9} \to 16_{ 6,10}$ & 231 & R~Dor & \\
     358.3442$^c$ &  $77_{ 8,70} \to 78_{ 5,73}$ & 2936 & R~Dor & ID uncertain\\
     358.4419$^c$ &  $46_{13,33} \to 47_{12,36}$ & 1408 & R~Dor & poss. U overlap\\
     360.7218$^a$ &  $20_{ 8,12} \to 21_{ 7,15}$ & 350 & both & I: very noisy\\
     360.8592$^b$ &  $51_{14,38} \to 52_{13,39}$ & 1698 & R~Dor & ID uncertain\\
\hline
\end{tabular}%}
\end{adjustbox}
\\{ References:} ($^a$) \cite{Lovas1985}; ($^b$) \cite{Belov1998}; ($^c$) \cite{Muller2005a} \& \cite{Muller2005}.
\end{table}

\subsubsection{Isotopologues of SO$_2$}

The \so2 isotopologue detections for both R~Dor and IK~Tau are listed in Table \ref{34so2linelist}. We detected several $^{34}$\so2 lines, mostly towards R~Dor. 
{ These are plotted for R~Dor in Fig. \ref{rdor34so2} for a 300~mas radius extraction aperture, since some of the lines are hidden in the noise for a larger extraction aperture.}
No lines of $^{33}$\so2 were detected. This is not unexpected due to the lower cosmic abundance of $^{33}$S and the hyperfine splitting that occurs due to the $^{33}$S nucleus. 

We list a few SO$^{17}$O lines towards R~Dor in Table \ref{34so2linelist}. These are included because these are the best identifications we presently have for these lines, however, they are by no means certain. 
%Combining the much lower abundances of $^{17}$O and $^{18}$O compared with $^{16}$O \citep[with $^{16}$O/$^{17}$O = 800 for R Dor and 1500 for IK Tau and $^{16}$O/$^{18}$O = 315 for R Dor and 650 for IK Tau, based on the \h2O modelling of][]{Danilovich2017}, and the break in symmetry of the V-shape of the molecules doubling the available energy levels in a given energy range, we would not expect to detect the SO$^{17}$O and SO$^{18}$O emission. 
We would not expect to observe SO$^{17}$O and SO$^{18}$O with the present sensitivity because of: (1) the much lower abundance of $^{17}$O and $^{18}$O --- where $^{16}$O/$^{17}$O = 800 in R Dor and 1500 in IK Tau, and $^{16}$O/$^{18}$O = 315 in R Dor and 650 in IK Tau on the basis of the H$_2$O models in \cite{Danilovich2017} --- and (2) the two oxygen nuclei have different masses in the singly substituted rare isotopic species, and therefore there are twice as many rotational levels which are populated in the rare isotopic species than in the main species (i.e. levels with $K_a + K_c = odd$ are permitted).
Furthermore, we only have tentative detections for S$^{18}$O lines towards R~Dor, and these lines are inherently brighter than SO$^{18}$O (there were no S$^{17}$O lines in the observed frequency range). \cite{Danilovich2017} find higher abundances of $^{18}$O than $^{17}$O from their \h2O results for both R~Dor and IK~Tau, so seeing SO$^{17}$O lines but not SO$^{18}$O lines does not fit with this. We conclude that the \so2 oxygen isotopologue lines are most likely misidentified and include them here only for completion.

\subsubsection{R Dor}\label{rdorso2text}

\begin{figure}
%\minipage{12cm}
\centering
\includegraphics[width=0.35\textwidth]{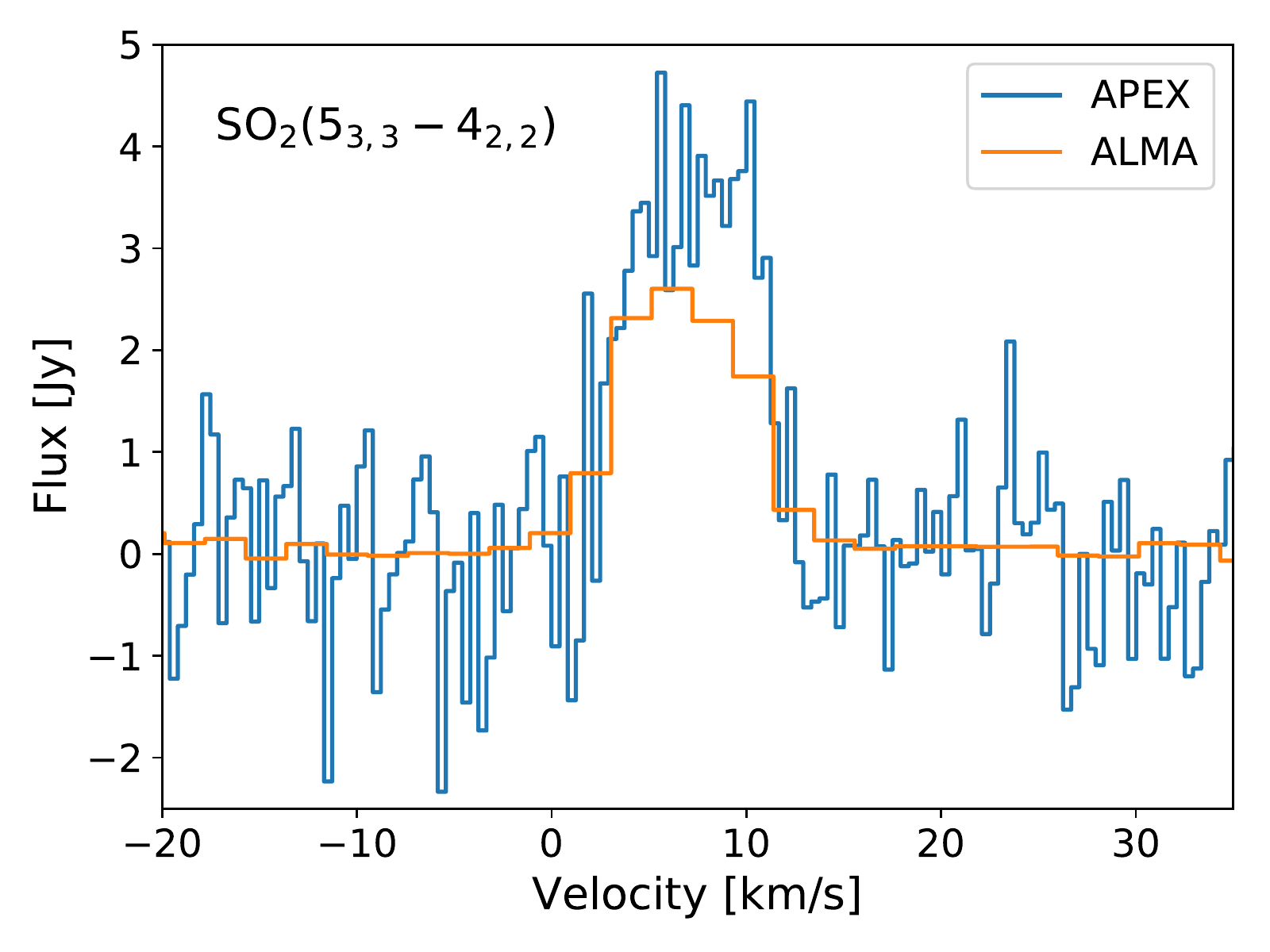}
\includegraphics[width=0.35\textwidth]{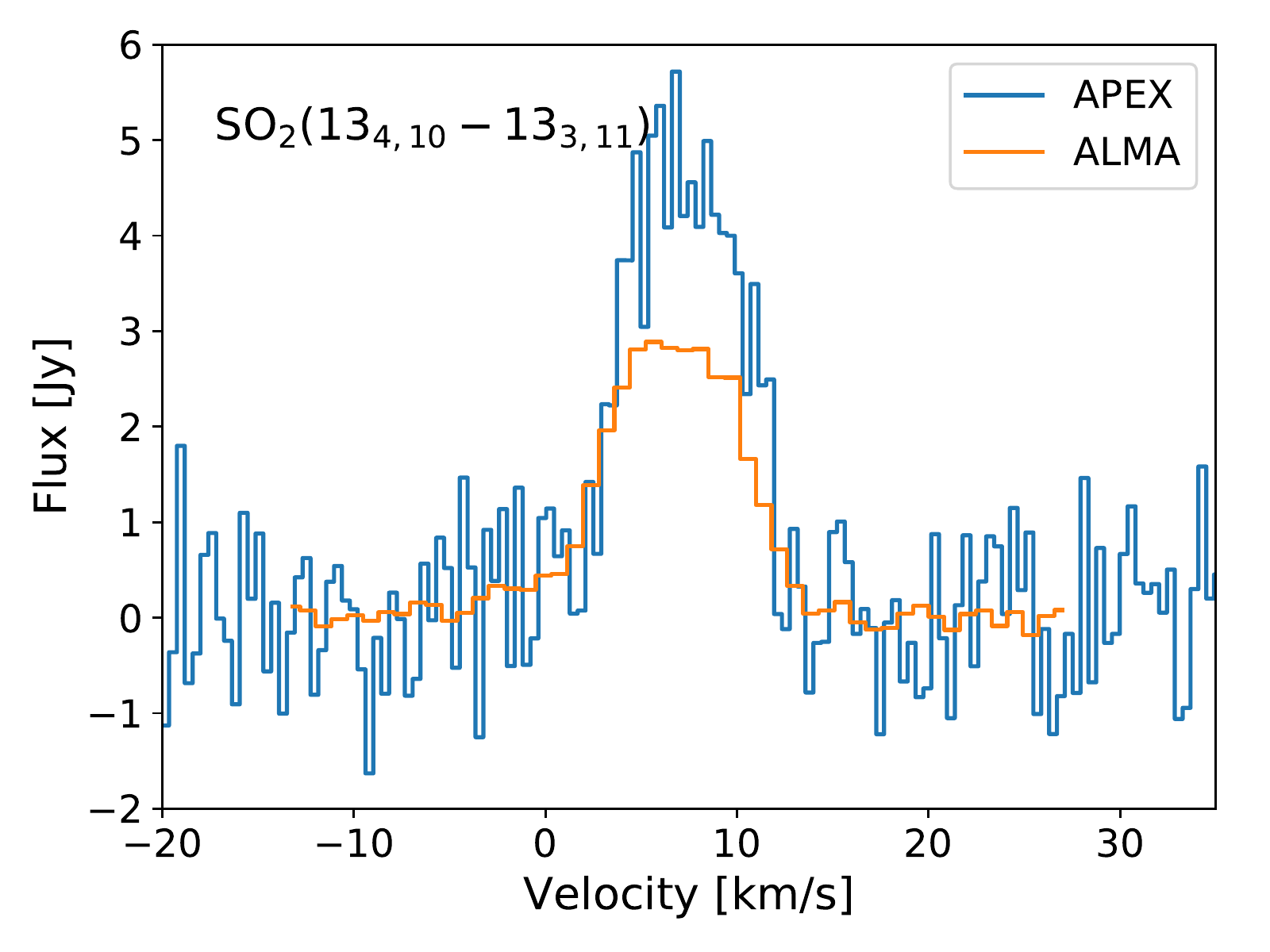}
\includegraphics[width=0.35\textwidth]{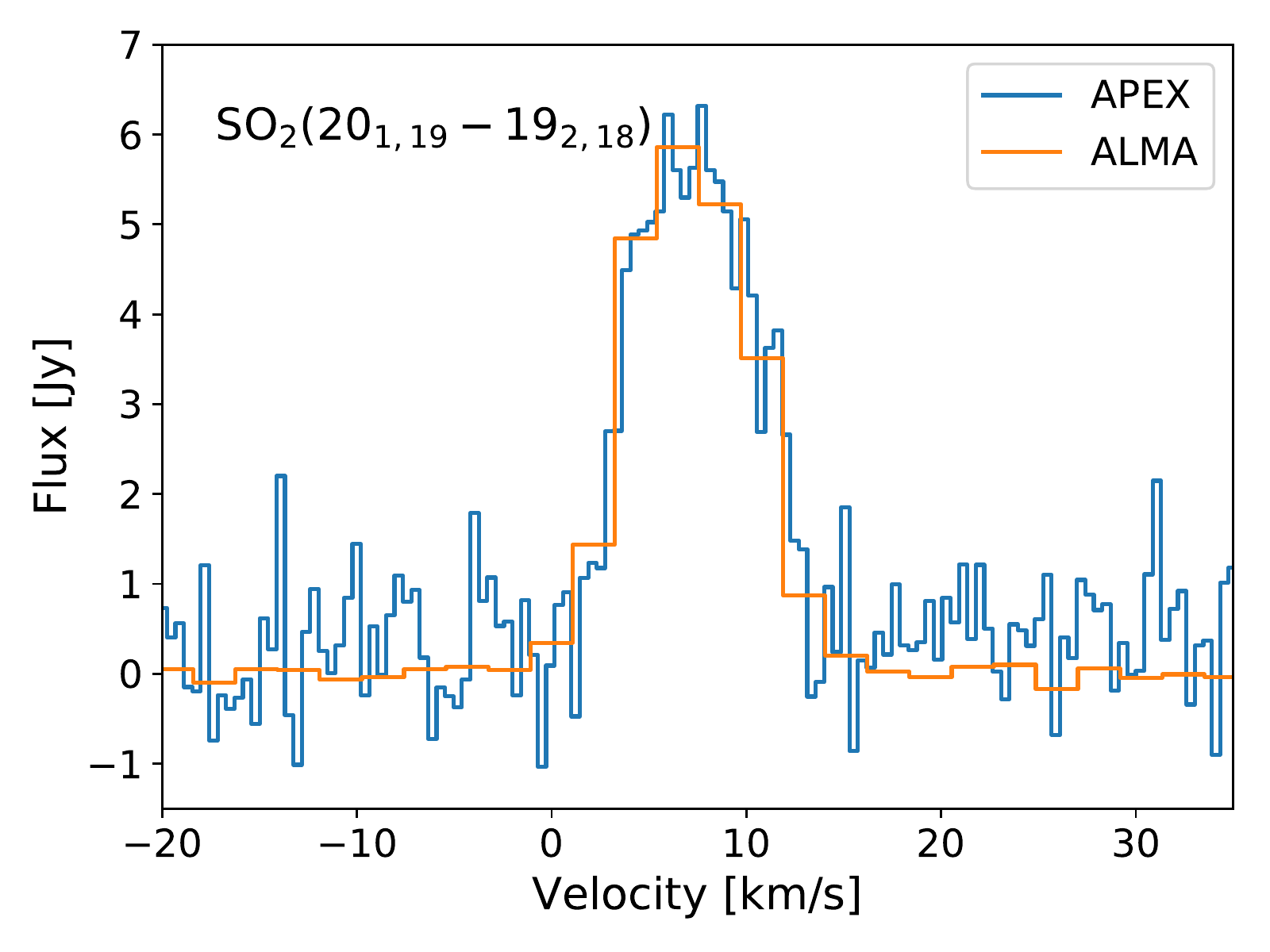}
\includegraphics[width=0.35\textwidth]{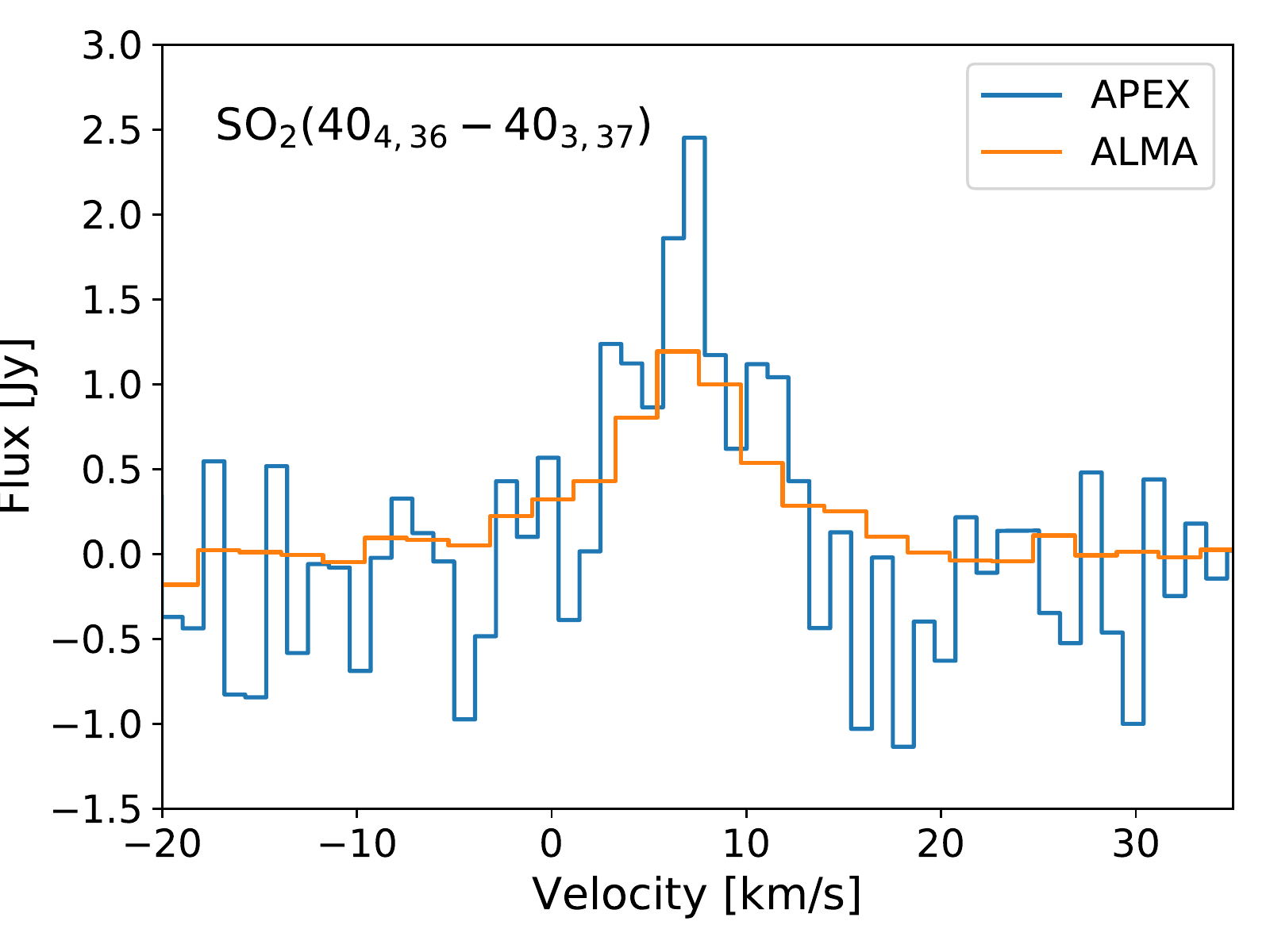}
%\endminipage
\caption{Comparison of the APEX (\textit{blue}) and ALMA (\textit{orange}) observations of four key transitions of \so2 towards R~Dor. Note that not all of the flux has been recovered in the lower-energy ALMA spectra. The ALMA spectra were extracted using a 5\arcsec{} radius aperture.}
\label{rdorso2lostflux}
\end{figure}

{ Our ALMA observations cover a large number of \so2 lines and a broad range of level energies are involved in producing these lines.}
Since we also expect the emitting regions of these lines to be spread out in the CSE, we check for any resolved out flux using four lines of different energies. In Fig. \ref{rdorso2lostflux} we show these lines and compare their total flux with the corresponding APEX observations \citep[originally presented in][]{Danilovich2016}. For the lowest energy lines, $(5_{3,3}\to 4_{2,2})$ with $E_\mathrm{up} = 36$~K and  $(13_{4,10}\to 13_{3,11})$ with $E_\mathrm{up} = 123$~K, we find that some of the large scale flux has indeed been resolved out. For the higher energy lines, $(20_{1,19}\to19_{2,18})$ with $E_\mathrm{up} = 199$~K and $(40_{4,36}\to40_{3,37})$ with $E_\mathrm{up} = 808$~K, all the flux has been recovered (although the noise in the latter APEX observation makes this difficult to be absolutely certain of).

\begin{figure*}
\centering
\includegraphics[width=\textwidth]{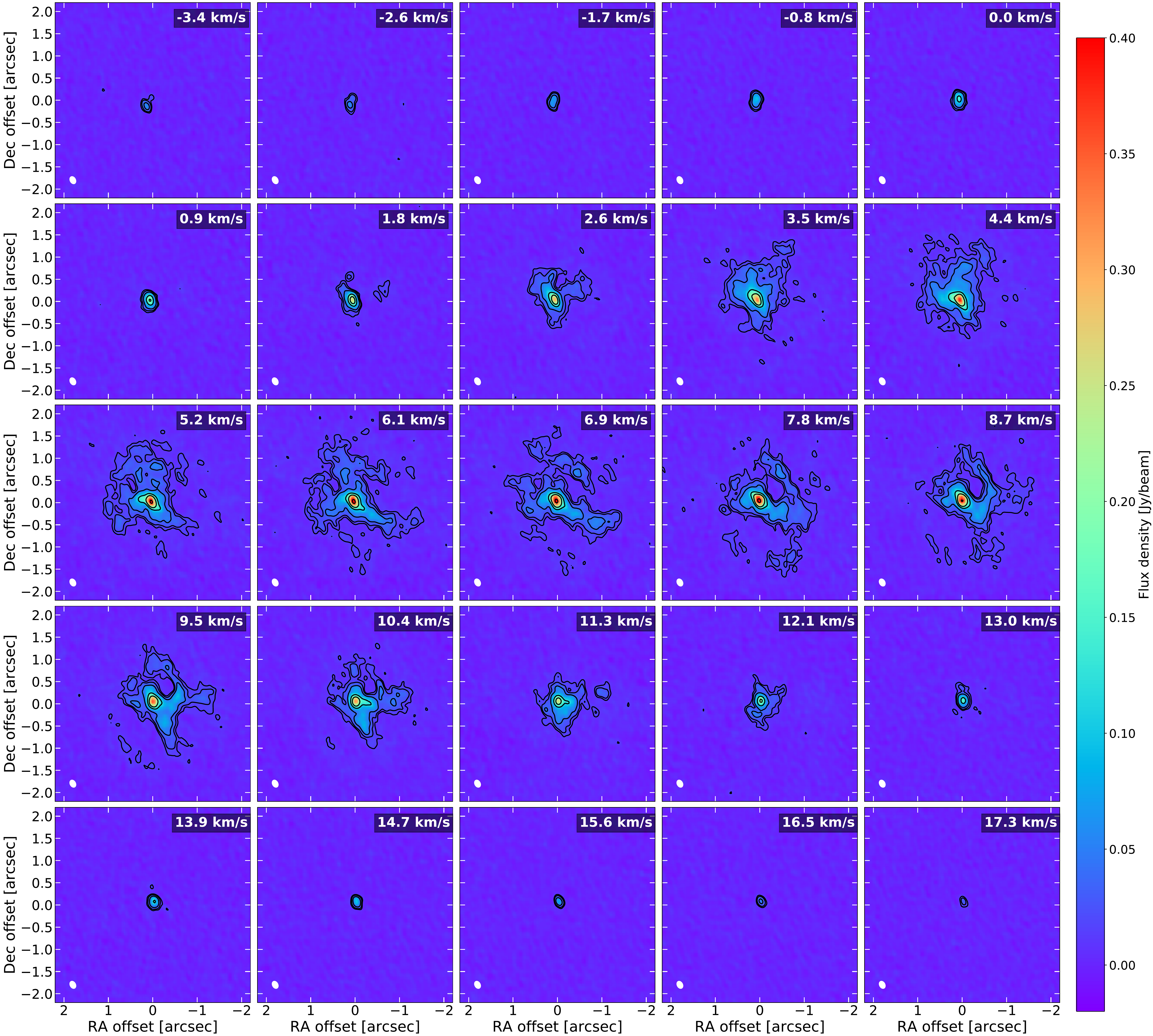}
\caption{R~Dor SO$_2$ $(20_{1,19}\to19_{2,18})$ channel maps. The contours show flux levels at 3, 5, 10, 30, 50, and 100 times the rms noise and the beam is shown in white in the bottom left hand corners of each channel plot. Plots are best viewed on a screen.}
\label{rdorso2chanmaps}
\end{figure*}

\begin{figure*}
\begin{center}
\includegraphics[width=0.85\textwidth]{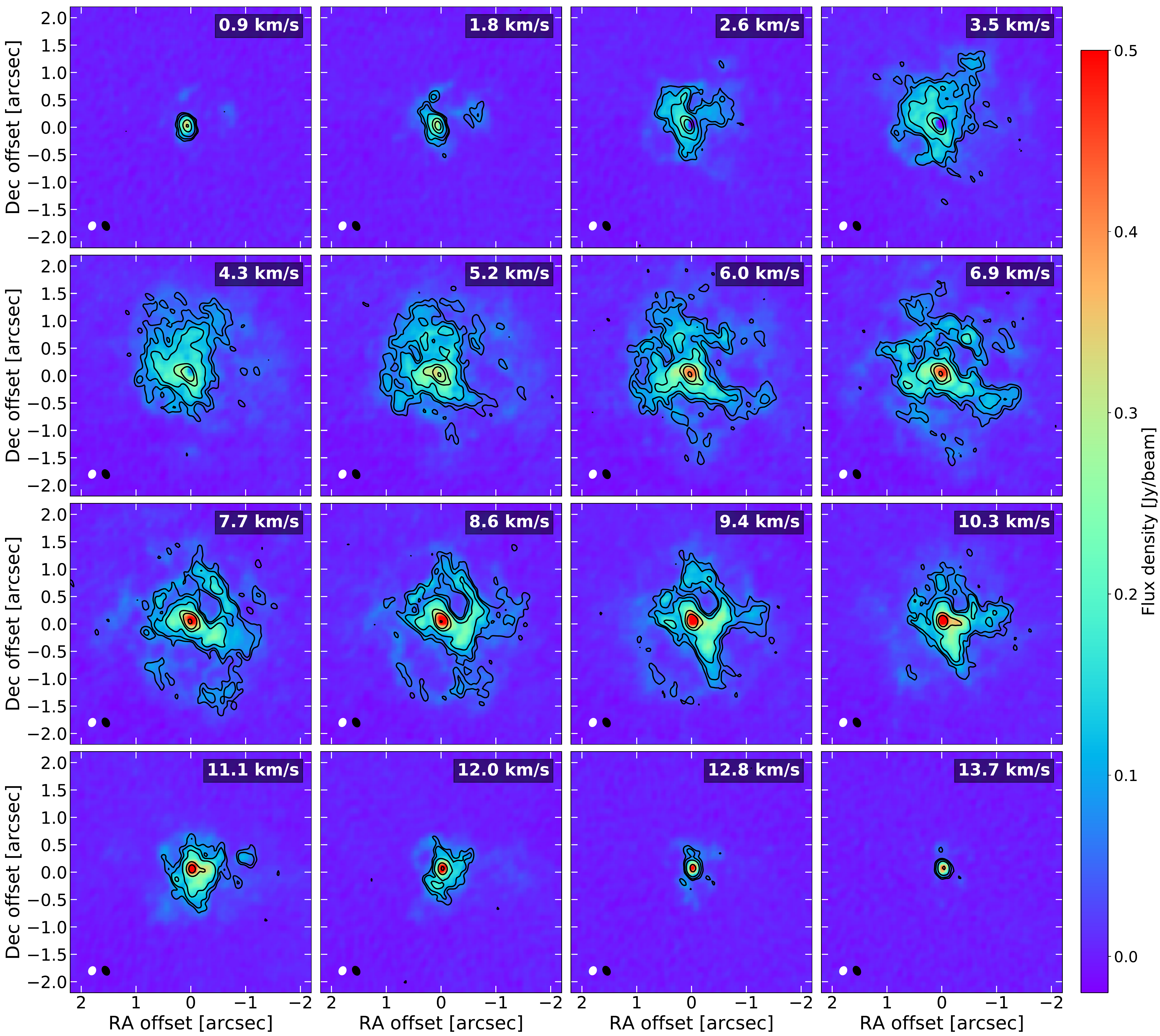}
\caption{Channel maps of SO and SO$_2$ towards R~Dor. The background colours show the SO ($8_8\to7_7$) transition, with the beam for those observations indicated in white in the lower left corners of each channel plot. The black contour plots show flux levels at 3, 5, 10, 30, 50, and 100 times the rms noise for the SO$_2$ ($20_{1,19} \to 19_{2,18}$) transition, with the beam for those observations shown in black in the bottom left hand corner of each channel plot. Plots are best viewed on a screen.}
\label{rdorso2onso}
\end{center}
\end{figure*}

In Fig. \ref{rdorso2chanmaps} we show the channel maps for the \so2 $(20_{1,19}\to19_{2,18})$ transition towards R~Dor. { Similar features can be seen as for the SO channel map shown in Fig. \ref{rdorsochannels} and described in Sect. \ref{so_overview}.} To check whether the structure is indeed the same --- since the SO channels maps show brighter and larger areas of emission --- we plotted the contours of SO$_2$ $(20_{1,19}\to19_{2,18})$ over the SO $(8_8 \to 7_7)$ emission in Fig. \ref{rdorso2onso}. As can be clearly seen, SO$_2$ does indeed trace out similar structures to { those seen in }the SO emission. 

\begin{figure*}
\includegraphics[width=\textwidth]{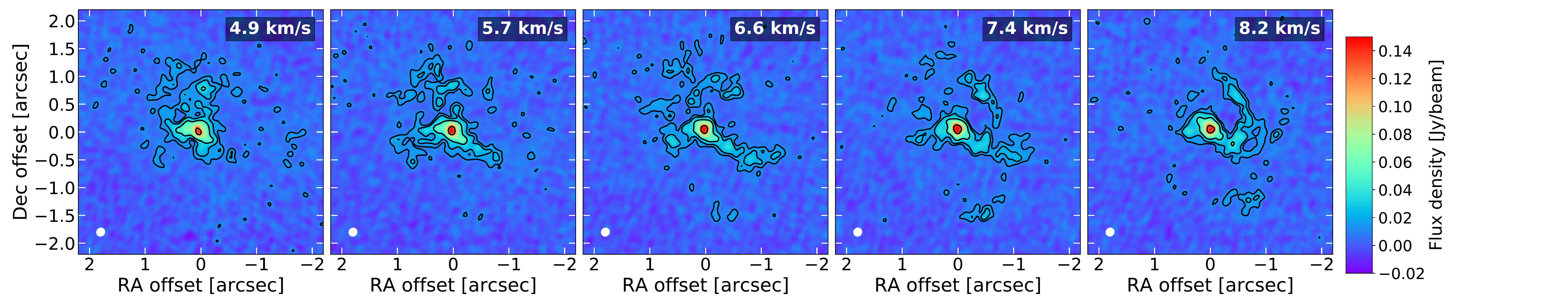}
\includegraphics[width=\textwidth]{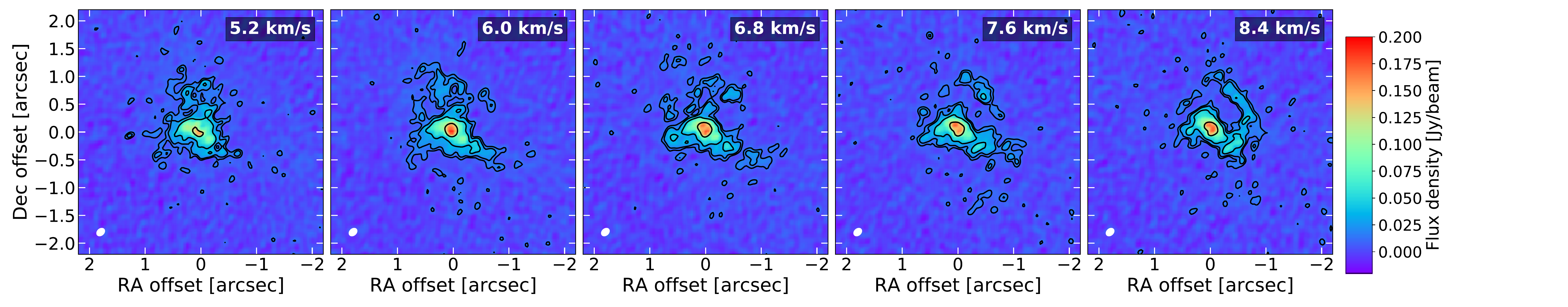}
\includegraphics[width=0.995\textwidth]{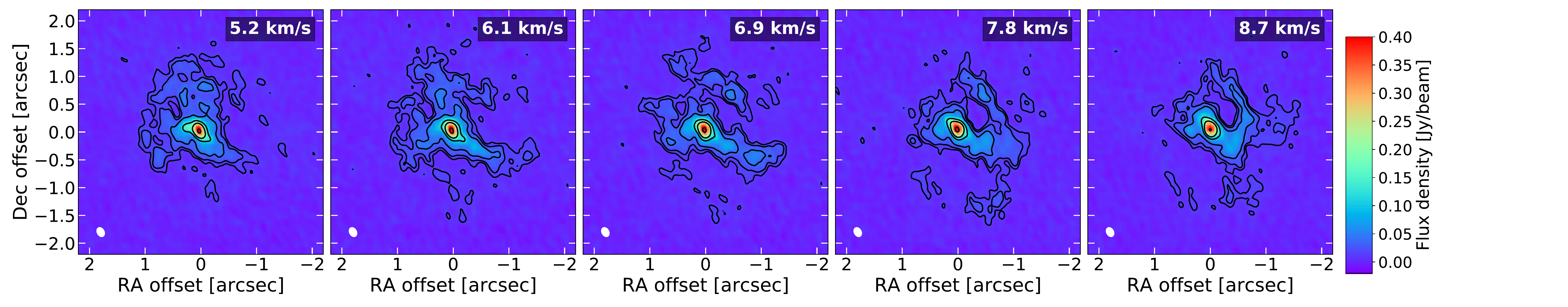}
\includegraphics[width=0.995\textwidth]{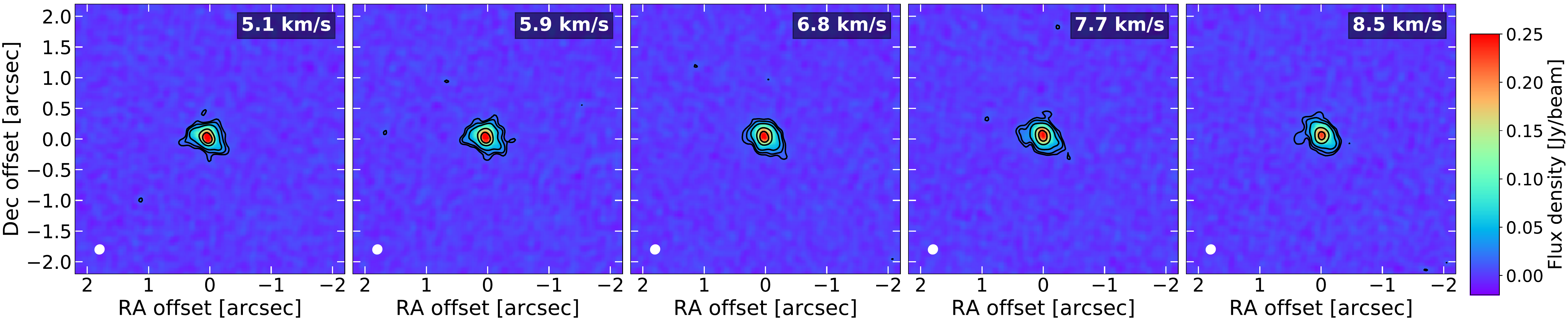}
\caption{The central channels of four SO$_2$ lines towards R Dor. From the lowest { level-}energy line in the top row to the highest { level-}energy line in the bottom row, we show the $(5_{3,3}\to 4_{2,2})$, $(13_{4,10}\to 13_{4,11})$, $(20_{1,19}\to19_{2,18})$, and $(40_{4,36}\to40_{3,37})$ lines. The contours show flux levels at 3, 5, 10, 30, 50, and 100 times the rms noise and the beams are shown in white in the bottom left hand corners of each channel plot. Note that for the two lowest { level-}energy lines some of the large scale flux has been resolved out. Plots are best viewed on a screen.}
\label{rdorso2centralchancomparison}
\end{figure*}

Although some of the flux is resolved out for the lower-energy SO$_2$ lines, it is still interesting to compare the distributions of the emission for different energy transitions. In Fig. \ref{rdorso2centralchancomparison} we plot the 5 central channels for each of the $(5_{3,3}\to 4_{2,2})$,  $(13_{4,10}\to 13_{4,11})$, $(20_{1,19}\to19_{2,18})$, and $(40_{4,36}\to40_{3,37})$ lines. Despite some of the larger scale emission being resolved out for the two lowest energy lines, it is clear that they follow similar spatial distributions to the $(20_{1,19}\to19_{2,18})$ line, tracing out some of the same { structures}. The highest energy $(40_{4,36}\to40_{3,37})$ line, however, is much more spatially confined, with all the emission coming from { within 0.5\arcsec{} of} the star. The trends seen for these four lines are consistently seen for the other SO$_2$ lines (that do not participate in overlaps { with any other lines}) listed in the upper part of Table \ref{so2linelist}; the lower energy lines have more extended emission while the higher energy lines exhibit more confined emission. 

The vibrationally excited lines all exhibit compact emission, as do the less favoured lines in the lower part of Table \ref{so2linelist}. The emission from both of these groups of lines is generally spatially unresolved and centred on the continuum peak. It is possible that two different causes lead to the same effect here. While the vibrationally excited lines are most likely emitted from regions close to the star, the less favoured lines with lower-energy levels could just be too faint in the outer regions of the star to be detected, giving the illusion that their emission is only coming from the central regions.

\subsubsection{IK Tau}\label{iktauso2text}

\begin{figure}
\begin{center}
\includegraphics[width=0.5\textwidth]{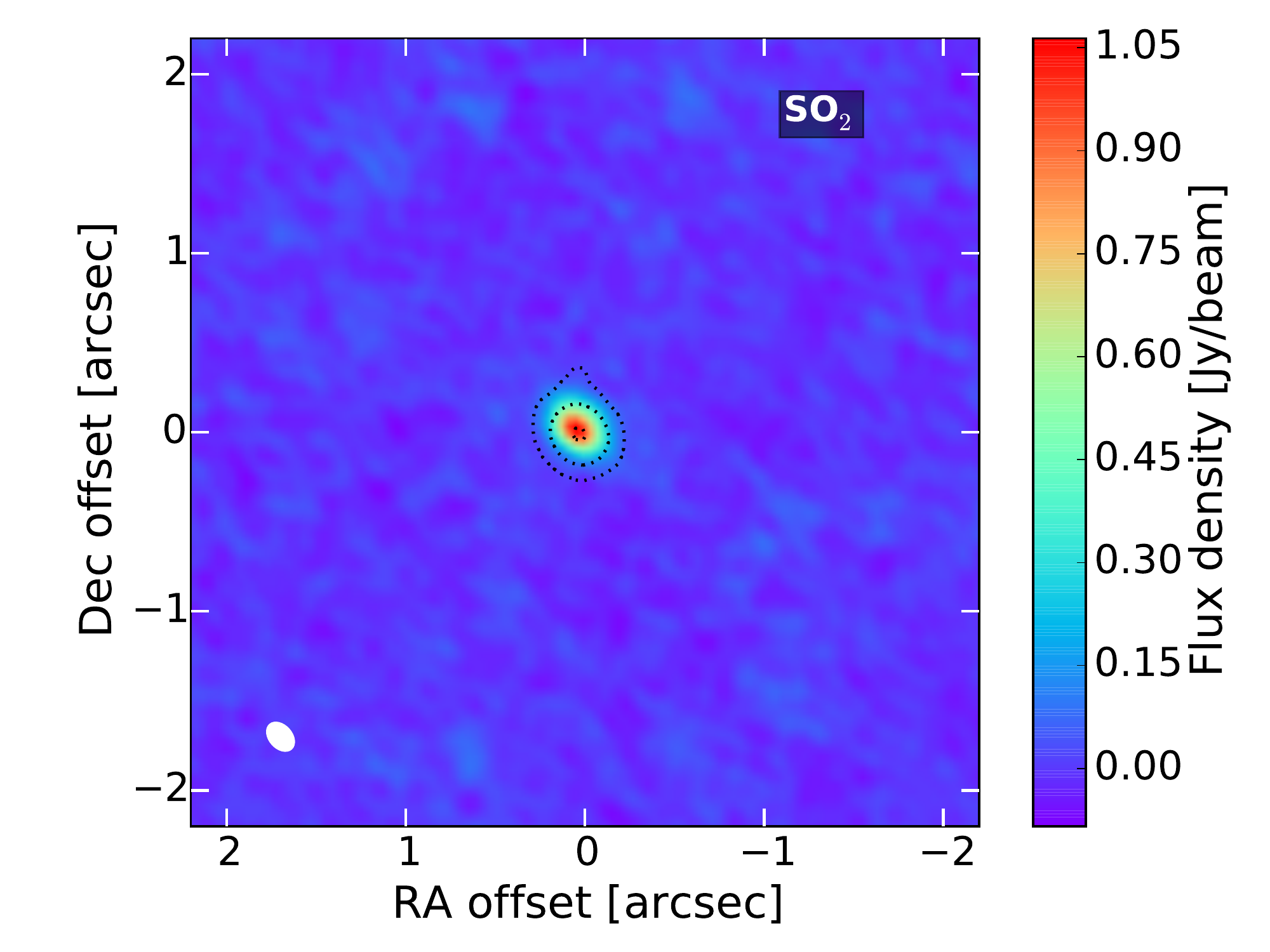}
\caption{The zeroth moment map of the \so2 ($20_{1,19}\to19_{2,18}$) line at 338.6118~GHz towards IK~Tau. The dotted black lines show the continuum flux at 1\%, 10\% and 90\% of the peak continuum emission and the restoring beam size is indicated by the white ellipse in the bottom left corner.}
\label{so2mom0iktau}
\end{center}
\end{figure}

All the \so2 lines detected towards IK~Tau are dominated by compact, spatially unresolved emission centred on the stellar continuum peak. For example, see the zeroth moment map of the ($20_{1,19}\to19_{2,18}$) line in Fig.~\ref{so2mom0iktau}. 
The spectra of the \so2 lines are characterised by profiles much narrower than expected based on the expansion velocity, with a half width of $\sim$10~\kms{} compared with the expansion velocity of 17.5~\kms{} found from single dish observations of CO \citep{Maercker2016}  and earlier detections of \so2 \citep{Omont1993,Kim2010,Decin2010,Danilovich2016,Velilla-Prieto2017}. They are also much narrower than the wide wings (up to expansion velocities of $\sim$25~\kms) observed for other molecules  towards IK~Tau in the same ALMA dataset by \cite{Decin2018}. Spectra of all the lines of \so2 lines observed with ALMA are shown in Fig.~\ref{iktauso2lines}, where we have also indicated 
the { LSR velocity \citep[$\upsilon_\mathrm{LSR}=34$~\kms,][]{Decin2018},} which corresponds well with the intensity peaks. 
In addition to a narrow central peak, many of the lines are wider at the base, more closely corresponding to the 
expansion velocity of 17.5~\kms derived from the single antenna observations. 

The only observation with ALMA that we were able to directly compare with an earlier single antenna observation was the ($5_{3,3}\to4_{2,2}$) line at 351.2572~GHz that was observed with APEX by \cite{Kim2010}. 
In Fig.~\ref{iktaulostflux} we compare this APEX observation with our observation with ALMA extracted for circular apertures with radii
of 0.8\arcsec{} and 5\arcsec{} centred on the continuum peak. 
Although the ($5_{3,3}\to4_{2,2}$) line is present in Fig.~\ref{iktauso2lines}, when it is compared with the  line observed with 
APEX it is apparent that most of the flux has been resolved out and less than 3\% of the flux has been recovered by ALMA. 
On the basis of the shapes of lines of other molecules observed with ALMA, and with the shapes of lines of \so2 lines observed
with single antennas \citep{Danilovich2016}, we conclude that a similarly large amount of flux has been resolved out for most of the lines of \so2 towards IK~Tau. 
A possible exception are lines from high lying levels such as those in the $v=1$ excited vibrational level, and perhaps even lines
from the highest levels in the ground vibrational state.

\section{Modelling and analysis}\label{sect:mod}

Both IK~Tau and R~Dor were included in the \citet{Danilovich2016} study of SO and \so2 molecules. That study involved finding the radial abundance profiles of both molecules around both stars. \textsl{Herschel}/HIFI data was available for both stars, but a much larger set of observations from ground-based single-dish telescopes (mostly APEX) was available for R~Dor, making the abundance determinations more reliable for R~Dor than for IK~Tau. Now, with the newly available ALMA observations, we ought to be able to confirm and possibly refine those earlier single-dish results. It should be noted, however, that the model used by \citet{Danilovich2016} is spherically symmetric and hence cannot take into account the asymmetric features seen in the ALMA observations (especially towards R~Dor). This and other issues are discussed in detail for R~Dor in Sect. \ref{rdoranalysis} and for IK~Tau in Sect. \ref{iktauanalysis}.

The modelling we perform in this section uses the same procedure as \citet{Danilovich2016} with modifications discussed in the text as relevant. We use a one-dimensional accelerated lambda iteration model, which assumes a spherically symmetric CSE with a smoothly accelerating wind. The stellar parameters of the models for R~Dor and IK~Tau are given in Table \ref{stellarparam}, as are the parameters of the molecular models found by \citet{Danilovich2016}. Alterations made to the molecular parameters in this work are discussed in the text.

\begin{table}
\caption{Stellar parameters for modelling.}\label{stellarparam}
\begin{center}
\begin{tabular}{lccc}
\hline\hline
 Property & Units  & IK Tau & R Dor \\
 \hline
Stellar luminosity, $L_*$ & L$_\odot$ 	&	7700	&	6500	\\
Distance, $D$& pc 	&	265	&	59		\\
LSR velocity, $\upsilon_\mathrm{LSR}$& \kms 	&	34	&	7	\\
Expansion velocity, $\upsilon_\infty$&  \kms 	&	17.5	&	5.7		\\
Stellar temperature, $T_*$ & K 	&	2100	&	2400\\
Model inner radius, $R_\mathrm{in}$ &$10^{14}$ cm	&	2.0	&	1.9		\\
Mass-loss rate, $\dot{M}$ & $10^{-7}$\spy 	&	 $50$ 	&	 $1.6$ 	\\
\hline
\multicolumn{4}{c}{SO parameters from \citet{Danilovich2016}} \\
\hline
Peak abundance, $f_{p}$ &\e{-6} 	&	 $1.0\pm0.2$ 	&	$6.7\pm0.9$ 		\\
$e$-folding radius, $R_{e}$ &\e{15} cm 	&	 - 	&	 $1.4\pm0.2$ 		\\
Radius at $f_p$, $R_{p}$ &\e{15} cm 	&	 $13\pm2$ 	&	 - 	\\
\hline
\multicolumn{4}{c}{SO$_2$ parameters from \citet{Danilovich2016}} \\
\hline											
Peak abundance, $f_{p}$ &\e{-6} 	&	0.86	&	5.0	\\
$e$-folding radius, $R_{e}$ & \e{15} cm	&	10	&	1.6	\\
\hline
\end{tabular}
\end{center}
\end{table}

%Background on \citet{Danilovich2016} abundances results for both R Dor and IK Tau.
%
%Something on the parameters and code.
%
%Spherically symmetric model, obviously can't take clumpiness properly into account. Assume an average radial abundance profile.
%
%Also mention here how the azimuthally radial profiles were extracted from the ALMA data.

\subsection{R Dor}\label{rdoranalysis}
\subsubsection{SO analysis}\label{rdorsoanalysis}

Plotted in Fig. \ref{rdorsospectra} are the spectra of the five lines of SO in the ground vibrational ground state 
observed towards R~Dor and extracted for apertures with small (75 mas), intermediate (300 mas) and large (1\arcsec) radii. 
The plots show how the line shapes and intensities change with increasing extraction radius. 
Spectra extracted with the small  (75~mas) aperture reveal broad wings and the ``blue hole'' feature a few \kms{} bluewards of the $\upsilon_\mathrm{LSR}$. 
For larger extraction radii, the extended emission at the central velocity channels dominates the line profile shapes, and although
the wings are still present they become less prominent with respect to the rest of the line profile. 
The wings are an indication of the departure from spherical symmetry, and cannot be modelled with a spherically symmetric description of the CSE.

\begin{figure*}
\begin{center}
\includegraphics[width=\textwidth]{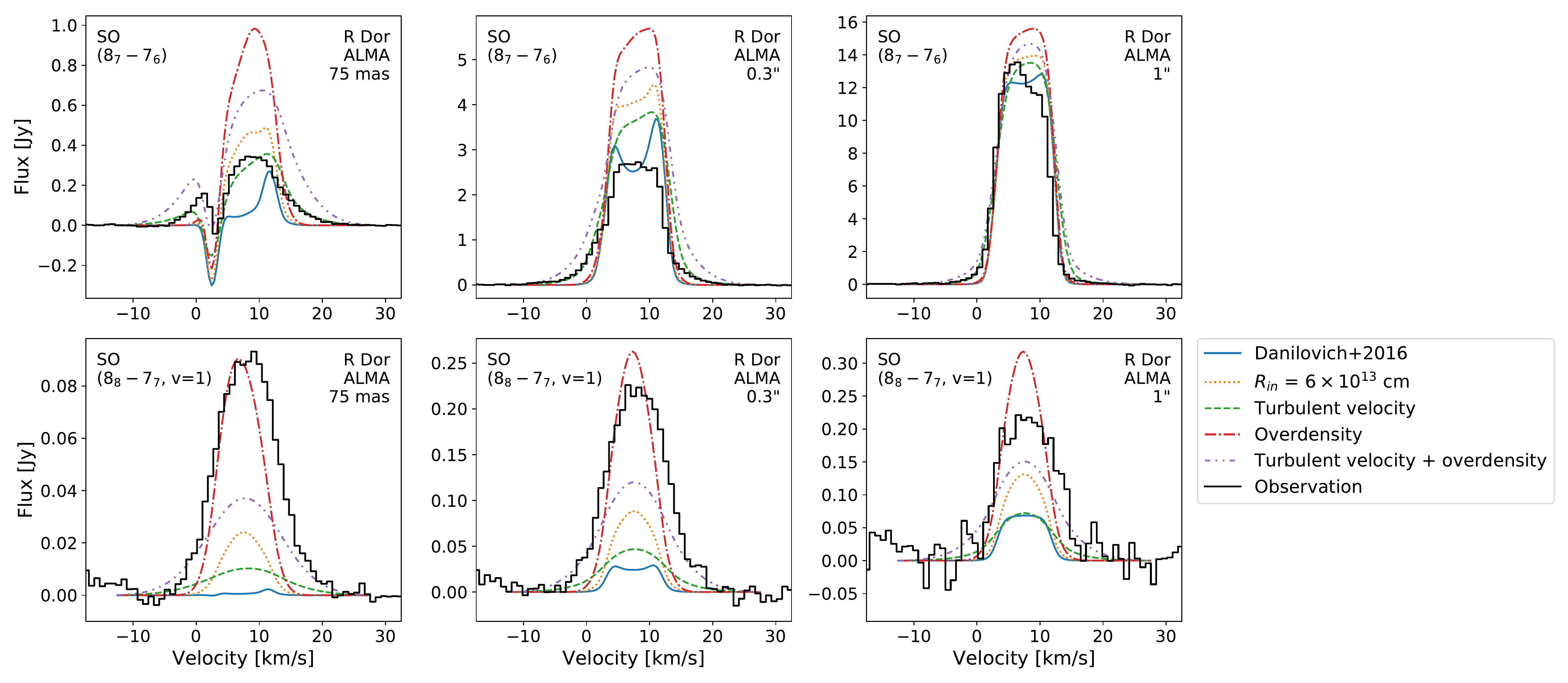}
\caption{The SO models for R~Dor discussed in the text (\textit{curves}) plotted against the ALMA spectra (\textit{black histograms}) for different spectral extraction radii. The $v=0$ ($8_7\to7_6$) line is shown along the top row and the $v=1$ ($8_8\to7_7$) line is shown along the bottom row.}
\label{rdorsooriginal}
\end{center}
\end{figure*}

\citet{Danilovich2016} modelled SO based on 17 lines observed using APEX and \textsl{Herschel}/HIFI. The final best-fit model in that study had a { radial} Gaussian abundance distribution for R~Dor, with a peak SO abundance of $f_p = 6.7$\e{-6} relative to \h2 and an $e$-folding radius of $R_e = 1.4$\e{15}~cm (Table \ref{stellarparam}). We start by comparing that SO model with the ALMA spectra extracted for an aperture radius of 1\arcsec. In that case the $v=0$ lines are in good agreement with the model, comparable to, or better than, the fits to the APEX lines used to originally find the model. The $v=1$ lines, however, are significantly under-predicted by the model and the observed lines are also a lot wider than the model lines (even excluding the line participating in an overlap with \so2). An example of this is shown in Fig. \ref{rdorsooriginal}, { plotted as a solid blue curve}. Comparing the model results with smaller spectral extraction apertures of 300~mas and 75~mas radii, the goodness of fit decreases for the smaller radii. For the $v=0$ lines, the line fits could be considered adequate aside from the high velocity wings, which become more prominent for most of these lines and are not at all reproduced by the model. For the 75~mas lines, the wings dominate the observed line profiles, making for the worst model fit. The small-aperture model lines are also a lot less bright than the observed lines, but this is mostly due to the inner radius of the model being set at too high a value ($1.9\times10^{14}$~cm $\approx0.2$\arcsec{} at 59~pc). The $v=1$ lines are under-predicted to an even greater extent for the smaller extraction apertures.
The above issues are shown for two example lines, one each of $v=0,1$, in Fig. \ref{rdorsooriginal}. { Note that the model lines have been ray-traced anew to match the ALMA observations (albeit with a higher velocity resolution), as have all subsequent model lines discussed here.}

Several refinements were made to the original model of \citet{Danilovich2016} with the goal of improving the fit to the ALMA lines. For the most part, the adjustments described below did not have a significant effect on the model fits to the single-dish lines, nor the ALMA $v=0$ lines extracted from a 1\arcsec{} radius region. 
Decreasing the inner radius to $6\times10^{13}$~cm ($\approx0.07$\arcsec{} at 59~pc and $\approx2R_*$) improved the fit to the smallest extraction radius lines, but did not significantly change the other results. 
To reproduce the wide wings seen in the $v=1$ lines and (most prominently) in the $v=0$ lines extracted at the smallest radius, we introduced a high turbulent velocity in the innermost regions in the CSE. Inner turbulent velocities of 11~\kms{} (decreasing to 1~\kms{} in the outer regions following $\upsilon_\mathrm{turb}(r)=1+0.15r^{-1}$) reproduced the line shapes of the 75~mas $v=0$ lines well. However, the $v=1$ lines were still significantly under-predicted, albeit now as broad as the observed lines. 
We had some success in reproducing the $v=1$ lines by introducing an overdense region close to the star (a density increase by a factor of 8 between $\sim 3~R_*$ and $4~R_*$). 
Combining this with the aforementioned increase in the inner turbulent velocity gave us good fits to the $v=1$ lines. However, this model over-predicted the the 75~mas $v=0$ lines. 
Listed in Table \ref{rdorsomods} are the refinements in our radiative transfer model, and plotted in Fig. \ref{rdorsooriginal} are
the calculated and observed spectra. 

Ultimately, we were unable to simultaneously reproduce the shape of the spectral lines taken for the smallest aperture and the intensities of the $v=1$ lines using a 1D model. Given the parameters that improved our fit, it seems that our observations are in agreement with the result of \cite{Homan2018} of a rotating disc around R~Dor, close to the star. { In that study, the authors found a disc in the region between 6~au and $\sim$25~au, with a scale height of 0.9~au and an inclination of $110^\circ$.}
Our increased turbulent velocity of 11~\kms{} could be thought of as a 1D approximation of the rotation of the disc (where the maximal disc velocity is 12~\kms, { equal to the Keplerian orbital velocity at 6~au}), while the overdensity, { which starts at the inner disc radius}, could be seen as a 1D approximation of the denser disc region. Of course, in 1D we cannot properly represent the disc and hence we plan to model the disc in three dimensions in a future paper. We hope to then reproduce the disc and extended SO emission simultaneously, since the model in \cite{Homan2018} only reproduced the $v=1$ SiO emission which mostly arose from the disc region.

\begin{table*}
\caption{R Dor SO parameters for model adjustments shown in Fig \ref{rdorsooriginal}.}\label{rdorsomods}
\begin{center}
\begin{tabular}{ll}
\hline\hline
 Model name & Details \\
 \hline
Danilovich+2016 & Model from \citet{Danilovich2016}, parameters listed in Table \ref{stellarparam}\\
$R_\mathrm{in}=6$\e{13}~cm & As above, but with the inner radius, $R_\mathrm{in}$ decreased to 6\e{13}~cm\\
Turbulent velocity & $R_\mathrm{in}=6$\e{13}~cm and with turbulent velocity described by  $\upsilon_\mathrm{turb}(r)=1+0.15r^{-1}$\\
Overdensity & $R_\mathrm{in}=6$\e{13}~cm and an increased \h2 number density by a factor of 8 between $\sim3R_*$ and $4R_*$\\
Turbulent velocity + overdensity & Both turbulent velocity and overdensity adjustments listed above\\
\hline
\end{tabular}
\end{center}
\end{table*}	

\subsubsection{SO$_2$ analysis}

In Table \ref{so2linelist}, we have marked some spatially unresolved lines with ``ID uncertain''. Lines falling into this category are all less favoured transitions, listed in the bottom section of Table \ref{so2linelist}. These lines, as well as most of the vibrationally excited lines in Table \ref{so2v1linelist} are not centred on the $\upsilon_\mathrm{LSR}=7\pm0.5$~\kms{} of R~Dor. In general, they are offset from the $\upsilon_\mathrm{LSR}$ by 1 to 2 \kms{} bluewards in the case of the vibrationally excited lines, and by 3 or 4 \kms{} bluewards in the case of the less favoured transitions. Examples of this behaviour are shown in Fig. \ref{rdorso2v1chanmaps} for a vibrationally excited line with an emission peak around 5~\kms{}, and in Fig. \ref{rdoroffcentreso2chanmaps} for a less favoured line with an emission peak around 3.5~\kms.
This velocity offset roughly corresponds to the location of the blue absorption feature seen in other lines, such as SO in Fig. \ref{rdorsochannels}, most of the lower-energy \so2 lines and CO, HCN, and SiO, as shown in \cite{Decin2018}, and the CS spectrum as shown in \cite{Danilovich2019}. As noted by \cite{Decin2018}, this absorption feature is { seen for observations where there is a large ratio between the stellar angular diameter and the angular beam size. It is the result of lines of sight passing through the star itself and is seen in other high resolution observations. For lower resolution images this absorption feature is masked by the brighter emission surrounding the star.}
However, when modelling CS, \cite{Danilovich2019} noted that the absorption feature given by the spherically symmetric model is offset from the observed blue hole by a few \kms. They attributed this to the rotating disc around R~Dor proposed by \cite{Homan2018}, which their 1D model could not properly take into account. The higher-energy emission we see in the same region could be because the dense and warm region of the disc is more likely to excite these lines than the cooler expanding regions of the CSE, { further from the star}. %A further discussion on the possible effects of the disc is given in Sect. \ref{rdorsoanalysis}.

\begin{figure*}
\begin{center}
\includegraphics[width=\textwidth]{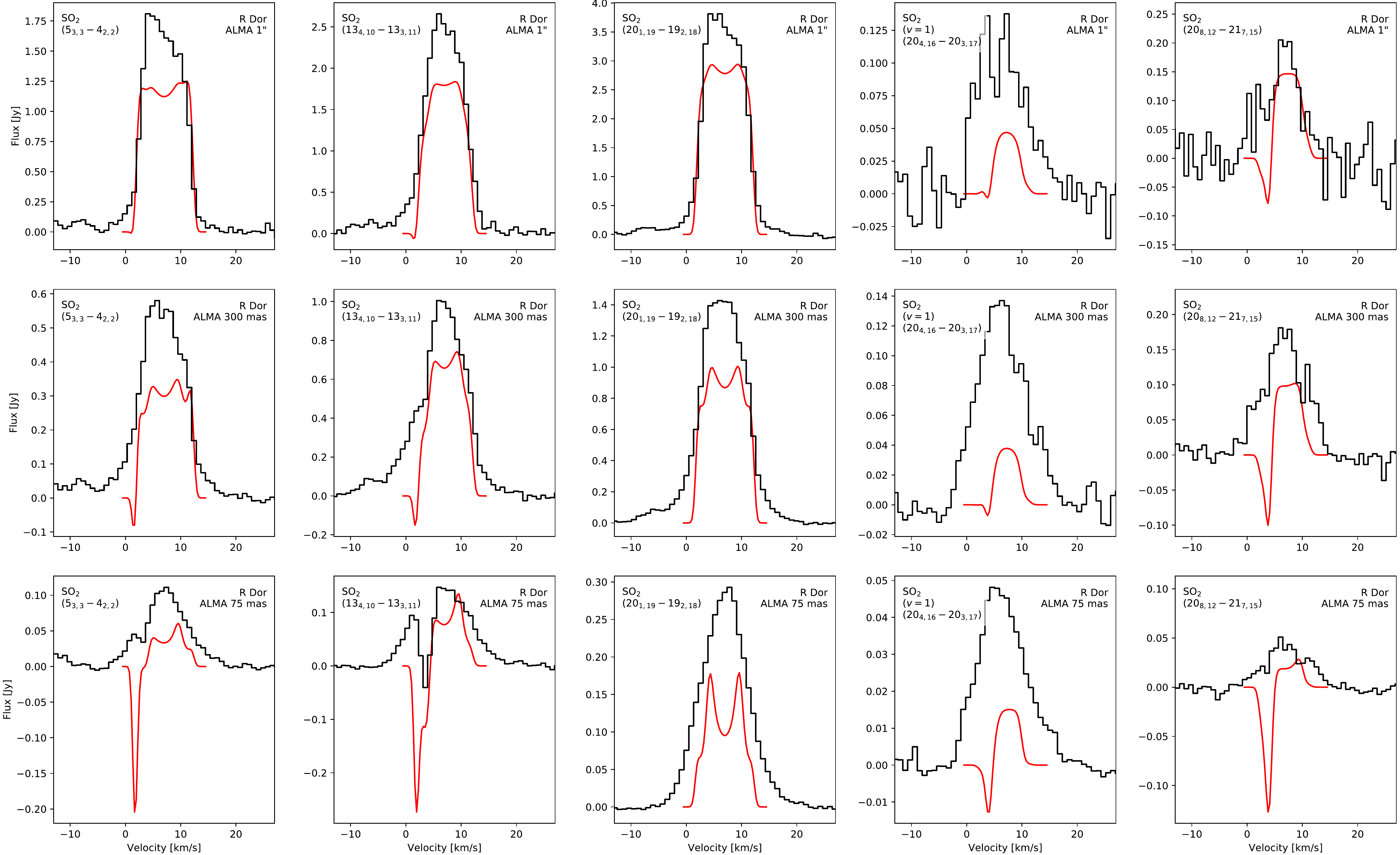}
\caption{The R~Dor \so2 model of \citet{Danilovich2016} (\textit{red curves}) plotted against the ALMA spectra (\textit{black histograms}) for different spectral extraction radii for the $(5_{3,3}\to 4_{2,2})$, $(13_{4,10}\to 13_{4,11})$, $(20_{1,19}\to19_{2,18})$, $v=1$ ($20_{4,16} \to 20_{3,17}$), and ($20_{ 8,12} \to 21_{ 7,15}$) \so2 lines.}
\label{rdorso2original}
\end{center}
\end{figure*}

Since we have indications of some flux being resolved out for some of the lower-energy \so2 lines, we do not expect the \cite{Danilovich2016} model to be a perfect fit to our ALMA observations. Comparing the $(5_{3,3}\to 4_{2,2})$, $(13_{4,10}\to 13_{4,11})$, and $(20_{1,19}\to19_{2,18})$ lines (excluding the  $(40_{4,36}\to40_{3,37})$ line because it is beyond the energy limit of the \cite{Danilovich2016} model), we find similar results as for our SO comparison, albeit with the model consistently under-predicting the ALMA lines by around 30\%, as seen in Fig. \ref{rdorso2original}. This suggests that a slightly higher abundance of \so2 is supported by the ALMA lines than by the predominantly APEX lines studied by \cite{Danilovich2016}. However, we run into the same wide wings for \so2 as for SO, and hence surmise that a 1D model cannot fully reproduce the effects of the disc in the inner regions on the \so2 emission. Referring to the model prediction for the $v=1$ ($20_{4,16} \to 20_{3,17}$) line, also shown in Fig. \ref{rdorso2original}, the very wide line is not reproduced by the model, as for the $v=1$ SO lines. The main difference for \so2 is that the model lines do not under-predict the emission as drastically, possibly due to optical depth effects (since the \so2 lines are in generally more optically thin). The less favoured ($20_{ 8,12} \to 21_{ 7,15}$) line, which we also compare with the model, follows similar trends to the brighter lines in the ground vibrational state.

\subsubsection{Isotopologues}\label{rdoriso}

If we have observations of transitions of different isotopologues with the same quantum numbers, it is possible to derive abundance ratios directly from optically thin emission lines (assuming no flux has been resolved out). To do this, we cannot take the direct ratios but need to account for differences in the line strengths\footnote{Since we have extracted the spectral lines from the ALMA cubes using identical extraction radii, we do not need to account for different beam filling factors. The purpose of the third factor of frequency (excluded here), commonly seen in isotopologue ratio calculations, is to take into account different beam sizes at different frequencies when observations are made using single-dish telescopes.} between isotopologues. Hence, to find the abundance ratio between two example isotopologues $^aX$ and $^bX$, we use:
\begin{equation}\label{lineratioeq}
^aX/^bX = \frac{I(^aX)}{I(^bX)} \left( \frac{\nu_{^bX}}{\nu_{^aX}} \right)^2
\end{equation}
where $I$ is the { integrated} line flux and $\nu$ are the frequencies of the transition for each isotopologue.

Using this method for R~Dor, and assuming that SO is a good indicator of S isotope abundances, we find $^{32}$S/$^{33}$S $=68\pm22$ {( when integrating over all the detected hyperfine components of $^{33}$SO)}. 
Using lines with the same quantum numbers for both SO and \so2 (and excluding lines that participate in overlaps) we find \mbox{$^{32}$S/$^{34}$S $=18.5\pm5.8$}, which is in  agreement with the result found by \citet{Danilovich2016} of $21.6\pm8.5$, from SO lines towards R~Dor observed by APEX and HIFI. Comparing our ALMA \ch{^{34}SO} lines with the model results and predictions of \citet{Danilovich2016}, we find them to be in good agreement, albeit with the same issues with wide wings that we encountered for \ch{^{32}SO} and \so2.
Finally, we find $^{33}$S/$^{34}$S $=0.17\pm0.02$ from the SO observations.

For S$^{18}$O the frequencies that fall inside our survey range cover the $N=9\to8$ lines, which were not observed for S$^{16}$O either in our ALMA survey nor by APEX in \citet{Danilovich2016}. As such, we cannot use Eq. \ref{lineratioeq} to find the abundance ratios. Such an endeavour would also be hampered by the very weak tentative detections that we have. Instead, we ran a radiative transfer model of S$^{18}$O, using a similar method as that used by \citet{Danilovich2016} for $^{34}$SO. We use rotational levels up to $N=30$ and for $v=0$, with level energies and Einstein A coefficients taken from CDMS \citep{Tiemann1974,Tiemann1982,Bogey1982,Lovas1992,Klaus1996}. Using a Gaussian abundance distribution and the same $e$-folding radius as \citet{Danilovich2016} found for SO (and neglecting the model adjustments made in Sect. \ref{rdorsoanalysis} since no wings or other features are seen for the comparatively faint S$^{18}$O lines), we find a peak S$^{18}$O abundance $\leq(2\pm0.5)\times10^{-8}$ relative to \h2, taking the tentative and nondetected lines as upper limits. 
%{ The model results are plotted with the observations detections in Fig. \ref{S18O}, for three different extraction apertures.} 
Combining this result with the S$^{16}$O abundance from \citet{Danilovich2016}, we find \ch{^{16}O/^{18}O} $\geq 335\pm 84$, which is in good agreement with the result of 315 found through the analysis of \h2O isotopologues towards R~Dor by \cite{Danilovich2017}. The good agreement of these results supports our tentative detections of \ch{S^{18}O}.
%I have a plot of the S18O model, but I don't think it's worth including.

\subsection{IK Tau}\label{iktauanalysis}
\subsubsection{SO analysis}

Plotted in Fig.~\ref{iktausospectra} are the spectra of the five lines of ground state SO observed towards IK~Tau extracted with small (80 mas), intermediate (320 mas) and large (800 mas) aperture radii. The fainter lines are observed more { clearly in the spectra extracted with} smaller apertures and are within the noise for the { spectrum extracted with the largest} aperture. 
The profile shapes of the three brightest lines change dramatically at the different aperture sizes: 
the lines observed with the smallest aperture are dominated by a narrow central peak, while the wings become increasingly more prominent at the larger apertures. 
This implies there is a significant amount of SO present at larger distances from the star and at higher velocities, which is consistent with the shell-like distribution found by \cite{Danilovich2016}.
Furthermore, the ratio of the peak intensity of the ($8_7\to7_8$) line to the other two bright lines shifts dramatically for the largest aperture, where the peak flux of the ($8_7\to7_6$) line is less than half that of the two other bright lines.
The peak flux of all three bright lines are similar at the smallest aperture.
We are unable to check whether flux has been resolved out for the ($8_7\to7_6$) or ($8_9\to7_8$) lines, therefore it is unclear whether this is a real phenomenon or the result of lost flux.

\begin{figure*}
\begin{center}
\includegraphics[width=0.515\textwidth]{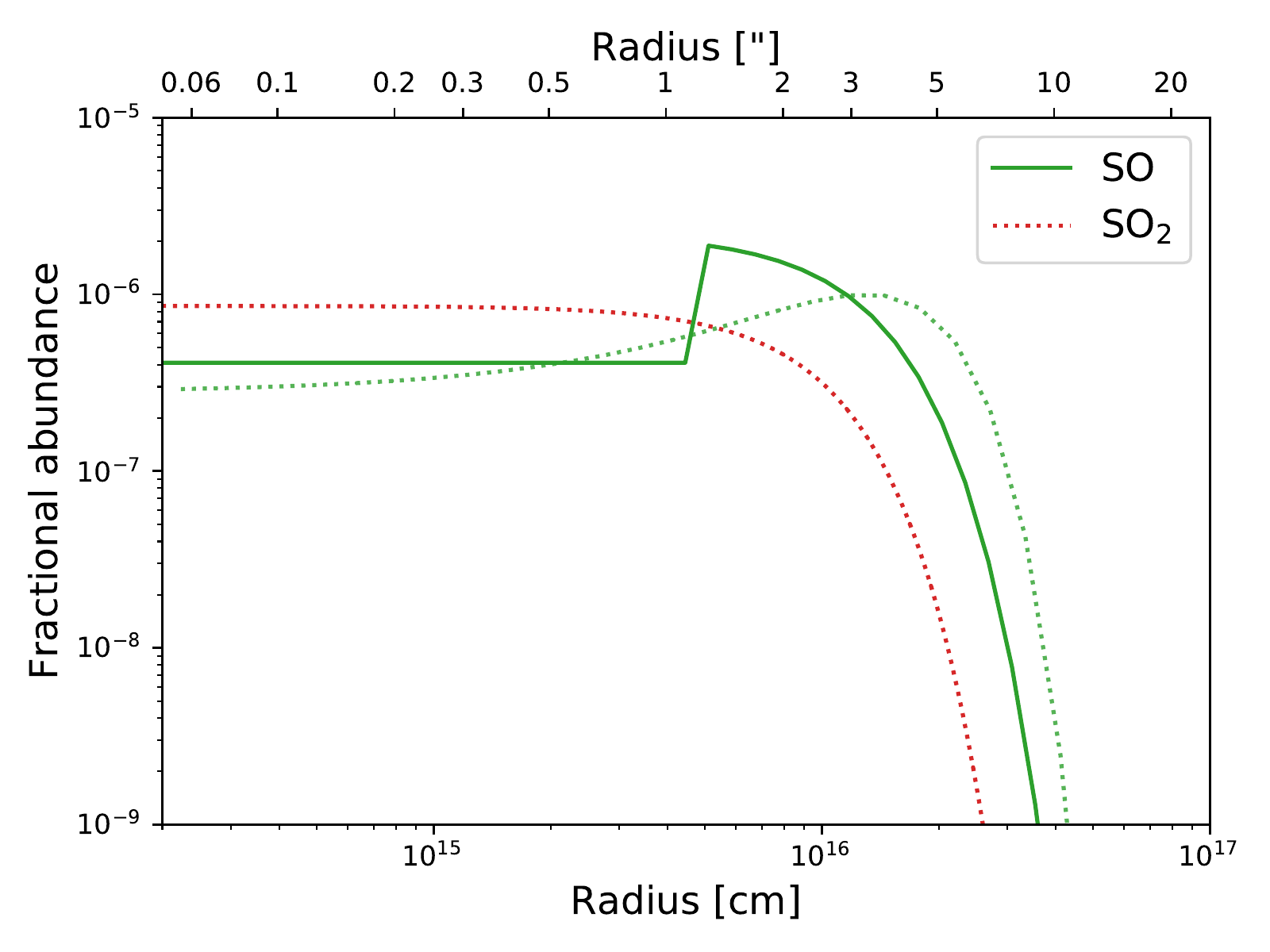}
\includegraphics[width=0.47\textwidth]{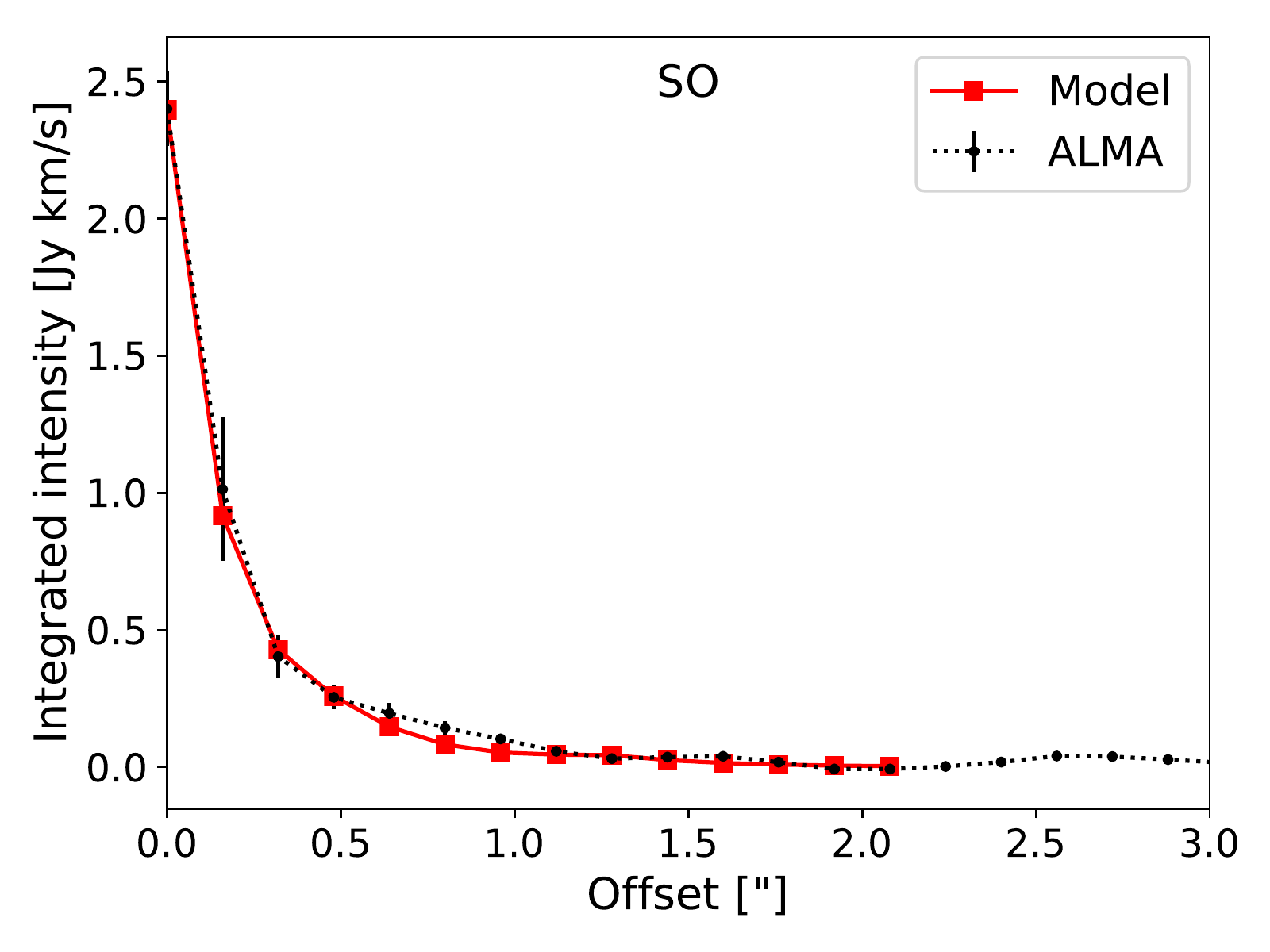}
\caption{\textit{Left:} the radial abundance profile for IK~Tau from our new SO model (solid green line) plotted with those for the SO (dotted green line) and \so2 (dotted red line) models of \citet{Danilovich2016}. \textit{Right:} The ALMA azimuthally averaged radial intensity distribution for IK~Tau (black dotted line) and the model radial intensity distribution (solid red line).}
\label{IKTauradabundance}
\end{center}
\end{figure*}

\begin{figure}
\includegraphics[width=\columnwidth]{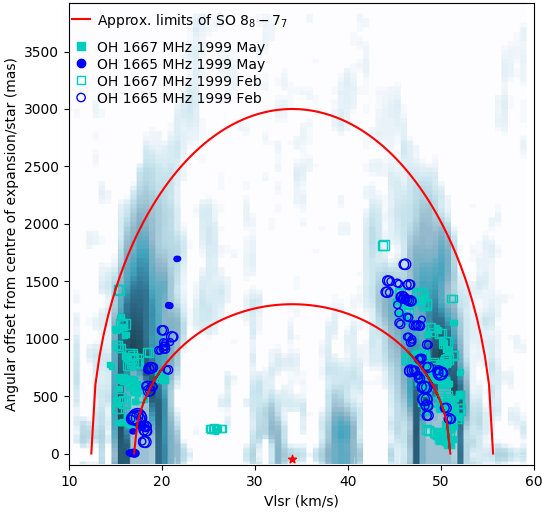}
\caption{ OH maser distribution for IK~Tau, showing: {\it background:} MERLIN observations of OH 1667 MHz maser emission at 200-mas resolution from 1993; {\it symbols:} components fitted to OH masers mapped using the EVN in 1999; {\it red lines:} the approximate location of the abundance peak of SO.}
\label{IKTaumasers}
\end{figure}

\citet{Danilovich2016} derived the abundance distribution of SO in IK~Tau from three lines observed with
\textsl{Herschel}/HIFI and seven lines with ground based single antennas and concluded:
 the radial abundance distribution of SO in IK~Tau is shell like, with a lower inner abundance and a peak farther out  
 in the CSE (see Table~\ref{stellarparam} and Fig.~\ref{IKTauradabundance}). { They described the radial abundance distribution with a Gaussian centred on the radius of the peak abundance.}
The abundance peak coincides with the peak in the OH abundance estimated from the $e$-folding radius of \h2O
and most likely is due to the formation of SO by the reaction of S with OH, rather than the formation of SO in the
inner wind by the reaction of SH with O.
{ Further evidence of the presence of OH at the relevant radii can be found from studies of OH masers.
\cite{Kirrane1987} mapped OH masers around IK Tau with MERLIN, fitting
shells with radii of $1\farcs3$ and $2\farcs5$ at 1667 and 1612~MHz,
respectively.  The corresponding expansion velocities were 19 and 17
km s$^{-1}$. IK Tau was re-observed in 1993 and 2001 using MERLIN with
additional, longer baselines and greater sensitivity, resolving a
complex structure with multiple arcs or possibly a bicone. Emission
extends out to $3\farcs3$ at 1667~MHz, between 12 -- 53 km s$^{-1}$
with two sets of peaks, the outer at an expansion velocity $\sim$17 km
s$^{-1}$ and the inner (in projection) $\sim5$ km s$^{-1}$, within
$0\farcs5$ of the centre of expansion.  The 1665 MHz line has a
similar but less extended distribution whilst, as expected, 1612~MHz
is found only at the larger distances. European VLBI Network (EVN) observations at 1665 and
1667~MHz in 1999 resolved-out the more extended emission but confirmed
that the brightest OH masers are found at a radius of $1^{\prime\prime}$--$1\rlap{.}^{\prime\prime}5$
(assuming a CSE with approximate reflection symmetry) (Richards,
private communication).  See Fig. \ref{IKTaumasers} and note that absence of masing
indicates unfavourable conditions, not necessarily absence of OH.
}

%\footnote{For evidence of the presence of OH in the envelope of IK~Tau, see \cite{Kirrane1987}.} 
The aim in the present work was to determine the abundance distribution of SO in IK~Tau more precisely, because the observations 
with ALMA were done over a very short time span of two months, they are more sensitive, and they are spatially resolved. 
We also include the HIFI lines presented in \citet{Danilovich2016} and lower-$J$ lines obtained using the APEX telescope as part of the observing programme first discussed in \cite{Danilovich2017a}. The properties of the APEX lines included in our modelling are given in Table \ref{iktauapexso}.
To facilitate a comparison between model and observations, we extracted an azimuthally averaged radial profile for SO ($8_8\to 7_7$) from the zeroth moment map of the ALMA data. We chose this transition because it is the only one which was observed with a single antenna,
thereby allowing us to establish whether flux has been resolved out. 
The uncertainties in the azimuthally averaged radial profile are due to fluctuations arising from clumpiness or asymmetries in the distribution of the emission, which are taken into account in the error bars \citep[see][]{Decin2018}.

We again used the same 1D modelling procedure described in \cite{Danilovich2016} and compared observed and modelled radial profiles as described in \cite{Brunner2018}. Testing the best IK~Tau SO model found by \cite{Danilovich2016}, we found that it generally underpredicted the ALMA emission. Rather than adjusting the Gaussian shell radial abundance distribution used by \cite{Danilovich2016}, which is difficult to do in a precise manner when comparing with ALMA radial profiles, we used a model with a constant inner abundance, $f_0$, with a step up in abundance to $f_1$ at some radius, $R_1$, then a Gaussian decline with some $e$-folding radius, $R_e$. Such an abundance distribution was more straightforward to adjust to the ALMA radial profile. The best fitting model we found had $f_0= 4.1\times 10^{-7}$, $R_1= 5\times 10^{15}$~cm, $f_1= (2.2\pm0.4)\times 10^{-6}$, and $R_e= (1.3\pm0.3)\times 10^{16}$~cm, where the uncertainties are given for a 90\% confidence interval.
%The model spectral lines are plotted with the ALMA observations and the the older single-dish spectra in Fig. \ref{iktauSOmodellines}. There it can notably be seen that 
The model predicts that some flux has been resolved out for the $(3_3\to2_3)$ line. As expected our spherically symmetric model cannot reproduce some of the asymmetries seen in the line profiles in Fig.~\ref{iktausospectra}. 

The radial profile of the model is plotted on the same scale as the azimuthally averaged radial profile from ALMA 
in the right hand panel of Fig.~\ref{IKTauradabundance}: shown in the panel on the left is our newly derived abundance profile and the profile in \cite{Danilovich2016}. 
Comparing the observed radial intensity plot with the radial abundance distribution, 
%(particularly the arcsecond units given along the top horizontal axis), 
it is apparent the observations with ALMA largely determine the value of $f_0$, but the single antenna observations  are needed to constrain the outer Gaussian portion of the radial abundance profile (see Table~\ref{iktauapexso}).
%The measurements with ALMA are reproduced equally well if we use the plotted profile or a similar one  but with $R_e$ twice as large, however the low lying lines observed with the single antennas are over predicted by a factor of about two for the larger $R_e$.
{ The ALMA data fits equally well whether we use the abundance profile plotted in Fig.~\ref{IKTauradabundance}, or a similar profile but with $R_e$ twice as large ($R_e=2.6\times 10^{16}$~cm). In that case, however, the low-energy single dish lines are over-predicted by a factor of approximately two.}

\subsubsection{SO$_2$ analysis}

As noted in Sect. \ref{iktauso2text}, a lot of \so2 flux was resolved out towards IK~Tau. In our ALMA observations, we are able to image structures only up to angular scales of 2\arcsec. {This means that the lost \so2 flux must be smooth and extend over 2\arcsec{} or more}. Since some of the observed \so2 lines have some faint emission surrounding the narrow central peak, we attempted to better see this extended emission by stacking, in the uv-plane, most of the $v=0$ \so2 lines. Of course, this method cannot recover flux that has been resolved out, but can improve sensitivity. For the stacking procedure we included all the \so2 lines detected towards IK~Tau that did not participate in overlaps, were not closely adjacent to any other lines, and had $J<30$ --- since the highest energy lines are expected to be compact, as we saw for R~Dor, and hence are not expected to contribute much diffuse emission further away from the star. We also selected only most favoured lines, since the less favoured lines are very compact for R~Dor so it was unclear whether they would contribute to diffuse emission towards IK~Tau. 

\begin{figure}
\centering
\includegraphics[width=\columnwidth]{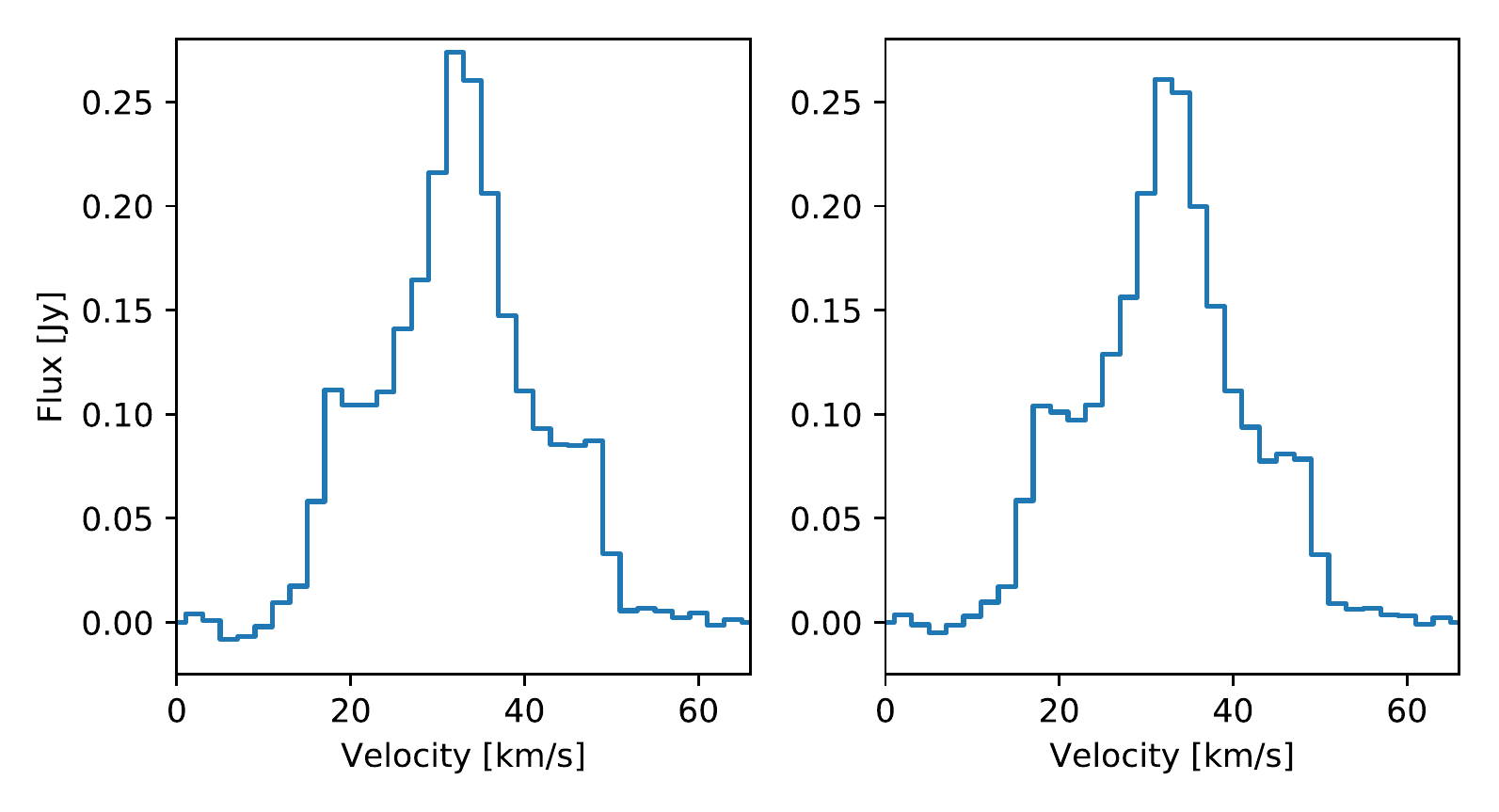}
\caption{Stacked IK~Tau \so2 plots. The \textit{left} plot shows the stacked spectrum with the normal baseline weighting while the \textit{right} plot shows the stacked spectrum with longest baselines given lower weightings so as to increase the sensitivity to large-scale flux. The spectra were extracted for a circular aperture with a 2\arcsec{} diameter.}
\label{iktaustack}
\end{figure}

Shown in Fig.~\ref{iktaustack} are two versions of the stacked spectrum. On the left is the full resolution spectrum, which makes full use of all the baselines and, on the right, the spectrum resulting from giving higher weightings to the shorter baselines to increase sensitivity to large-scale flux. 
{ The differences between the two plots are subtle, but the version with normal baseline weightings retrieves slightly more flux, especially of the smaller structures such as in the central peak and the flux at the highest and lowest velocities (e.g. see Fig. \ref{iktausochannels}). The stacked spectrum with lower weightings for the longest baselines does not seem to show an increase in (larger scale) flux.}
The broad component in these emission lines can be seen much more clearly in the stacked spectrum than in the individual lines (shown in Fig. \ref{iktauso2lines}), although it is still not as bright as the central flux peak. The width of this broad component is in very good agreement with the expansion velocity of 17.5~\kms{} determined from single dish CO observations \citep{Decin2010,Maercker2016}. Unfortunately, the majority of the extended emission is still not discernible in the stacked channel maps, so we cannot confirm the true extent of the \so2 emission around IK~Tau.

% lines included in stacking: 338.306, 338.6118, 340.3164, 341.2755, 345.449, 346.6522, 348.3878, 350.8628, 351.2572, 351.8739, 355.0455, 356.7552, 357.1654, 357.2412, 357.3876, 357.6718, 357.8924, 358.2156, 359.1512, 359.7707

Since most of the \so2 flux is resolved out for most of the lines towards IK~Tau (with the possible exception of some of the highest energy and vibrationally excited lines, which we are unable to check in the absence of single-dish observations), we cannot extract much information through radiative transfer modelling. In comparing our observations with the model from \cite{Danilovich2016} the most notable result is the large amount of flux that was resolved out. To truely check the validity of the abundance profile used in that work and to conclusively determine the spatial extent of \so2 emission towards IK~Tau, spatially resolved observations that recover flux at larger scales are needed. The possible spatial distribution of \so2 around IK~Tau is discussed further in Sect. \ref{discussion}.

\subsubsection{Isotopologues}\label{iktauiso}

An examination of the channel maps of the less abundant SO isotopologues reveals that the extended emission seen for the main isotopologue towards IK~Tau is not present for $^{34}$SO. It is unlikely that this emission has been resolved out by ALMA since we have no reason to expect the $^{34}$SO emission to be larger in spatial extent than the $^{32}$SO emission. Instead, we suggest that the extended emission is so weak for $^{34}$SO that it has not been detected above the noise level. The same is true of $^{33}$SO, where the central emission is also much weaker than for the more abundant two isotopologues. In light of this, we advise caution when interpreting the following isotopologue ratios that we calculate from the ALMA data. For $^{32}$SO/$^{33}$SO we find $241\pm185$, for $^{32}$SO/$^{34}$SO we find $\sim42$, and for $^{33}$SO/$^{34}$SO we find $\sim0.1$. The two ratios involving $^{32}$S agree within the uncertainties with the results found by \cite{Danilovich2019} from SiS towards IK~Tau, while our $^{33}$SO/$^{34}$SO is about half of the $^{33}$S/$^{34}$S found in that study from SiS. These results indicate that the $^{32}$S and $^{34}$S isotopes are traced equally well by SO as by SiS.
For a more thorough analysis of S isotopologue ratios in IK~Tau's CSE we direct the interested reader to the isotopologue analysis in \cite{Danilovich2019} based on SiS from the same ALMA dataset.

\section{Discussion}\label{discussion}

\subsection{The co-location of SO and SO$_2$}\label{coloc}
% R Dor
\subsubsection{R Dor}

The two sulphur oxides SO and \so2 are co-located and trace out the same wind structures in the CSE of R~Dor 
(see Fig.~\ref{rdorso2onso} and  Sect.~\ref{rdorso2text}).
Although the SO features appear a little more extended, { with SO emission extending up to 0.5\arcsec{} further from the continuum peak than the \so2 emission}, { we cannot simply conclude that the \so2 emission is more compact since it is intrinsically less bright and hence weaker}
extended emission might not be detectable with the present sensitivity.
The co-location of SO and \so2 suggests that both molecules are formed in similar conditions and that one is not fully consumed in the production of the other.
The same density features traced by SO and \so2 are also seen in CO \citep{Decin2018}, despite a considerable amount of flux being resolved out for that molecule. For the HCN emission \citep{Decin2018}, the most central features (within $\sim0.5$\arcsec{} of the continuum peak) are different to those of SO and \so2, but there is a lot of similarity between the medium-scale features (around 0.5--1.5\arcsec) of HCN and SO and \so2. In contrast the SiO emission \citep{Decin2018} is smooth and does not trace out the same density structures. From this we can conclude that for a low mass-loss rate AGB star such as R~Dor, SO and \so2 are good tracers of density features, particularly in cases where CO may not be available. %In contrast, SiO is a poor tracer of density features.

% IK Tau
\subsubsection{IK Tau}

Since very little of the \so2 emission towards IK~Tau is detected by ALMA, we cannot directly make the same comparisons between SO and \so2 as we do for R~Dor. We note, however, that the stacked \so2 spectra shown in Fig. \ref{iktaustack} closely resemble the general shape of the SO spectra shown in Fig. \ref{iktausospectra} (albeit with less recovered emission for velocities further away from the $\upsilon_\mathrm{LSR}$). 
The stacked IK~Tau spectrum excluded the highest energy lines, since these are not expected to have much extended emission (as demonstrated in Fig. \ref{rdorso2centralchancomparison} for R~Dor). 

\begin{figure}
\centering
\includegraphics[width=0.5\textwidth]{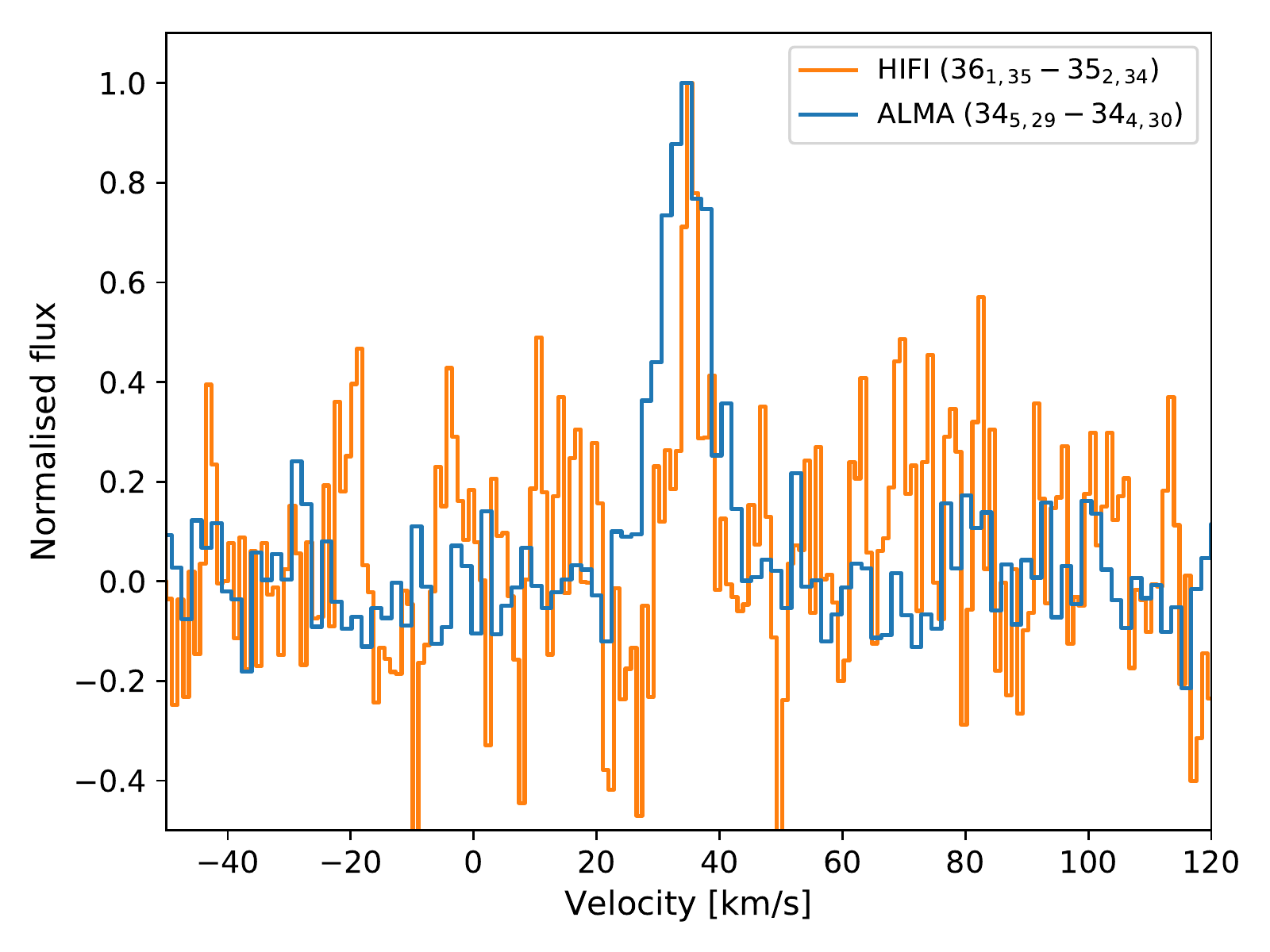}
\caption{\so2 lines of similar upper level energy, as observed with \textsl{Herschel}/HIFI (orange) and ALMA (blue). Both are normalised to facilitate the comparison of line shapes.}
\label{almahifi}
\end{figure}

When it comes to examining the line shapes of the higher-energy \so2 lines, we can look to the observations presented in \cite{Danilovich2016}, which include \so2 lines with a range of energies. The lower energy lines are all approximately the width of the \so2 ($5_{3,3}\to4_{2,2}$) APEX line plotted in Fig. \ref{iktaulostflux}, but the highest energy line in that study was \so2 ($36_{1,35}\to35_{2,34}$), observed with \textsl{Herschel}/HIFI, which has a much narrower line profile and an upper level energy of 606~K. To see whether this line shape is comparable with the higher energy lines observed with ALMA, we find the ALMA line with the upper level energy closest to the HIFI line, which is \so2 ($34_{5,29}\to 34_{4,30}$) with an upper level energy of 612~K, and plot the two normalised lines together in Fig. \ref{almahifi}. As shown there, the HIFI line appears narrower than the ALMA line, although that could be partly due to the lower signal-to-noise for the HIFI observation. The similarity between the two lines does suggest, however, that there may be less flux resolved out for the high energy ALMA line than for the lower-energy lines, since the line shape may not have been altered by the loss of flux. Unfortunately, even if this is the case, we still do not have adequate data to run comprehensive radiative transfer models for \so2 towards IK~Tau.

% IK Tau SO2 similar to SO?

The modelling results of \cite{Danilovich2016} found centrally peaked abundance distributions for both SO and \so2 for the low mass-loss rate stars in their sample, and SO abundance distributions that peaked further out in the wind for the higher mass-loss rate stars. However, they did not have sufficient \so2 observations to conclusively determine the shape of the abundance distribution for any of the higher mass-loss rate stars, and hence assumed a centrally peaked Gaussian distribution. Although our ALMA observations for IK~Tau do not allow us to check the shape of the abundance distribution, we will now consider the possibility that the \so2 distribution around IK~Tau is similar to the SO distribution, as is seen for the two sulphur oxides towards R~Dor. The shell-like peak in SO distribution for IK~Tau manifested itself in the channel maps as relatively smooth diffuse emission (see Fig. \ref{iktausochannels}). While we determined that there was no flux resolved out for at least one SO line (see Fig. \ref{iktaulostflux}), we can see from the channel maps that the smooth (albeit slightly noisy) features are close to 2\arcsec{} in diameter in the central channels. Since the largest angular scales for SO are close to the largest resolvable scale, if the \so2 emission was a bit smoother or a bit larger, then it would indeed be resolved out. Hence, a shell-like \so2 abundance distribution is consistent with the observations. 

Alternatively, consider that, for a given abundance, \so2 emission is intrinsically weaker than SO emission --- certainly when comparing with the bright SO line plotted in Fig. \ref{iktaulostflux}. If the \so2 and SO emission were distributed similarly, the diffuse \so2 emission may not be detectable with the present sensitivity. To check the plausibility of this hypothesis, we examined the flux density for $0.25\times0.25$\arcsec{} circular apertures centred 0.5\arcsec{} from the continuum peak in each of the four cardinal directions for the central channel of the SO ($8_8\to7_7$) line. We found that, in general, the enclosed flux density of these regions is about five times smaller than the flux density of the same size region centred on the continuum peak.
If we assume \so2 behaves analogously and examine the ($20_{1,19}\to19_{2,18}$) line in a similar way, we find that the flux density in a $0.25\times0.25$\arcsec{} region centred on the continuum peak is close to five times the rms noise. Hence, if a similar ratio between peak and diffuse emission is found for \so2 as for SO, the diffuse emission would not be detectable in the present dataset. Ergo, even if no flux of \so2 was resolved out, we might not have detected \so2 with the current sensitivity, on the assumption that the distributions of \so2 and SO are similar.

\subsubsection{Results from chemical modelling}

To better determine whether we should expect the \so2 distribution around IK~Tau to be similar to the SO distribution, we turn to chemical modelling. \cite{Danilovich2016} found that the SO and \so2 distributions for the low mass-loss rate stars (such as R~Dor) did not agree with chemical models such as those produced by \cite{Willacy1997}. For the higher mass-loss rate stars (such as IK~Tau) there was partial agreement with chemical models for SO. 

\begin{figure*}
\centering
\includegraphics[width=0.49\textwidth]{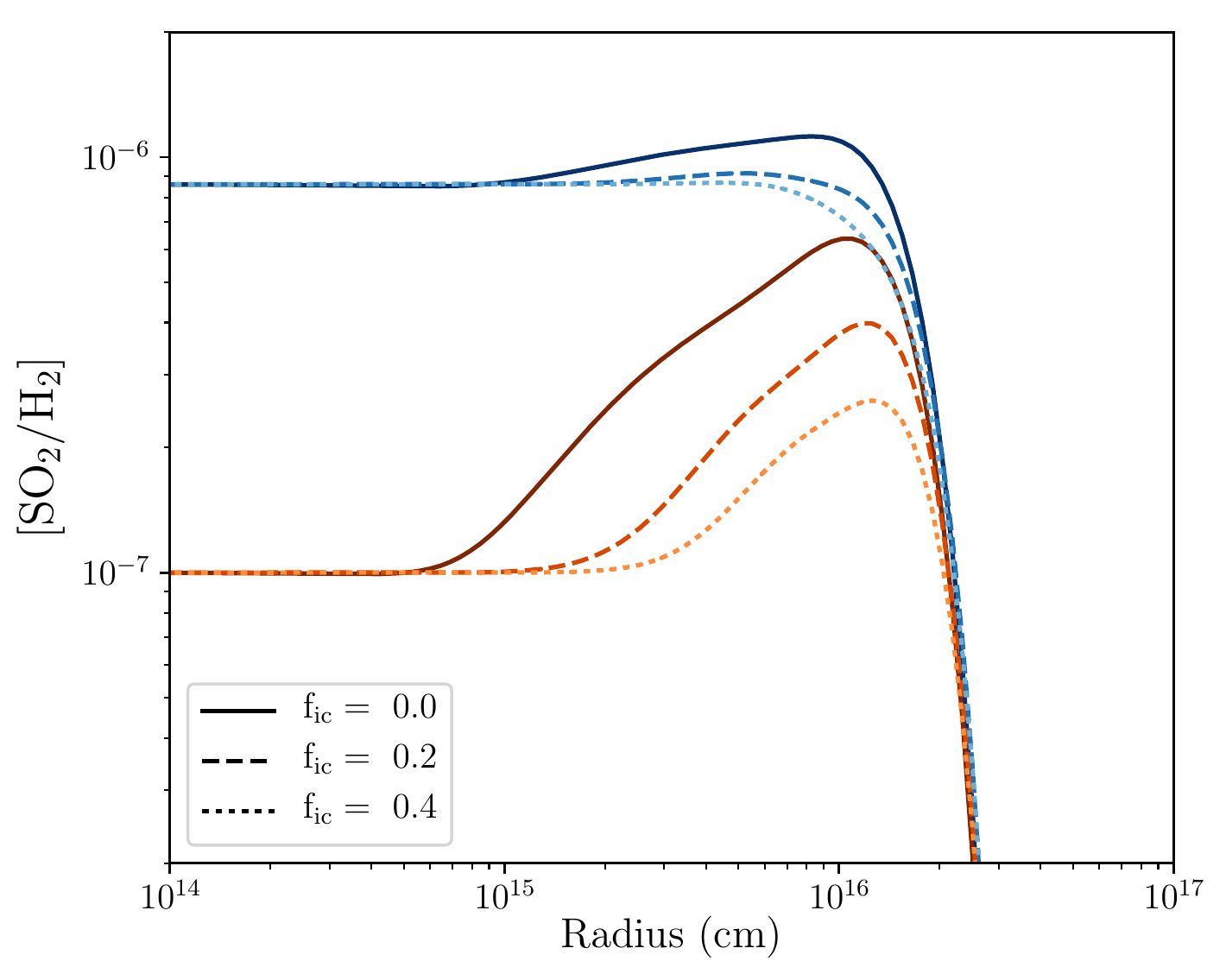}
\includegraphics[width=0.49\textwidth]{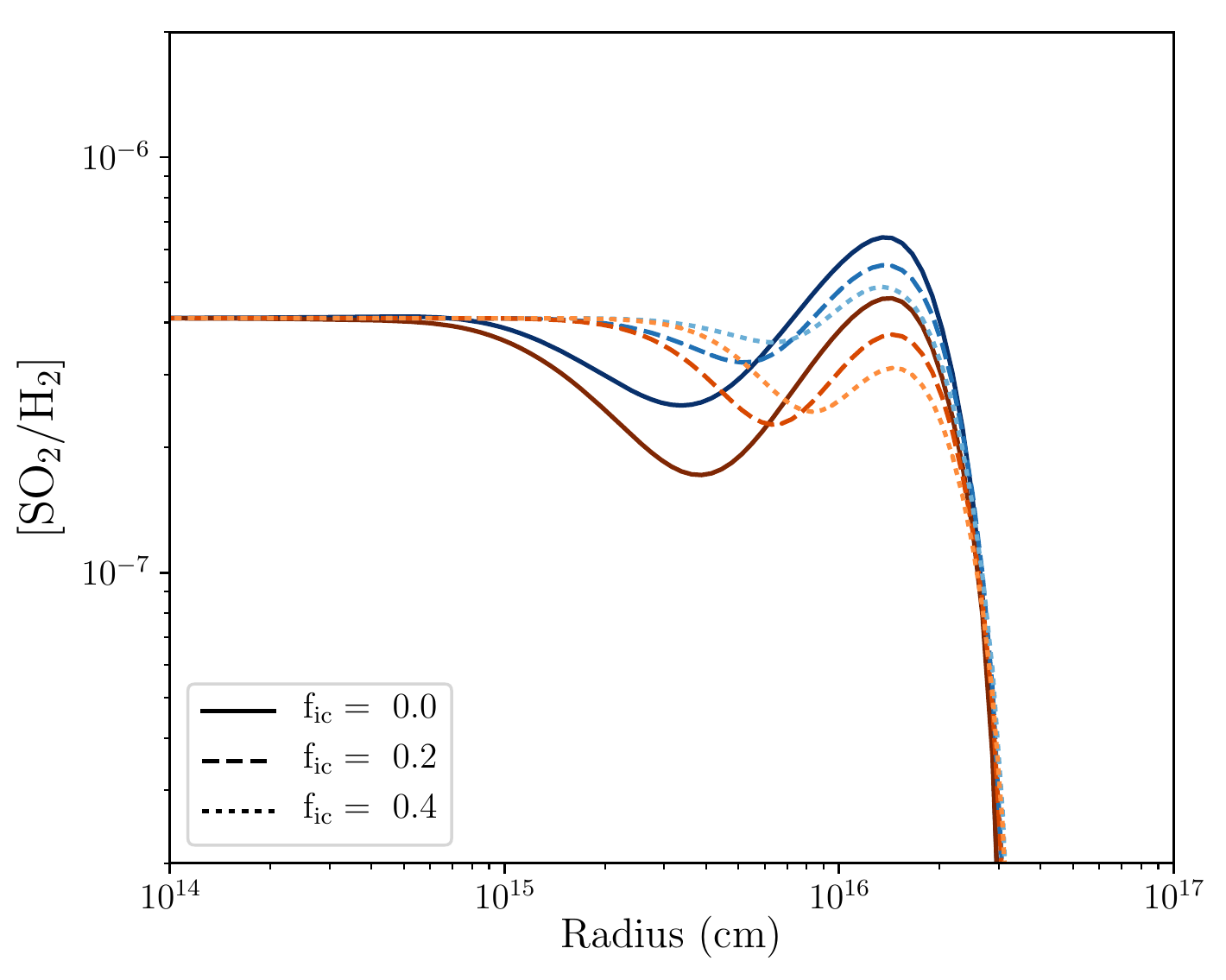}
\caption{The results of chemical models, taking different clumping parameters in the circumstellar envelope into account. \so2 results shown on the left and SO results shown on the right. In both cases, solid, dashed and dotted lines indicate different values of $f_\mathrm{ic}$, the amount of material in the inter-clump medium; the blue set of lines are for models with inner \so2 abundance of 8.7\e{-7} and the red set of models have an inner \so2 abundance of 1\e{-7} relative to \h2. The fraction of the total volume occupied by clumps is 0.2 for all models shown.}
\label{chemmod}
\end{figure*}

Updated chemical models from \cite{Van-de-Sande2018b} and \cite{Van-de-Sande2019} consider, in addition to the usual reactions, the effects of a clumpy circumstellar medium (by means of a statistical porosity formalism) and the role of stellar UV photons, respectively. The inclusion of stellar photons in the chemical model did not have an effect on the \so2 distribution. Testing the effects of a clumpy outflow, we found that, depending on the fraction of the total volume occupied by clumps and the amount of material in the inter-clump medium, the models predict an \so2 abundance profile reminiscent of the observationally-derived SO abundance profile shape. This was even more readily achieved if the inner abundance of \so2 (an input parameter to the model) was reduced to 1\e{-7} relative to \h2, rather than kept at the higher value of 8.7\e{-7} found by \cite{Danilovich2016}. Some example models showing this effect are shown in Fig. \ref{chemmod}, where we also include SO distributions for the same models. While the SO distributions do not perfectly agree with our observational results, they are qualitatively similar in most cases and suggest an increase in abundance at the same radial location as for \so2. Until more sensitive observations of \so2 are obtained, on the basis of the models it is reasonable to assume the distributions of \so2 and SO are similar in higher mass-loss rate oxygen-rich AGB stars.

\subsection{Isotopologues}\label{isotopediscussion}
% isotopes

In sections \ref{rdoriso} and \ref{iktauiso} for R~Dor and IK~Tau, respectively, we determined isotopologue ratios from our ALMA observations. Since the data quality is higher for R~Dor, we are able to draw more conclusions for that star than for IK~Tau. Rather than the S isotopologue ratios found here for IK~Tau, we consider the results of \citet{Danilovich2019} based on SiS observations to be more reliable.

Comparing our S isotope results with those for other AGB stars, we note that the R~Dor $^{33}$S/$^{34}$S ratio is in agreement, within uncertainties, with the ratio found for IK~Tau by \citet{Danilovich2019} using SiS observations. However, the $^{32}$S/$^{34}$S and $^{32}$S/$^{33}$S ratios are significantly lower than those found for IK~Tau. The $^{32}$S/$^{34}$S we find here is about half and the $^{32}$S/$^{33}$S is about a third of those found by \citet{Danilovich2019} towards IK~Tau (note the larger uncertainties for both stars on the latter ratio). The $^{32}$S/$^{33}$S for R~Dor is also about half that of the solar ratio given by \citet{Asplund2009}, while the $^{32}$S/$^{34}$S and $^{33}$S/$^{34}$S ratios are very close to the solar ratio. This might reflect that \ch{^{33}S} is the least abundant of the three isotopes studied here, and the less precise ratios result 
from lines with lower signal-to-noise. { We note that while $^{32}$S and $^{34}$S are primarily produced through explosive nucleosynthesis during Type II supernovae, the abundance of $^{33}$S may increase during the AGB phase via the slow neutron capture process \citep{Anders1989,Hughes2008}.}
While \citet{Danilovich2019} and \cite{Decin2010} concluded IK~Tau may have a lower metallicity than the Sun, the general consistency of our R~Dor results with the solar abundances suggests the metallicity is close to solar. { This is not a surprising result, since R~Dor is in close proximity to the Sun and hence was likely formed from a nebula enriched by early supernovae to a similar extent as the Sun.}

\section{Conclusions}

We presented and analysed ALMA observations of SO and \so2 and their isotopologues towards the oxygen-rich AGB stars R~Dor and IK~Tau. We note that the brightest lines --- which should be preferentially observed if investigating these molecules --- are those with the following quantum number designations. For SO, an allowed transition $N_J \to N'_{J'}$ is intrinsically bright if $N-J = N'-J'$. For \so2, the brightest transitions $J_{K_a,K_c} \to J_{K_a',K_c'}'$ are those which have $\Delta J = 0, -1$ and $\Delta K_a, \Delta K_c = -1,+1$.

For R~Dor we found that the new, spatially resolved observations of SO and \so2 agreed well with earlier models based on single-dish observations, and diverged mainly when the effects of the compact, rotating disc close to the central star dominated the emission.
The observations with ALMA confirm SO and \so2 trace out the same density structures in the CSE in R~Dor. 
%A comparison of the spatial and abundance distributions of SO and \so2 with those of other molecules, confirms that SO and \so2 are good indicators of underlying structures in low mass-loss rate oxygen-rich CSEs.
%We also examined the detections of isotopologues for SO and \so2 in the sensitive ALMA dataset. For R~Dor we calculated $^{32}$S/$^{33}$S $= 68\pm 22$ based on SO observations, $^{32}$S/$^{34}$S $= 18.5 \pm 5.8$ based on both SO and \so2 observations and $^{33}$S/$^{34}$S $= 0.17 \pm 0.02$ based on SO. On the basis of two tentatively identified lines of S$^{18}$O and an estimate of the upper limit for the abundance calculated with our radiative transfer model, we obtained a fractional abundance of $\leq (2\pm0.5)\times 10^{-8}$ and 
% $^{16}$O/$^{18}$O $\geq 335\pm84$ that is in agreement with earlier estimates.

For IK~Tau, we found that the ALMA observations of SO agreed with the general results from single-dish observations, and we were able to refine these further based on the spatially resolved and sensitive ALMA observations. For \so2 towards IK~Tau, we ascertained that the majority of the flux was resolved out for ALMA, especially for the low-energy reference line of ($5_{3,3}\to 4_{2,2}$). 
%This precluded us from testing any \so2 models against the new observations. 
Earlier results assume a centrally-peaked Gaussian abundance distribution for \so2, but our results are compatible with a shell-like abundance distribution similar in shape to the SO distribution. Some recently-developed chemical models agree with this hypothesis, but spatially resolved observations with a larger resolvable scale are needed to confirm.

%The less sensitive observations of SO towards IK~Tau resulted in a higher uncertainty in the derived isotopic ratios than in R~Dor. 
%Owing to the large amounts of flux resolved out for the lines of \so2, we were unable to derive isotopic ratios for \so2
%in IK~Tau.
%From measurements of the isotopologues of SO, we found $^{32}$S/$^{33}$S $= 241\pm 185$, 
%$^{32}$S/$^{34}$S $\sim 42$, and $^{33}$S/$^{34}$S $\sim 0.1$. 
%The first two ratios are in agreement with earlier determinations derived from observations of SiS, but the latter ratio differs 
%by a factor of two, possibly owing to the noisy faint emission from the less abundant isotopologues.

%Overall, the presented ALMA observations of SO and \so2 around a low mass-loss rate and a high mass-loss rate oxygen-rich AGB star confirm the significantly different SO abundance distributions between these two classes of AGB stars. Our results agree with the earlier single-dish results wherein the fractional abundance distribution of SO peaks close to the star for low mass-loss rate stars and peaks further out in the envelope for high mass-loss rate stars. Although our \so2 data was incomplete for the higher mass-loss rate star, it is possible that the same is true for \so2.

Our observations of SO and \so2 around the two prototypical oxygen-rich AGB stars --- R~Dor with a low mass-loss rate 
and IK~Tau with a high mass-loss rate --- confirm the abundance distributions of SO in these two classes of AGB stars
differ significantly.
{ We conclude that an adequate set of SO observations could indeed be used as a secondary diagnostic of mass-loss rate in cases of uncertainty.}
%The earlier conclusion that the fractional abundance distribution of SO peaks close to the star for low mass-loss rate stars and peaks farther out in the envelope for high mass-loss rate stars which was derived from single antenna observations is confirmed by our observations of R~Dor and IK~Tau with ALMA.
Although our measurements of \so2 in the high mass-loss rate star with ALMA are incomplete, it appears the abundance
distribution of \so2 might also follow a shell-like abundance distribution, similar to that of SO.

\section*{Acknowledgements}

TD and MVdS acknowledge support from the Research Foundation Flanders (FWO) through grants 12N9917N \& 12X6419N, respectively.
LD acknowledges support from the ERC consolidator grant 646758 AEROSOL and the FWO Research Project grant G024112N.
CAG acknowledges partial support from NSF grant AST-1615847.
This paper makes use of the following ALMA data: ADS/JAO.ALMA2013.0.00166.S. ALMA is a partner- ship of ESO (representing its member states), NSF (USA) and NINS (Japan), together with NRC (Canada) and NSC and ASIAA (Taiwan), in cooperation with the Republic of Chile. The Joint ALMA Observatory is operated by ESO, AUI/NRAO and NAOJ.
%Based on observations made with APEX under programme IDs O-097.F-9318 and O-098.F-9305. APEX is a collaboration between the Max-Planck-Institut f\"ur Radioastronomie, the European Southern Observatory, and the Onsala Space Observatory. 
Based on observations with the Atacama Pathfinder EXperiment (APEX) telescope under programme IDs O-097.F-9318 and O-098.F-9305. APEX is a collaboration between the Max Planck Institute for Radio Astronomy, the European Southern Observatory, and the Onsala Space Observatory. Swedish observations on APEX are supported through Swedish Research Council grant No 2017-00648.
HIFI has been designed and built by a consortium of institutes and university departments from across Europe, Canada and the United States under the leadership of SRON Netherlands Institute for Space Research, Groningen, The Netherlands and with major contributions from Germany, France and the US. Consortium members are: Canada: CSA, U. Waterloo; France: CESR, LAB, LERMA, IRAM; Germany: KOSMA, MPIfR, MPS; Ireland, NUI Maynooth; Italy: ASI, IFSI-INAF, Osservatorio Astrofisico di Arcetri-NAF; The Netherlands: SRON, TUD; Poland: CAMK, CBK; Spain: Observatorio Astron\'omico Nacional (IGN), Centro de Astrobiolog\'ia (CSIC-INTA). Sweden: Chalmers University of Technology -- MC2, RSS \& GARD; Onsala Space Observatory; Swedish National Space Board, Stockholm University -- Stockholm Observatory; Switzerland: ETH Zurich, FHNW; USA: Caltech, JPL, NHSC.

%%%%%%%%%%%%%%%%%%%%%%%%%%%%%%%%%%%%%%%%%%%%%%%%%%



%\bibliographystyle{mnras}

%\begin{thebibliography}{99}
%\end{thebibliography}

\bibliographystyle{mnras}
%\bibliography{../../AGBpapers}
\bibliography{ALMASOSO2}

%%%%%%%%%%%%%%%%%%%%%%%%%%%%%%%%%%%%%%%%%%%%%%%%%%

%%%%%%%%%%%%%%%%% APPENDICES %%%%%%%%%%%%%%%%%%%%%

\appendix

\section{Additional SO plots and data}\label{app:soplots}

\begin{figure*}
\centering
\includegraphics[width=\textwidth]{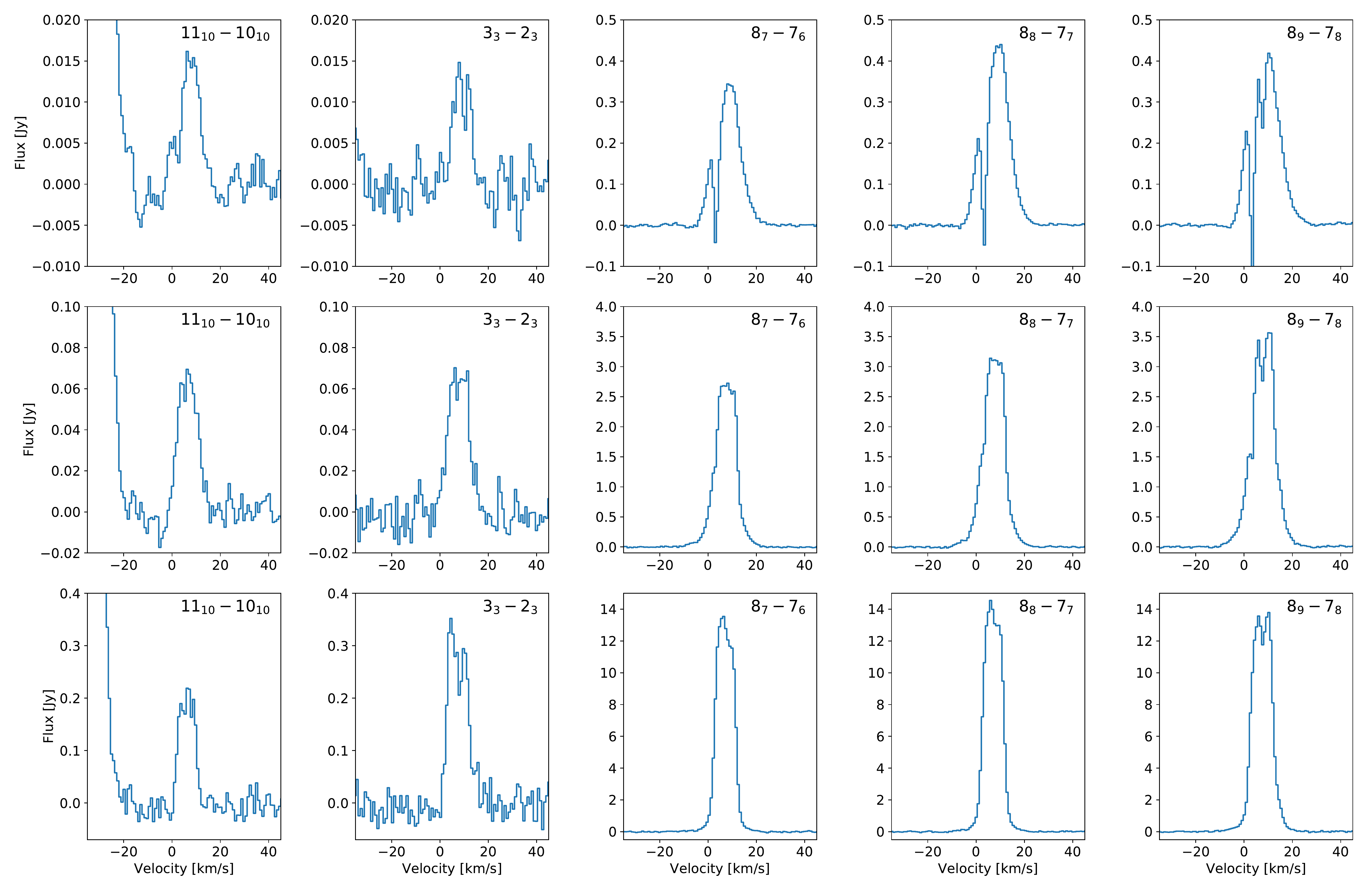}
\caption{ALMA SO transitions towards R~Dor, extracted for different apertures centred on the continuum peak. \textit{Top}: 75 mas spectra; \textit{middle}: 300 mas spectra; \textit{bottom}: 1\arcsec{} spectra.}
\label{rdorsospectra}
\end{figure*}

%\begin{figure*}
%\begin{center}
%\includegraphics[width=\textwidth]{rdor-s18o0.pdf}
%\caption{Observations (\textit{black histograms}) and model results (\textit{red curves}) for S$^{18}$O towards R~Dor. The radius of the extraction aperture is listed in the top right corner of each plot.}
%\label{S18O}
%\end{center}
%\end{figure*}

\begin{figure*}
\centering
\includegraphics[width=\textwidth]{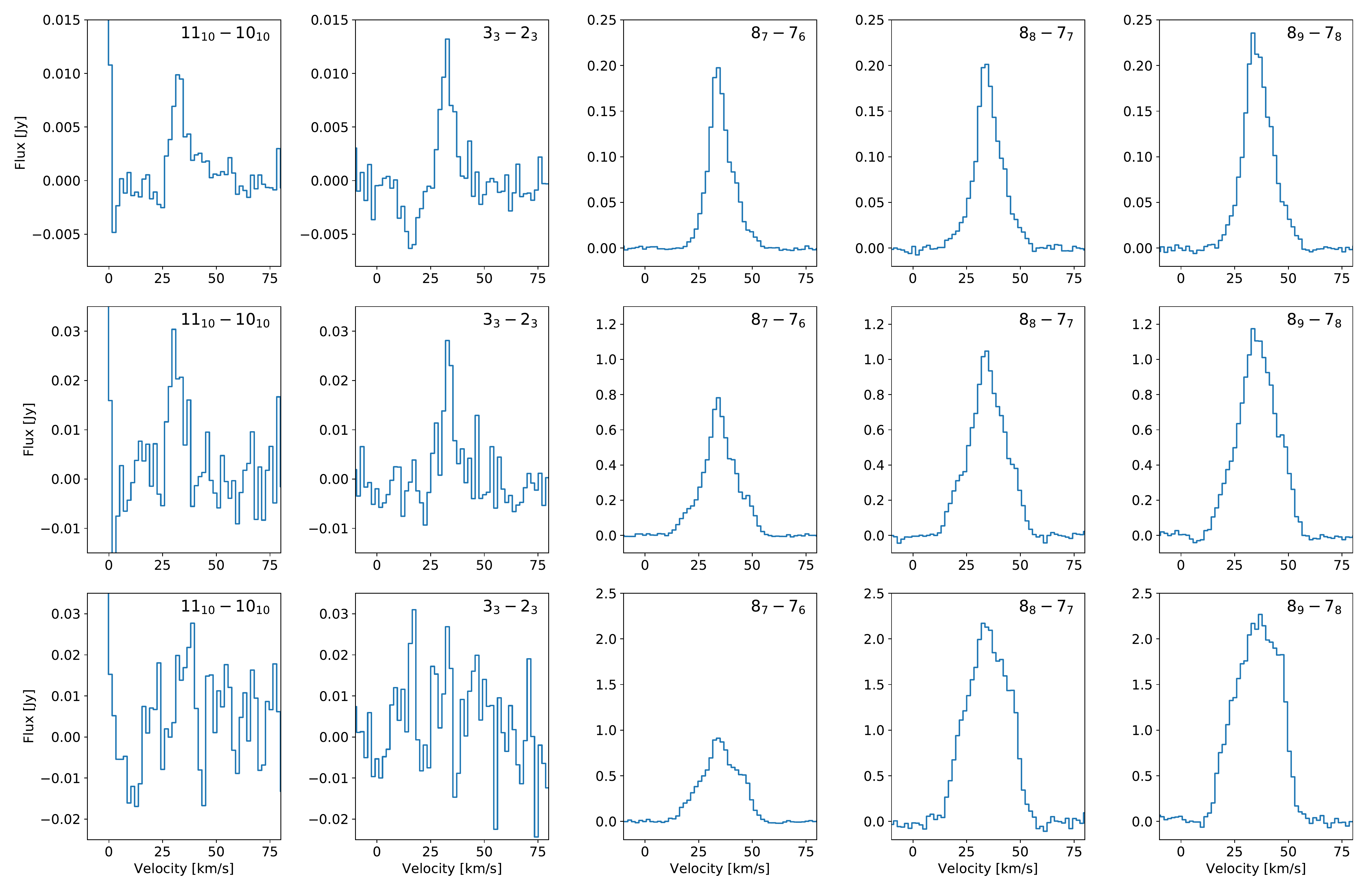}
\caption{ALMA SO transitions towards IK~Tau, extracted for different apertures centred on the continuum peak. \textit{Top}: 80 mas spectra; \textit{middle}: 320 mas spectra; \textit{bottom}: 800 mas spectra.}
\label{iktausospectra}
\end{figure*}

\begin{table}
\caption{SO lines observed with APEX towards IK~Tau}\label{iktauapexso}
\begin{center}
\begin{tabular}{ccccc}
\hline\hline
 Frequency & Line  & $E_\mathrm{up}$ & $\theta_\mathrm{HPBW}$ & $I_\mathrm{mb}$ \\
 $\mathrm{[GHz]}$ & & [K] & [$\arcsec$]& [K~\kms]\\
 \hline
178.605$^a$ & $4_5\to3_4$ & 24 &35	& 3.2	\\
206.176$^a$ & $5_4\to4_3$ & 39 & 30	& 1.3 \\
219.949$^a$ & $5_6\to4_5$ & 35 & 28	&3.7	\\
251.826$^a$ & $6_5\to5_4$ & 51 & 25	&1.2	\\
301.286$^a$ & $7_7\to6_6$ & 71 & 21	& 1.1	\\
304.078$^a$ & $7_8\to6_7$& 62 & 21	& 2.6	\\
344.311$^a$ & $8_8\to7_7$ & 88 & 18 & 1.2\\
\hline
\end{tabular}
\end{center}
\textbf{References:} ($^a$) Measured frequencies from \cite{Clark1976}.
\end{table}	

\section{Additional SO$_2$ plots and data}\label{app:so2plots}

\begin{table}
\caption{SO$_2$ ($\nu_2=1$) vibrationally excited lines detected with ALMA}\label{so2v1linelist}
\centering
\begin{tabular}{ccrcl}
\hline\hline
 Frequency & Line  & $E_\mathrm{up}$ & Star & Notes\\
 $\mathrm{[GHz]}$ & & [K] & & \\
 \hline
     335.1285$^a$ &  $20_{ 4,16} \to 20_{ 3,17}$ & 980 & both &  \\
     335.4777$^b$ & $44_{5,39} \to 44_{4,40}$ & 1732 & both\\
     336.0324$^b$ & $38_{5,33} \to 38_{4,34}$ & 1496 & R Dor \\
     336.7607$^a$ &  $20_{ 1,19} \to 19_{ 2,18}$ & 945 & both &  \\
     337.3499$^b$ &  $57_{ 6,52} \to 56_{ 7,49}$ & 2365 & R~Dor & very weak, ID uncertain \\
     337.8925$^b$ & $21_{2,20} \to 21_{1,21}$ & 966 & both & SO, $^{34}$SO$_2$, TiO$_2$ overlap\\
     338.3487$^b$ &  $ 4_{ 3, 1} \to  3_{ 2, 2}$ & 778 & R~Dor &  \\
     338.3764$^b$ &  $ 8_{ 2, 6} \to  7_{ 1, 7}$ & 789 & R~Dor &  \\
     342.4359$^a$ &  $23_{ 3,21} \to 23_{ 2,22}$ & 1022 & both &  \\
     343.9238$^a$ &  $24_{ 2,22} \to 23_{ 3,21}$ & 1039 & both &  \\
     344.6137$^b$ &  $28_{ 2,26} \to 28_{ 1,27}$ & 1138 & both &  \\
     344.9742$^b$ &  $40_{ 4,36} \to 40_{ 3,37}$ & 1555 & both &  \\
     346.3653$^b$ &  $34_{ 3,31} \to 34_{ 2,32}$ & 1328 & both & \so2 ($v=1$) overlap \\
     346.3792$^a$ &  $19_{ 1,19} \to 18_{ 0,18}$ & 914 & both & \so2 ($v=1$) overlap \\
     346.5918$^a$ &  $18_{ 4,14} \to 18_{ 3,15}$ & 943 & R~Dor &  \\
     347.9918$^a$ &  $13_{ 2,12} \to 12_{ 1,11}$ & 839 & R~Dor &  \\
     351.2900$^b$ &  $36_{ 5,31} \to 36_{ 4,32}$ & 1426 & both & \so2 overlap \\
     351.9824$^a$ &  $16_{ 7, 9} \to 17_{ 6,12}$ & 994 & R~Dor & very weak\\
     354.6242$^b$ &  $46_{ 5,41} \to 46_{ 4,42}$ & 1819 & both &  \\
     354.8000$^a$ &  $16_{ 4,12} \to 16_{ 3,13}$ & 911 & both & \h2O overlap \\
     357.0872$^b$ &  $ 5_{ 3, 3} \to  4_{ 2, 2}$ & 782 & both & R: $^{34}$\so2 overlap \\
     &&&& I: very weak\\
     357.6026$^a$ &  $20_{ 0,20} \to 19_{ 1,19}$ & 931 & R~Dor &  \so2 overlap\\
      &&&& I: SiS $v=3$ overlap\\
     358.8719$^a$ &  $21_{ 8,14} \to 22_{ 7,15}$ & 1119 & R~Dor &  \\
     360.1332$^b$ &  $14_{ 4,10} \to 14_{ 3,11}$ & 883 & R~Dor &  \\
\hline
\end{tabular}
{ References:} ($^a$) \cite{Lovas1985}; ($^b$) \cite{Muller2005a} \& \cite{Muller2005}.
\end{table}	

\begin{figure}
\centering
\includegraphics[width=0.5\textwidth]{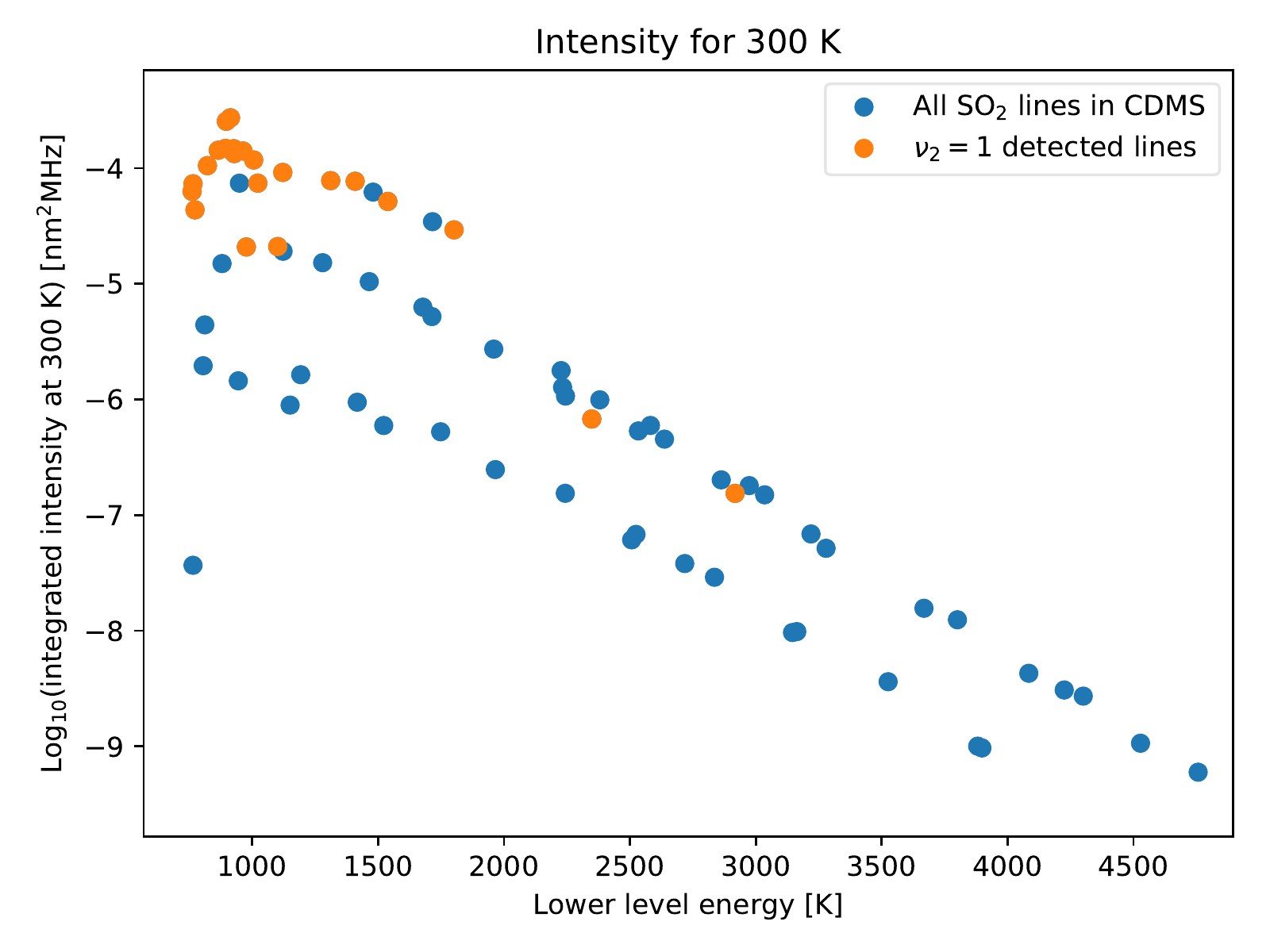}
\caption{Calculated integrated intensities at 300~K versus the lower state energy for all the $\nu_2=1$ \so2 lines in CDMS that fall within the frequency range of our spectral line scan. The detections indicated are for R~Dor.}
\label{so2EvsIv1}
\end{figure}

\begin{figure*}
\begin{center}
\includegraphics[width=\textwidth]{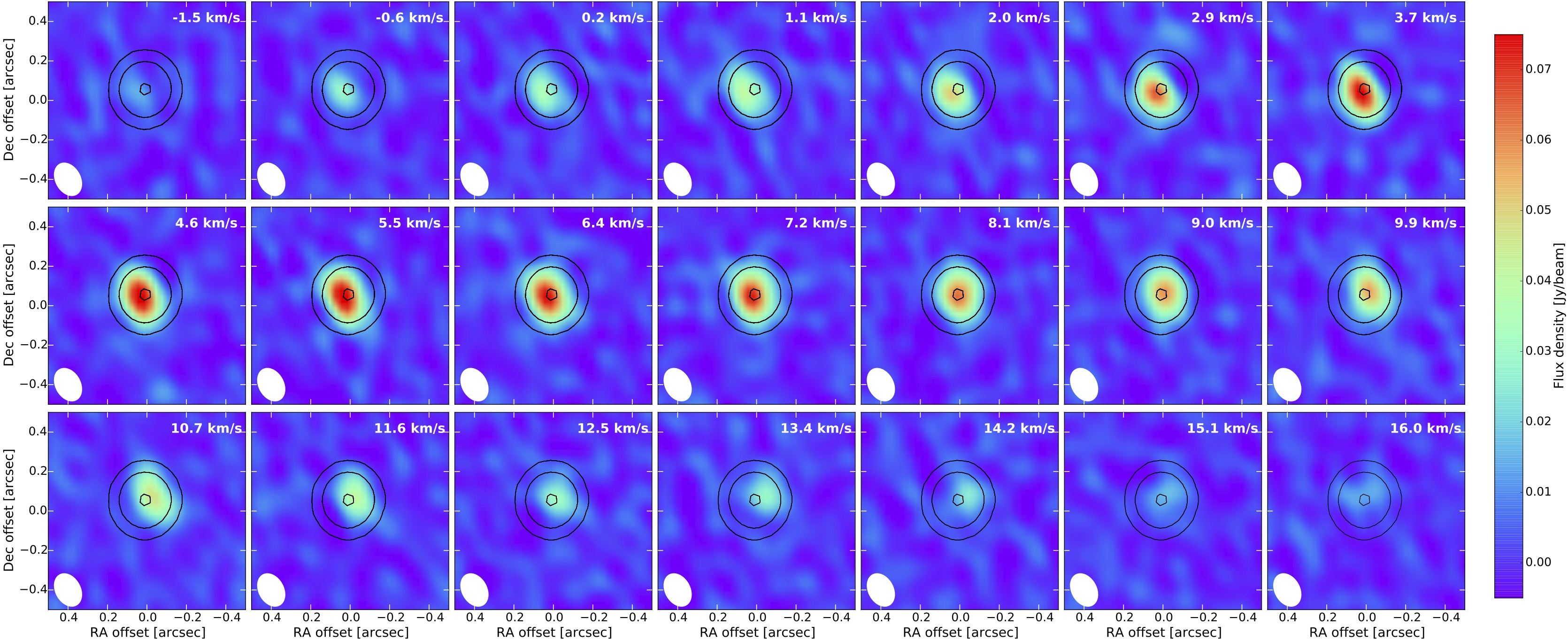} %final is pdf
\caption{R~Dor channel maps for SO$_2$ $v=1$ ($20_{4,16} \to 20_{3,17}$) at 335.1285 GHz. The solid black lines indicate the continuum emission levels at 1, 10 and 90\% of the continuum peak flux. The beam is indicated in white in the bottom left corner of each channel plot. Plots are best viewed on a screen.}
\label{rdorso2v1chanmaps}
\end{center}
\end{figure*}

\begin{figure*}
\begin{center}
\includegraphics[width=\textwidth]{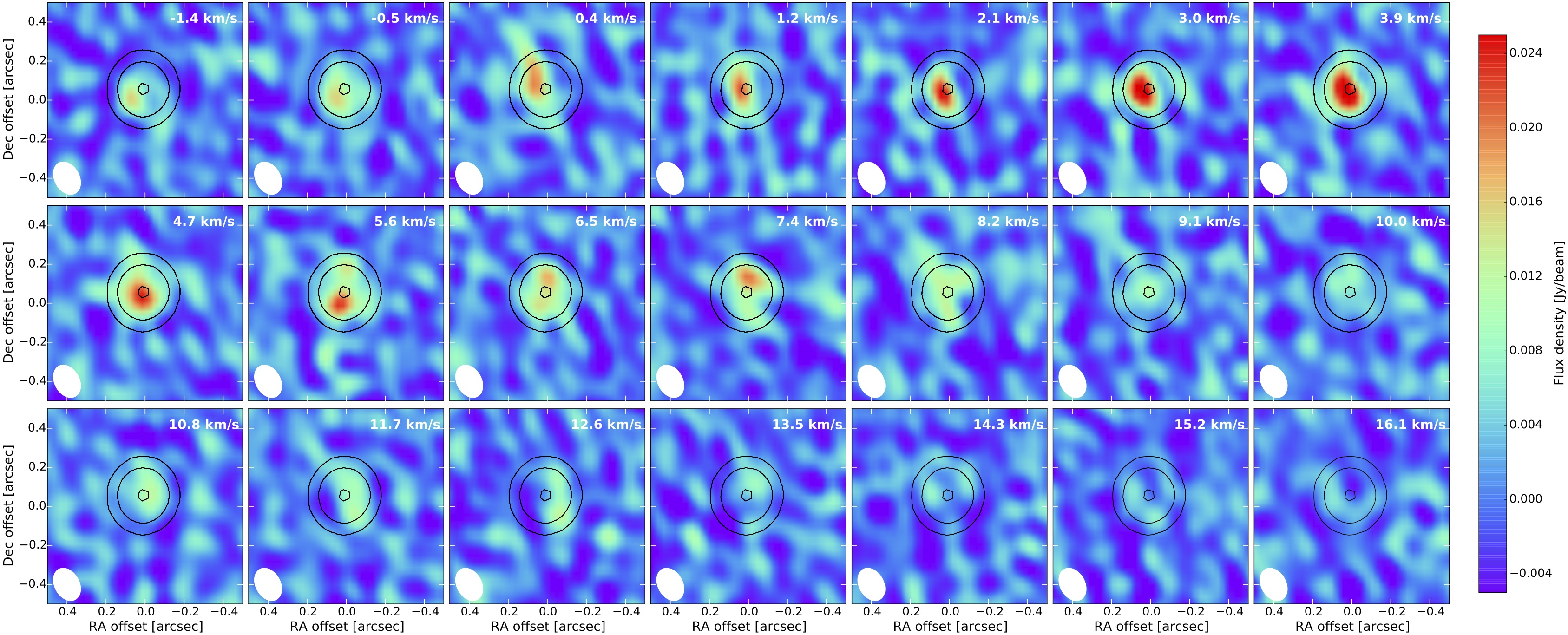} %final is pdf
\caption{R~Dor channel maps for SO$_2$ ($29_{ 5,25} \to 30_{ 2,28}$) at 335.7732 GHz. The solid black lines indicate the continuum emission levels at 1, 10 and 90\% of the continuum peak flux. The beam is indicated in white in the bottom left corner of each channel plot. Plots are best viewed on a screen.}
\label{rdoroffcentreso2chanmaps}
\end{center}
\end{figure*}

\begin{table}
\caption{SO$_2$ isotopologue lines detected with ALMA. Only $^{34}$\so2 lines with confident IDs are included. All oxygen isotopologue IDs are uncertain.}\label{34so2linelist}
\centering
\begin{tabular}{ccccl}
\hline\hline
 Frequency & Line  & $E_\mathrm{up}$ & Star & Notes\\
 $\mathrm{[GHz]}$ & & [K] & &\\
 \hline
\multicolumn{2}{c}{$^{34}$\so2\phantom{slkdfsdf}}\\
     342.2316$^a$ &  $20_{ 1,19} \to 19_{ 2,18}$ & 198 & R~Dor & \\
     342.3320$^a$ &  $12_{ 4, 8} \to 12_{ 3, 9}$ & 110 & R~Dor & \\
     344.5810$^c$ &  $19_{ 1,19} \to 18_{ 0,18}$ & 168 & both & \\
     347.4831$^b$ &  $28_{ 2,26} \to 28_{ 1,27}$ & 391 & R~Dor & \\
     348.1175$^b$ & $19_{ 4,16} \to 19_{ 3,17}$ & 213 & R~Dor & TiO overlap\\
     352.0829$^b$ & $21_{4,18}\to21_{3,19}$ & 251 & IK Tau & poss. U overlap\\
     353.9499$^c$ &  $40_{ 4,36} \to 40_{ 3,37}$ & 807 & R~Dor & \\
     354.2776$^c$ &  $34_{ 3,31} \to 34_{ 2,32}$ & 581 & R~Dor & \\
     356.2224$^c$ &  $25_{ 3,23} \to 25_{ 2,24}$ & 320 & R~Dor & \\
     357.1022$^c$ &  $20_{ 0,20} \to 19_{ 1,19}$ & 185 & R~Dor & \so2 ($v=1$) overlap \\
      &&&& I: $^{30}$Si$^{34}$S overlap\\
     357.4977$^c$ &  $32_{ 5,27} \to 32_{ 4,28}$ & 547 & R~Dor & \\
\hline
\multicolumn{2}{c}{SO$^{17}$O\phantom{asdf}} \\
     345.4221$^d$ & $ 5_{ 3, 3} \to  4_{ 2, 2}$ & 35 & R~Dor & \\
     351.7014$^d$ & $35_{ 5,30} \to 35_{ 4,31}$ & 627 & R~Dor & \\
     361.5113$^d$ & $33_{ 2,31} \to 34_{ 1,34}$ & 499 & R~Dor & \\
     361.9885$^d$ & $26_{ 3,24} \to 26_{ 2,25}$ & 341 & R~Dor & \\
\hline
\end{tabular}
{ References:} ($^a$) \cite{Lovas1985}; ($^b$) \cite{Belov1998}; ($^c$) \cite{Muller2005}; ($^d$) \cite{Muller2001} \& \cite{Muller2005}
\end{table}	

\begin{figure*}
\centering
\includegraphics[width=\textwidth]{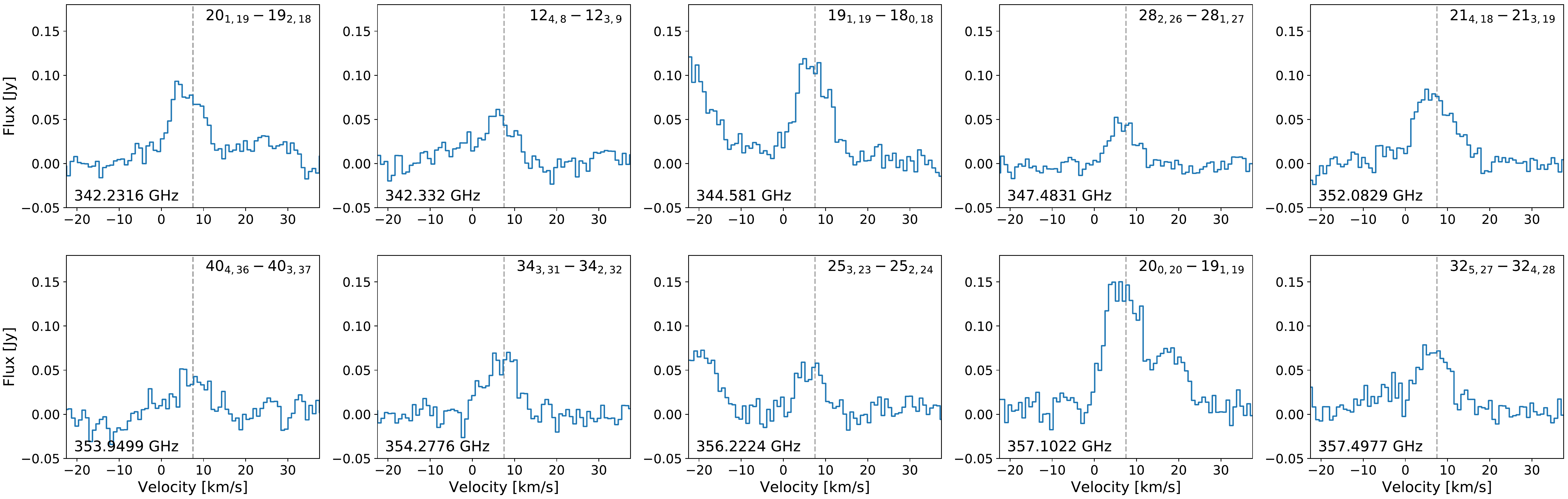}
\caption{$^{34}$\so2 spectra observed towards R~Dor with ALMA. 
The spectral lines have been extracted with a 300~mas circular aperture centred on the stellar continuum peak. 
The quantum numbers are in the top right corner and the frequencies in the bottom left corner. 
The $\upsilon_\mathrm{LSR}$ of 7.5~\kms{} is indicated by the grey line. The additional peak to the red of the 357.1022~GHz line is the 357.0872~GHz vibrationally excited $^{32}$\so2 line.}
\label{rdor34so2}
\end{figure*}

%\begin{landscape}
\begin{figure*}
\begin{center}
\includegraphics[width=\textwidth]{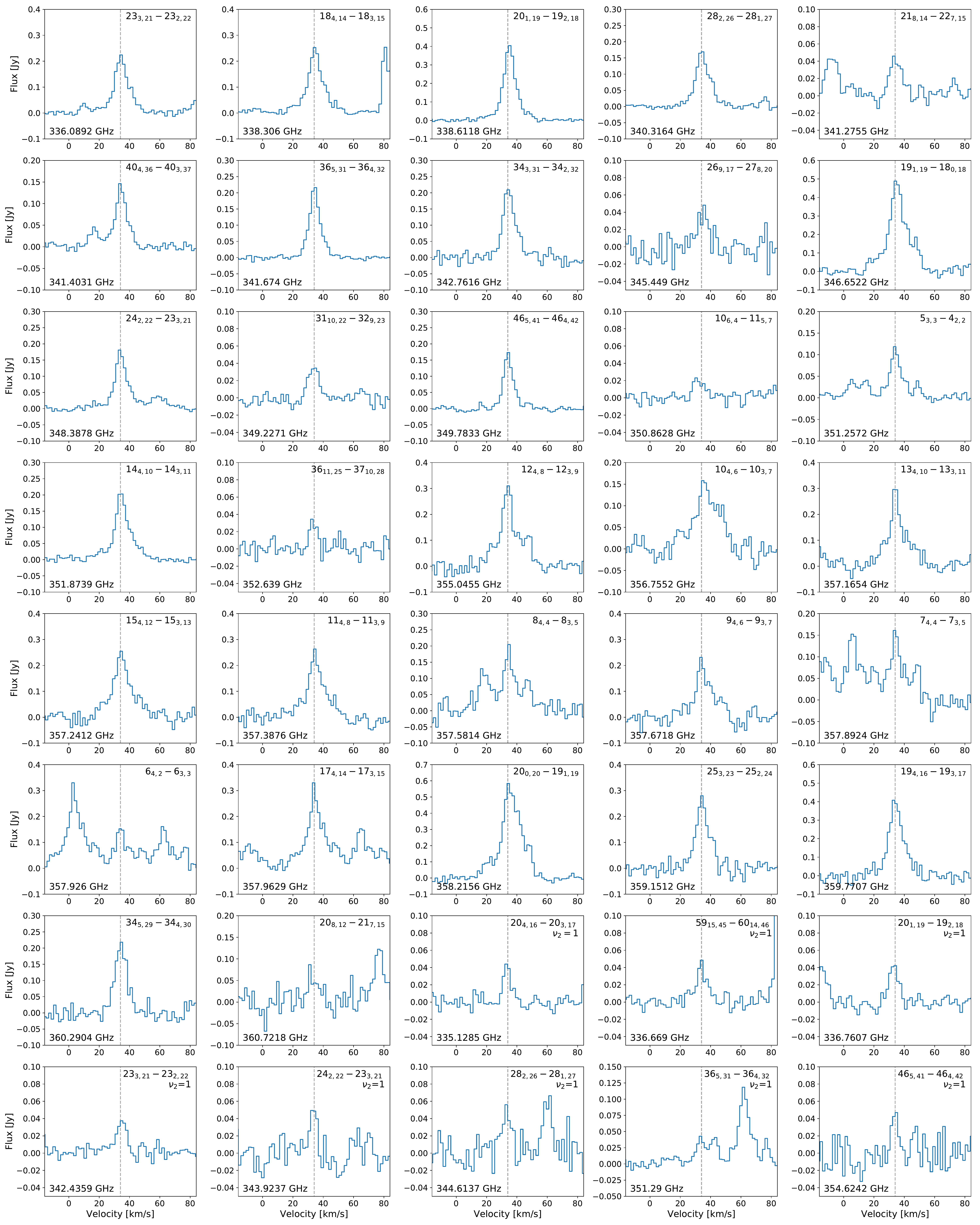}
\caption{\so2 spectra observed towards IK~Tau with ALMA. 
The spectral lines have been extracted with a 320~mas circular aperture centred on the stellar continuum peak. 
The quantum numbers are in the top right corner and the frequencies in the bottom left corner. 
The $\upsilon_\mathrm{LSR}$ of 34~\kms{} is indicated by the grey line.
}
\label{iktauso2lines}
\end{center}
\end{figure*}
%\end{landscape}

%\section{More plots}
%
%\begin{figure*}
%\begin{center}
%\includegraphics[width=\textwidth]{iktau-so0.pdf}
%\caption{The emission lines from our new SO model (red lines) plotted against single dish data \protect\citep[see][for details]{Danilovich2016} and ALMA spectral lines (black histograms). The selected ALMA spectra are extracted for a 1\arcsec{} radius aperture.}
%\label{iktauSOmodellines}
%\end{center}
%\end{figure*}

%\begin{figure}
%\begin{center}
%\includegraphics[width=0.5\textwidth]{iktau-so-positions.pdf}
%\caption{The radial profile...
%\todo[inline]{finish this caption}}
%\label{iktauSOradprof}
%\end{center}
%\end{figure}

%%%%%%%%%%%%%%%%%%%%%%%%%%%%%%%%%%%%%%%%%%%%%%%%%%

% Don't change these lines
\bsp	% typesetting comment
\label{lastpage}
\end{document}